\documentclass[12pt]{iopjournal}

\usepackage[T1]{fontenc}
\usepackage[utf8]{inputenc}
\usepackage[english]{babel}

\usepackage[version=4]{mhchem}
\usepackage{siunitx}
\usepackage{graphicx}
\usepackage{booktabs}
\usepackage{longtable}
\usepackage{amsmath,amssymb,mathtools}
\usepackage{makecell}
\usepackage{subcaption}

\usepackage{hyperref}
\hypersetup{colorlinks=true,linkcolor=blue,citecolor=blue,urlcolor=blue}

\begin{document}

\title{Hydrogen storage in pristine and Janus transition-metal dichalcogenide monolayers: electronic origins, coverage effects, and finite-temperature stability}

\author{Fl\'avio Bento de Oliveira$^{1}$\orcid{0000-0002-7411-0099},Gabriel Elyas Gama Ara\'ujo$^{1}$\orcid{0009-0005-4675-8807} and Andreia Luisa da Rosa$^1$\orcid{0000-0002-2780-6448}}
\affil{$^{1}$Federal University of Goi\'as, Institute of Physics, Campus Samambaia, 74960600 Goi\^ania, Goi\'as, Brazil}
\email{andreialuisa@ufg.br}

	
\begin{abstract}
Hydrogen adsorption and storage in two-dimensional transition-metal
dichalcogenides are governed by a subtle interplay between electronic
structure, lattice geometry, and surface chemistry. Here, we present a
systematic first-principles study of hydrogen adsorption on pristine
and Janus MX$_2$ and MSSe monolayers (M = Ni, Pd, Pt; X = S, Se),
combining density-functional theory calculations with
finite-temperature \textit{ab initio} molecular dynamics
simulations. Orbital-resolved electronic-structure analysis reveals
that hydrogen binding strength is controlled primarily by the
contribution of metal $d$ states near the Fermi level, which increases
from Pd and Pt to Ni.  Chalcogen substitution from S to Se further
modulates adsorption by raising the valence-band manifold and
enhancing surface polarizability.  Structural polymorphism plays a
decisive role: metallic 1T phases promote stronger single-molecule
adsorption due to enhanced electronic screening, whereas the more open
trigonal-prismatic coordination of the 2H phase provides greater
accessible volume and configurational freedom at high hydrogen
coverage. Janus functionalization breaks out-of-plane symmetry and
introduces chemically inequivalent S- and Se-terminated surfaces,
leading to side-dependent adsorption energies, hydrogen-layer
thicknesses, and spatial distributions. While Janus asymmetry enables
fine tuning of adsorption strength, hydrogen uptake remains dominated
by molecular physisorption across all systems. Finite-temperature
\textit{ab initio} molecular dynamics simulations at 300~K demonstrate
that hydrogen remains strictly molecular even at high loadings, with
no dissociation or irreversible chemisorption. Among the materials
studied, 2H-NiSSe and 2H-PdSSe exhibit the most favorable combination
of moderate adsorption energies, stable multilayer hydrogen
configurations, and robust thermal stability, whereas Pt-based Janus
systems display overly strong confinement and reduced
reversibility. These results establish clear design principles for
hydrogen storage in two-dimensional dichalcogenides, showing that
optimal performance arises from intermediate physisorption rather than
maximal binding strength and highlighting Janus Ni- and Pd-based
systems as promising platforms for reversible molecular hydrogen
storage.
\end{abstract}

\section{Introduction}

The development of efficient and reversible hydrogen storage materials
remains a central challenge in condensed matter physics and materials
science, driven by the need for sustainable energy carriers compatible
with low-carbon technologies \cite{Schlapbach2001,Crabtree2004}.
Molecular hydrogen possesses an exceptionally high gravimetric energy
density. However, its practical use is limited by the lack of materials
that can adsorb and release H$_2$ under ambient conditions with
sufficient capacity, reversibility, and kinetic accessibility
\cite{Zuttel2003,Graetz2009}. Achieving this balance requires adsorption
energies that are neither too weak to retain hydrogen nor too strong to
impede desorption, placing constraints on candidate materials
\cite{Bhatia2006}.

Two-dimensional materials have emerged as promising platforms for
hydrogen storage due to their high surface-to-volume ratio, tunable
electronic structure, and structural flexibility
\cite{Novoselov2005,Geim2007}. In particular, transition-metal
dichalcogenides (TMDs) exhibit a rich combination of chemical diversity,
polymorphism, and electronic behavior, ranging from semiconducting to
metallic \cite{Manzeli2017,Chhowalla2013}. These properties make TMDs
attractive model systems for exploring how electronic structure and
lattice geometry govern molecular adsorption at the atomic scale.

Hydrogen adsorption on pristine TMD monolayers is typically dominated by
physisorption, with binding strengths that depend sensitively on the
transition-metal center, chalcogen species, and structural phase
\cite{Chhowalla2013,Manzeli2017}. Metallic phases, such as the octahedral 1T
polymorph, tend to enhance adsorption through increased electronic
screening and polarization effects, whereas semiconducting 2H phases
often exhibit weaker interactions \cite{Chhowalla2013}. At the same time,
stronger binding does not necessarily translate into improved storage
performance, since excessively strong adsorption can suppress
reversibility and limit hydrogen mobility at finite temperature
\cite{Bhatia2006}.

Beyond composition and phase, symmetry engineering offers an additional
route to tuning adsorption behavior. Janus TMDs, in which the two chalcogen layers are chemically distinct,
break out-of-plane mirror symmetry and generate an intrinsic dipole moment
perpendicular to the monolayer \cite{Zhang2017Janus,Zhang2020JanusReview,Dong2021JanusReview}.
 This asymmetry creates chemically
inequivalent adsorption surfaces and modifies charge redistribution,
local electrostatic fields, and adsorption energetics
\cite{Lu2017Janus,Guan2018Janus}. Janus functionalization has therefore
been proposed as a strategy to enhance hydrogen binding and improve
storage characteristics without introducing defects or chemical
dopants.

Despite growing interest in Janus dichalcogenides, several fundamental
questions remain unresolved. First, the relative importance of metal
$d$-orbital character, chalcogen chemistry, and structural phase in
governing hydrogen adsorption has not been systematically disentangled.
Second, most existing studies focus on single-molecule adsorption,
providing limited insight into high-coverage behavior relevant for
practical storage \cite{Manzeli2017}. Finally, finite-temperature
stability and reversibility—essential criteria for realistic
operation—are often inferred indirectly rather than explicitly
addressed through dynamical simulations.

In this work, we address these issues through a comprehensive
first-principles investigation of hydrogen adsorption on pristine and
Janus MX$_2$ and MSSe monolayers (M = Ni, Pd, Pt; X = S, Se). Using
density-functional theory combined with orbital-resolved electronic
analysis, we identify the microscopic electronic factors that control
hydrogen binding across different metals, chalcogens, and structural
phases. By systematically increasing hydrogen coverage up to multilayer
regimes, we examine how adsorption strength, configurational entropy,
and surface geometry jointly determine storage capacity and stability.
Finite-temperature \textit{ab initio} molecular dynamics simulations at
300~K are employed to assess thermal robustness, molecular integrity,
and reversibility under realistic conditions.

Our results reveal that efficient hydrogen storage in two-dimensional
dichalcogenides does not arise from maximal adsorption strength, but
from an optimal balance between electronic screening, geometric
openness, and molecular mobility. In particular, Janus Ni- and Pd-based
systems in the 2H phase emerge as promising platforms for reversible
molecular hydrogen adsorption, while Pt-based systems exhibit overly
strong confinement that compromises reversibility. These findings
establish clear physical design principles for hydrogen storage in
low-dimensional materials and clarify the realistic capabilities and
limitations of Janus engineering in physisorption-dominated systems.
	
\section{Computational details}

The VASP (Vienna Ab-initio Simulation Package) computational code
\cite{Kresse1996,Kresse1999} was used to calculate the
electronic and atomic properties. The kinetic energy cutoff is
\SI{400}{\electronvolt} and the converged electronic k-grid sampling
is taken to be $4\times4\times1$. Our calculations are based on the
projector augmented wave (PAW)
\cite{Blochl1994,Kresse1999} method with a PBE
exchange-correlation functional \cite{Perdew1996} in the
GGA approximation \cite{Perdew1996}. The
convergence criteria for the total energy is
$10^{-4}$\,\si{\electronvolt}. For calculations with supercell sizes of $(1\times1)$ a $(10\times10\times1)$ k-point sampling is used. For $(2\times2)$ supercells a (2x2x1) k-point mesh is adopted.
AIMD calculations are performed
using the Andersen thermostat at \SI{300}{\kelvin}
\cite{Andersen2008}. Time steps of \SI{1}{\femto\second} are
adopted for a total simulation time of \SI{5}{\pico\second}. For  $(2\times2)$ supercells the $\Gamma$ point only is used. 

\section{Results and discussion}

\subsection{Electronic structure and orbital origin of phase-dependent properties}

\begin{figure}[!ht]
\centering
\includegraphics[width=0.6\columnwidth,clip=true]{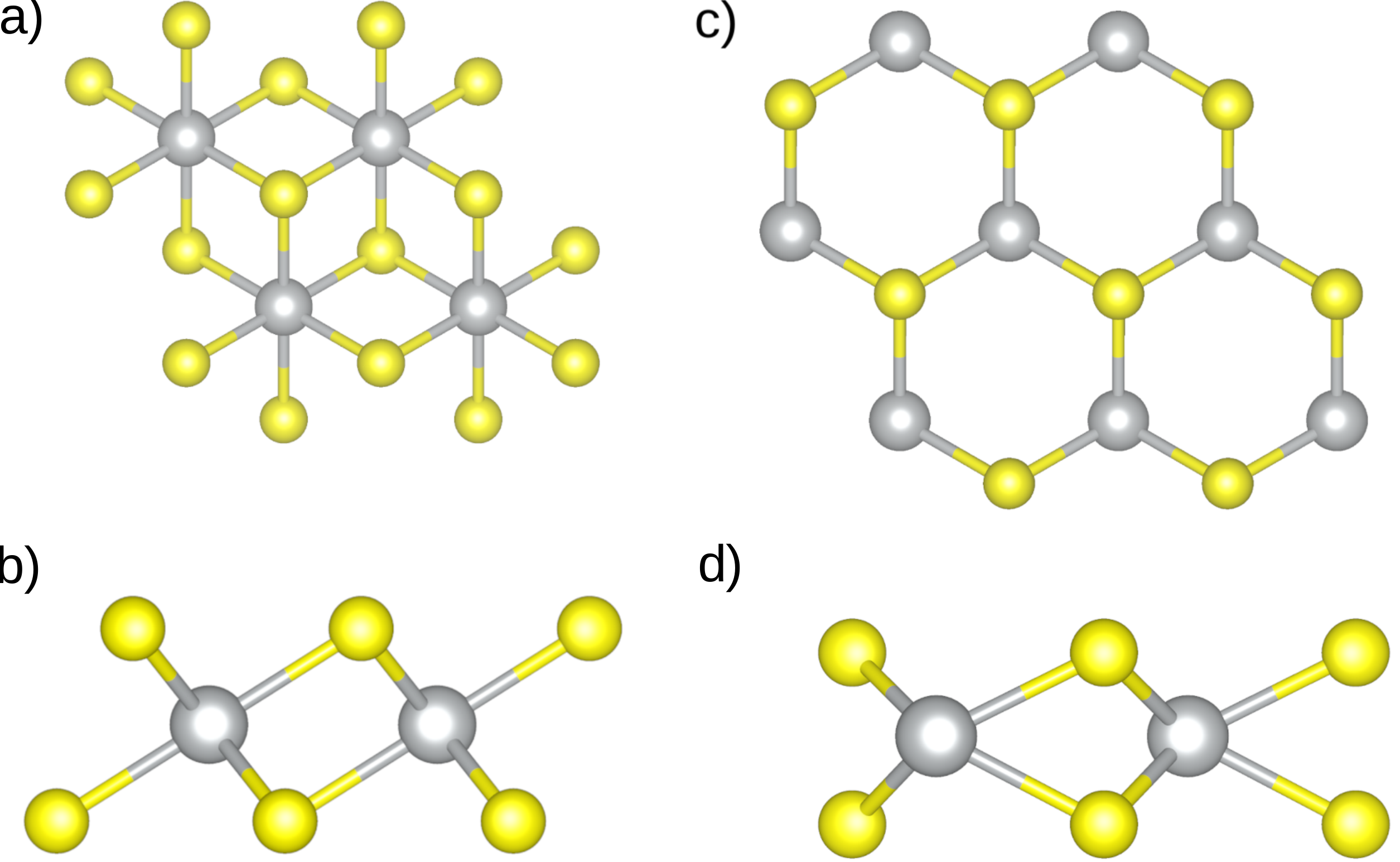}
\caption{Geometry of  pristine MX$_2$ (M = Pd, Pt and Ni; X= S, Se). a) on top view 1T-phase, b) on top view 2H-phase, c) side view 1T-phase and d) side view 2H-phase. Yellow (grey) spheres are sulfur (metal) atoms.}
\label{fig:mx2relaxedstructures}
\end{figure}

\begin{figure}[!ht]
\centering
\includegraphics[width=0.6\linewidth,clip=true]{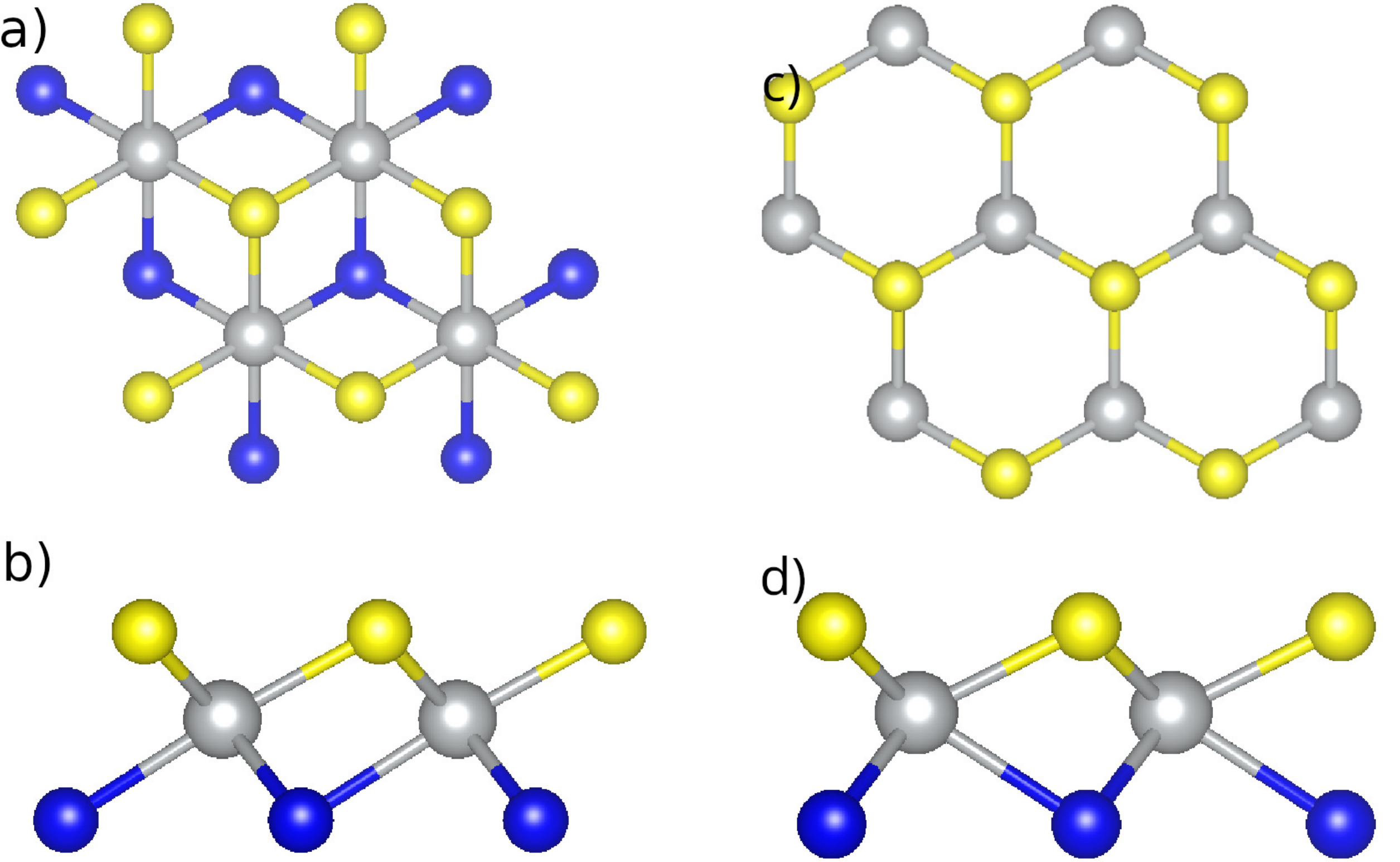}
\caption{\label{fig:relaxationMSSe}Janus structures of MSSe (M = Ni, Pd, Pt).
(a) on top view of 1T, (b) side view of 1T,
  (c) on top view of 2H, and (d) side view of 2H polytypes. Yellow, grey and blue
spheres represent sulfur, metal, and selenium atoms, respectively.}
\end{figure}

\begin{figure}[!ht]
		\centering
		\begin{subfigure}[b]{0.32\columnwidth}
			\subcaption[]{}
			\includegraphics[width=\columnwidth,clip=true]{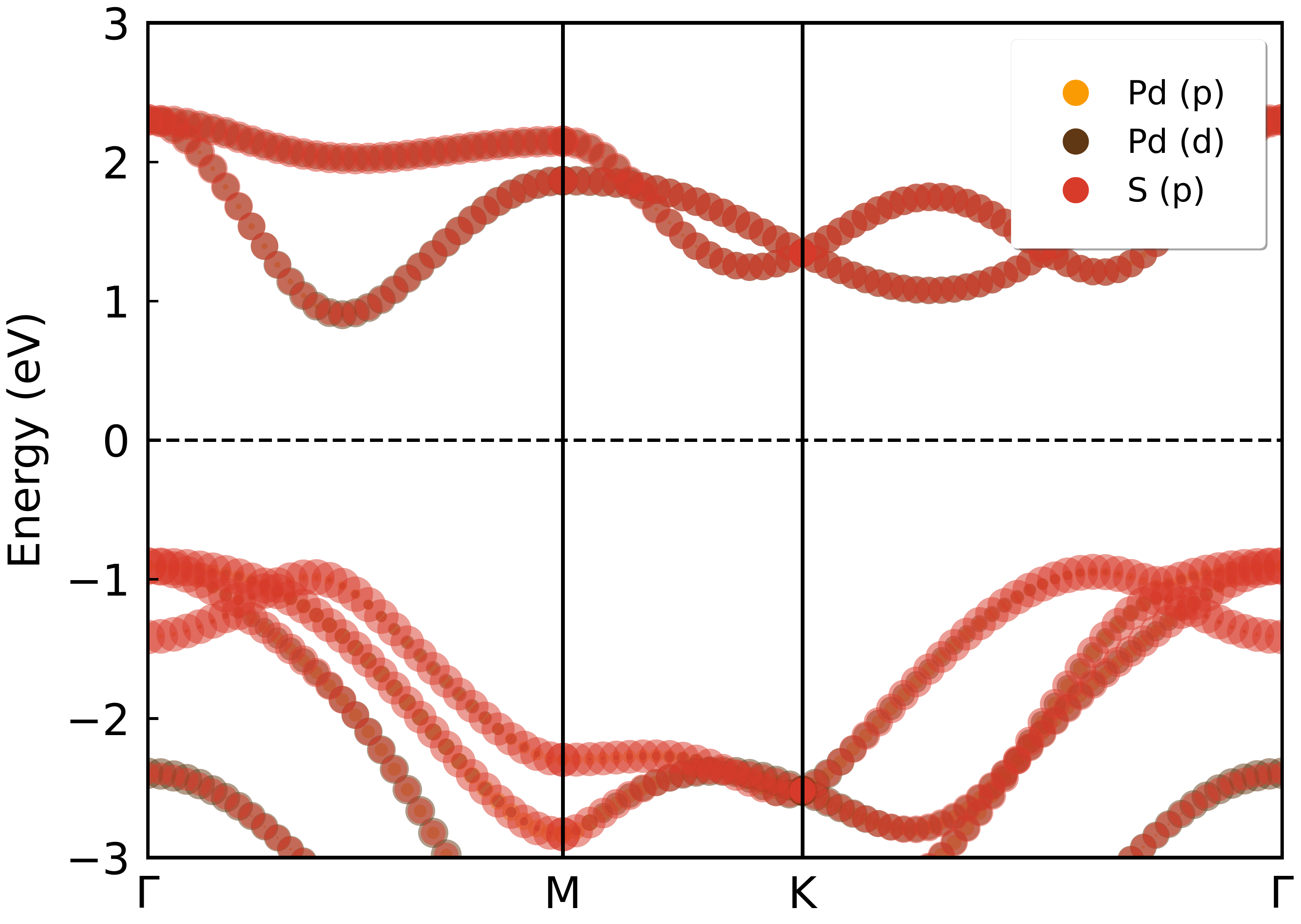}
		\end{subfigure}\hfill
		\begin{subfigure}[b]{0.32\columnwidth}
			\subcaption[]{}
			\includegraphics[width=\columnwidth,clip=true]{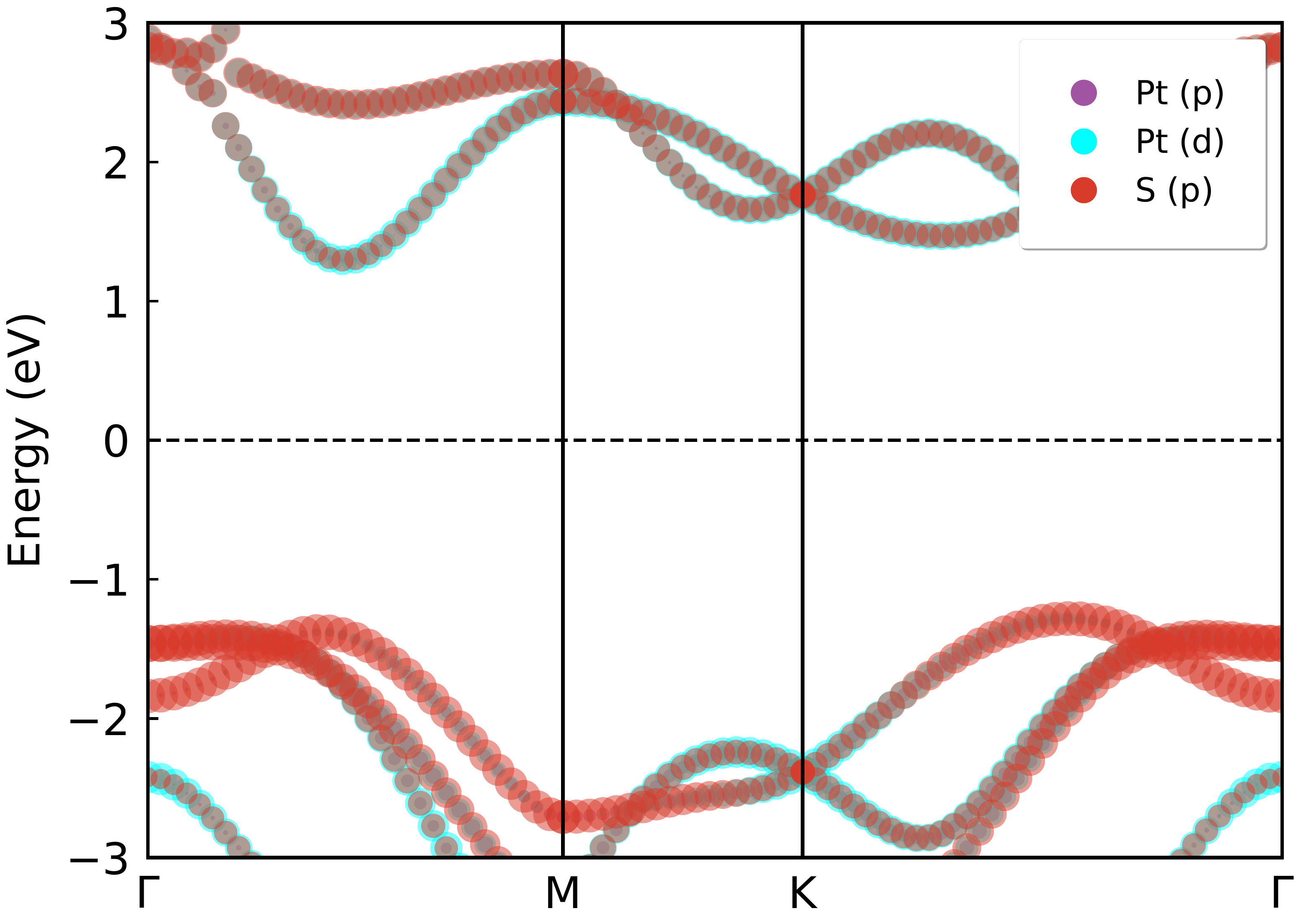}
		\end{subfigure}\hfill
		\begin{subfigure}[b]{0.32\columnwidth}
			\subcaption[]{}
			\includegraphics[width=\columnwidth,clip=true]{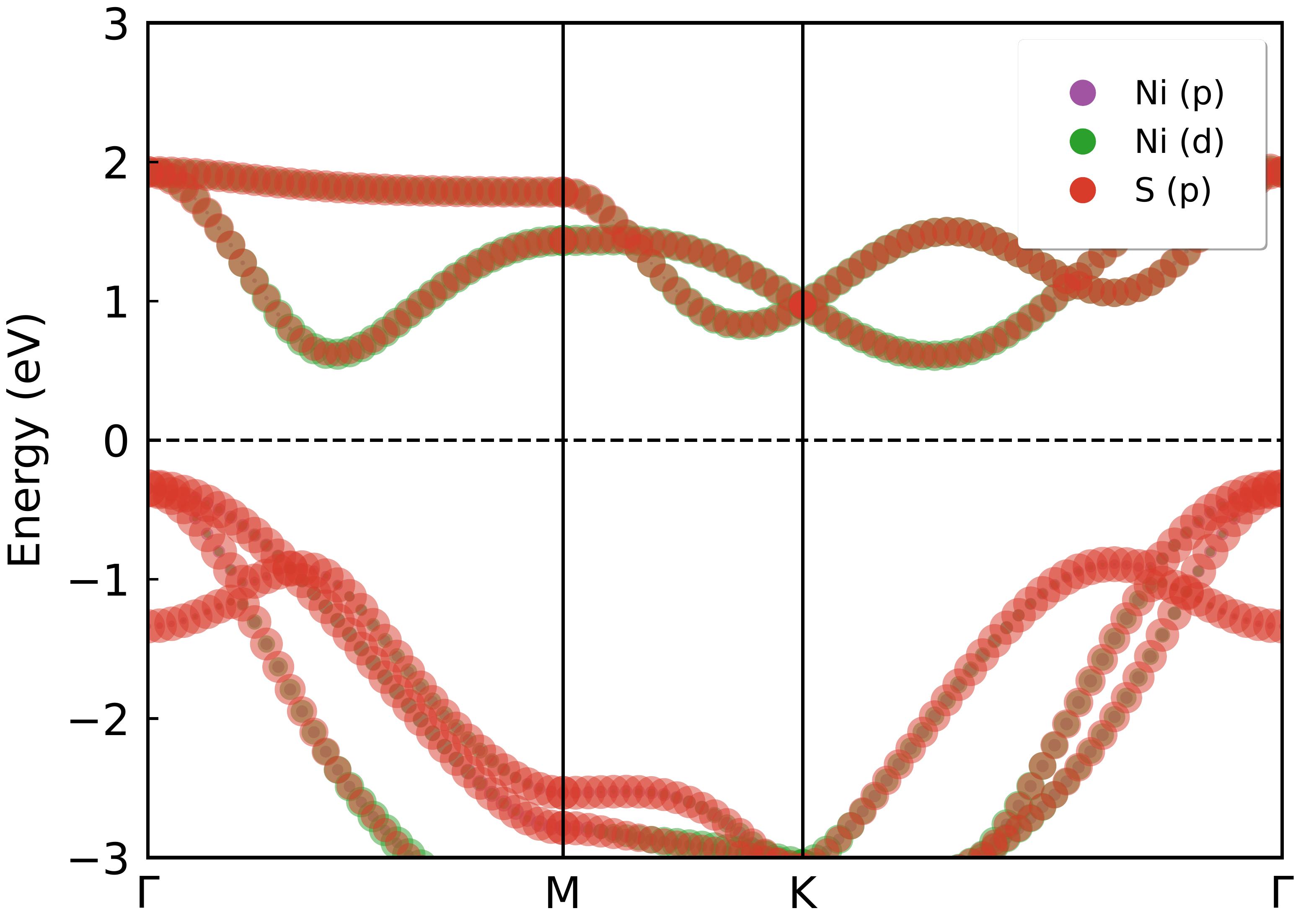}
		\end{subfigure}
		\\
		\begin{subfigure}[b]{0.32\columnwidth}
			\subcaption[]{}
			\includegraphics[width=\textwidth,clip=true]{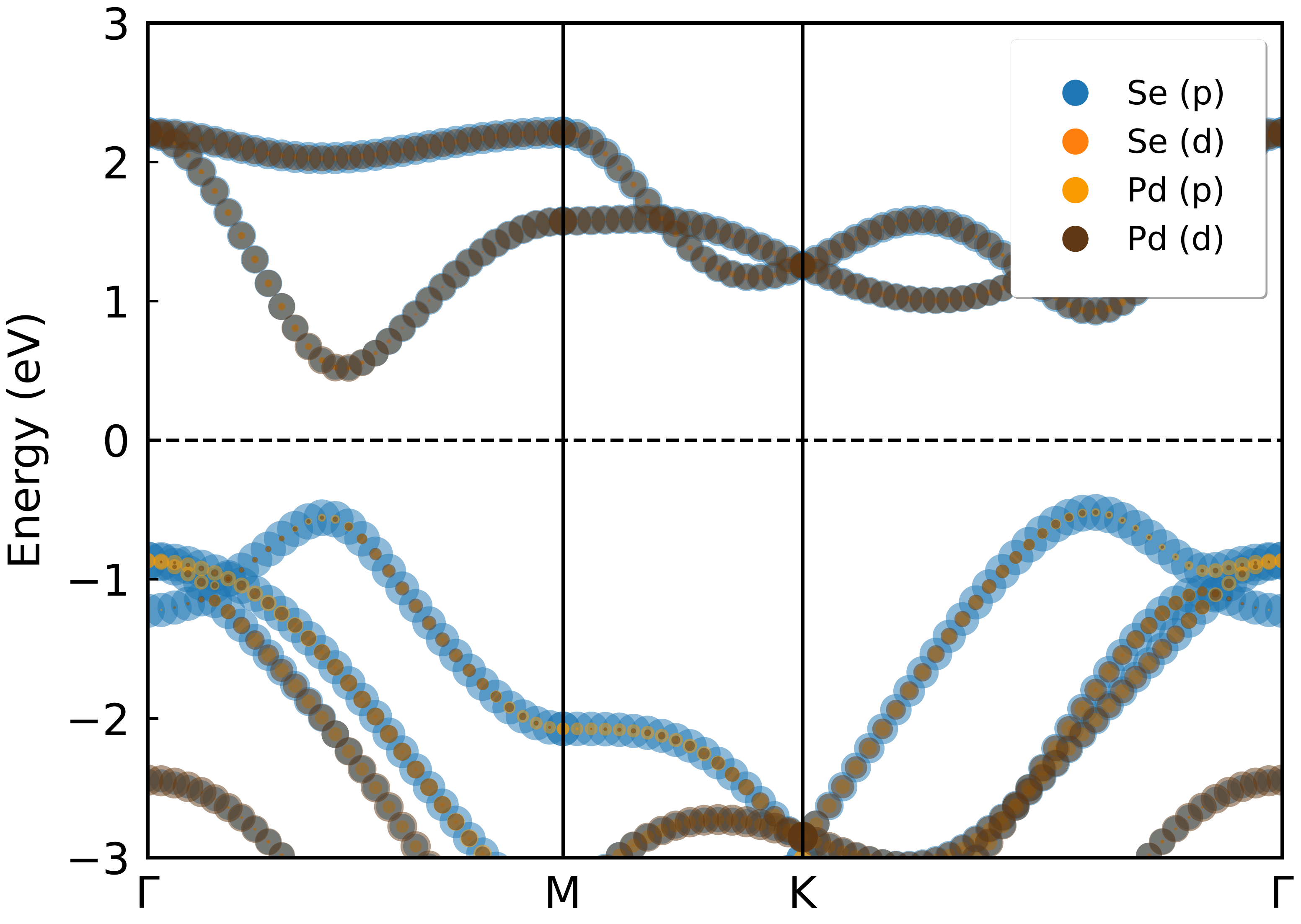}
		\end{subfigure}\hfill
		\begin{subfigure}[b]{0.32\columnwidth}
			\subcaption[]{}
			\includegraphics[width=\textwidth,clip=true]{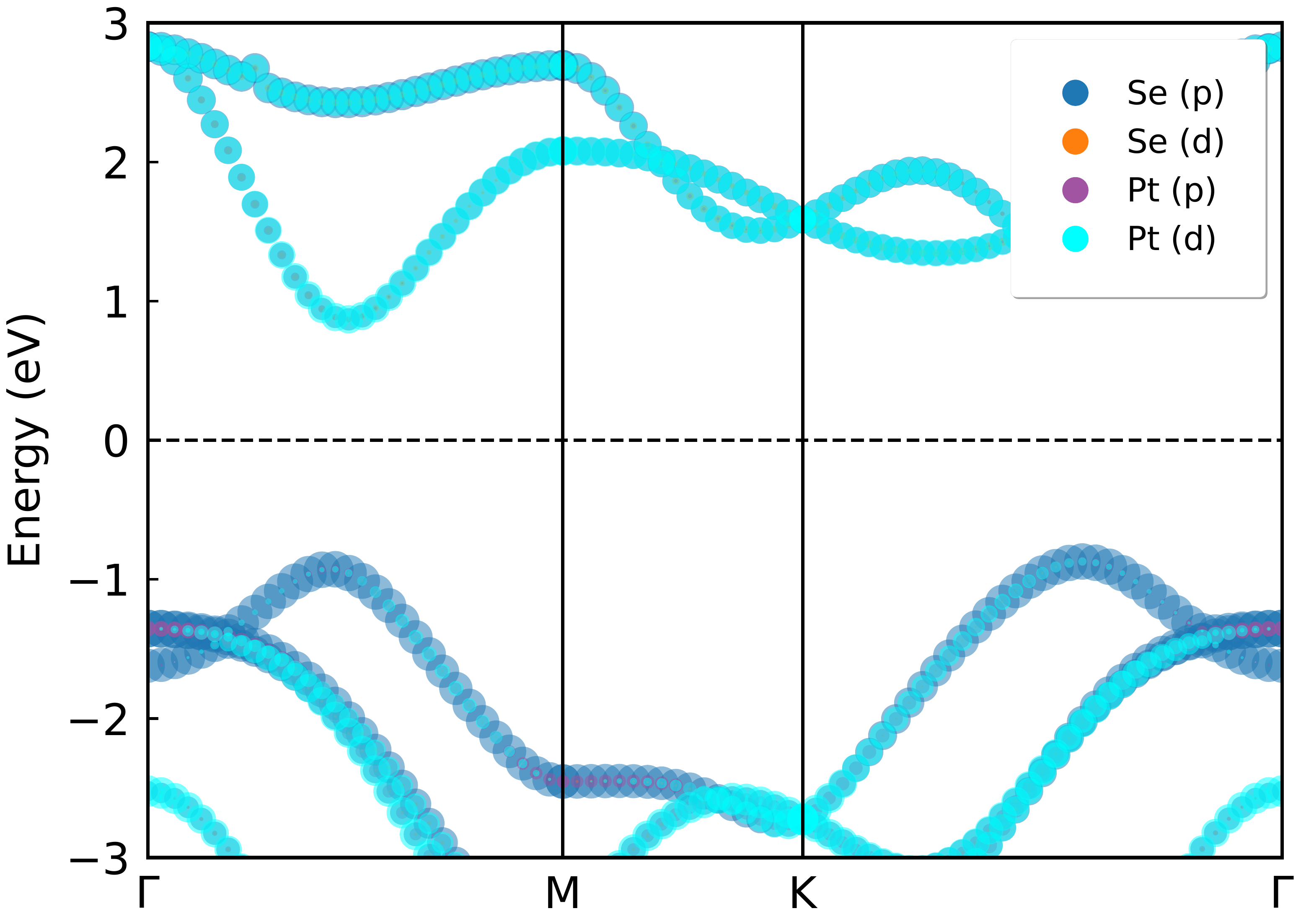}
		\end{subfigure}\hfill
		\begin{subfigure}[b]{0.32\columnwidth}
			\subcaption[]{}
			\includegraphics[width=\columnwidth,clip=true]{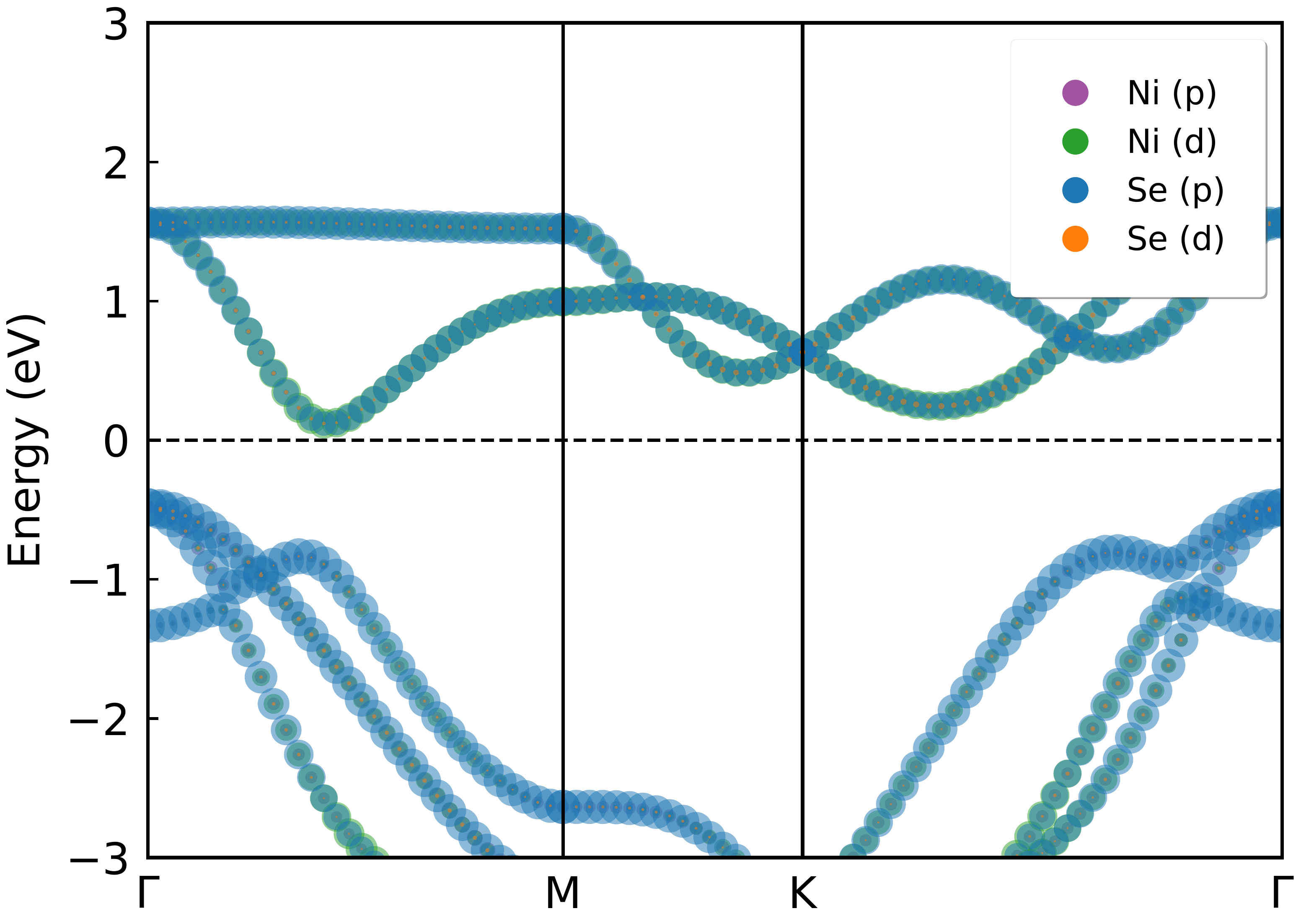}
		\end{subfigure}
		\\
		\begin{subfigure}[b]{0.32\columnwidth}
			\subcaption[]{}
			\includegraphics[width=\columnwidth,clip=true]{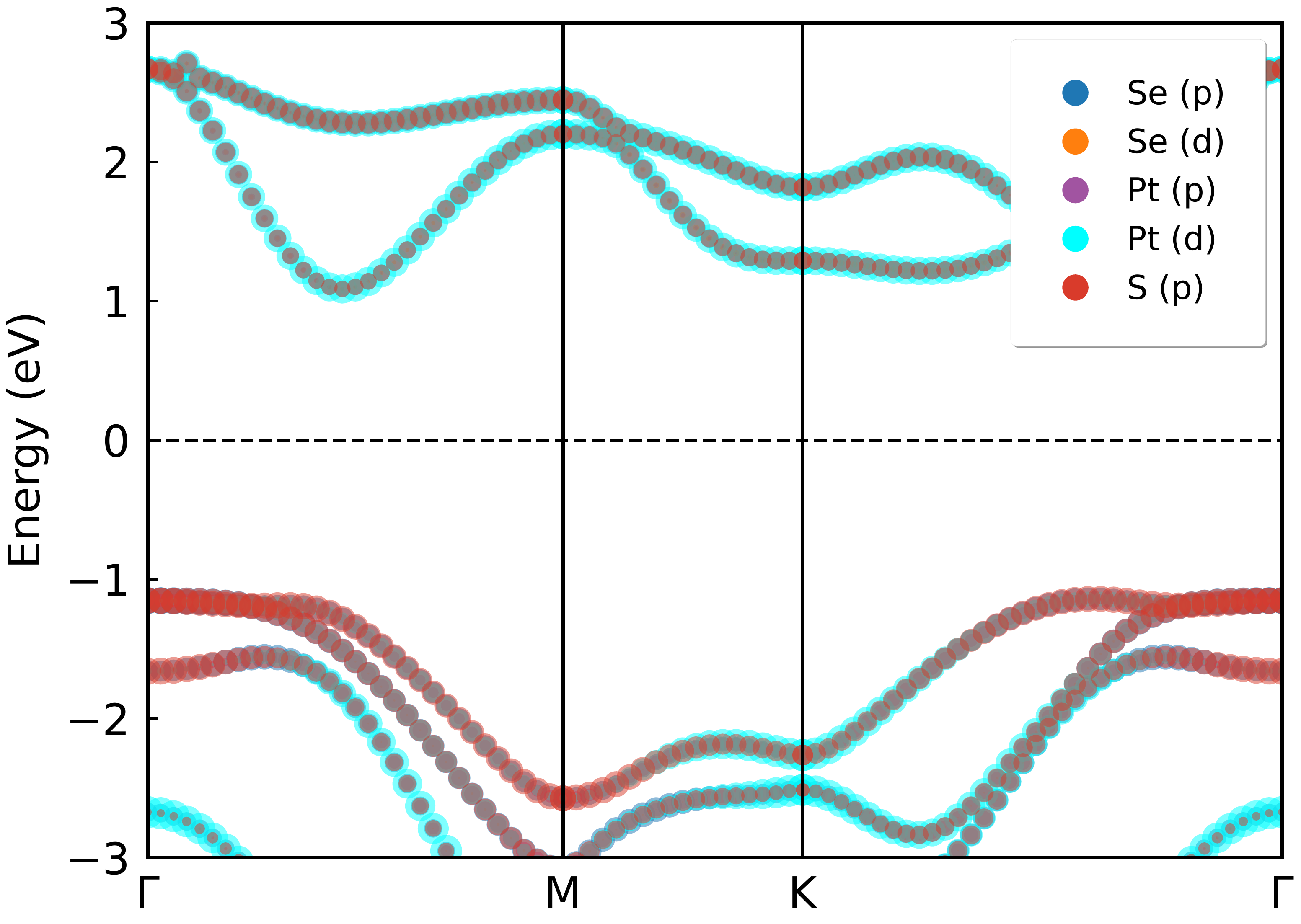}
		\end{subfigure}\hfill
		\begin{subfigure}[b]{0.32\columnwidth}
			\subcaption[]{}
			\includegraphics[width=\columnwidth,clip=true]{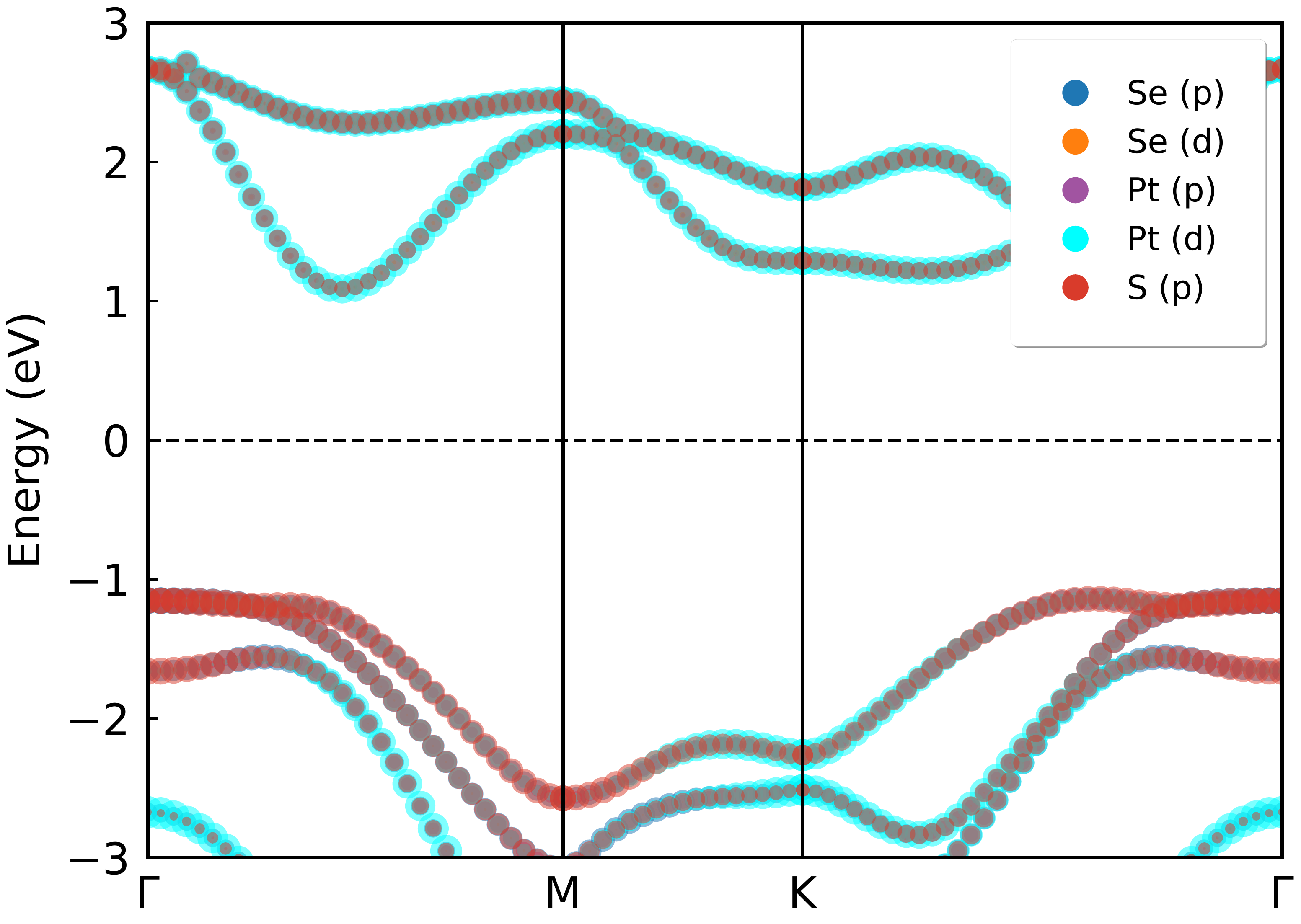}
		\end{subfigure}\hfill
		\begin{subfigure}[b]{0.32\columnwidth}
			\subcaption[]{}
			\includegraphics[width=\columnwidth,clip=true]{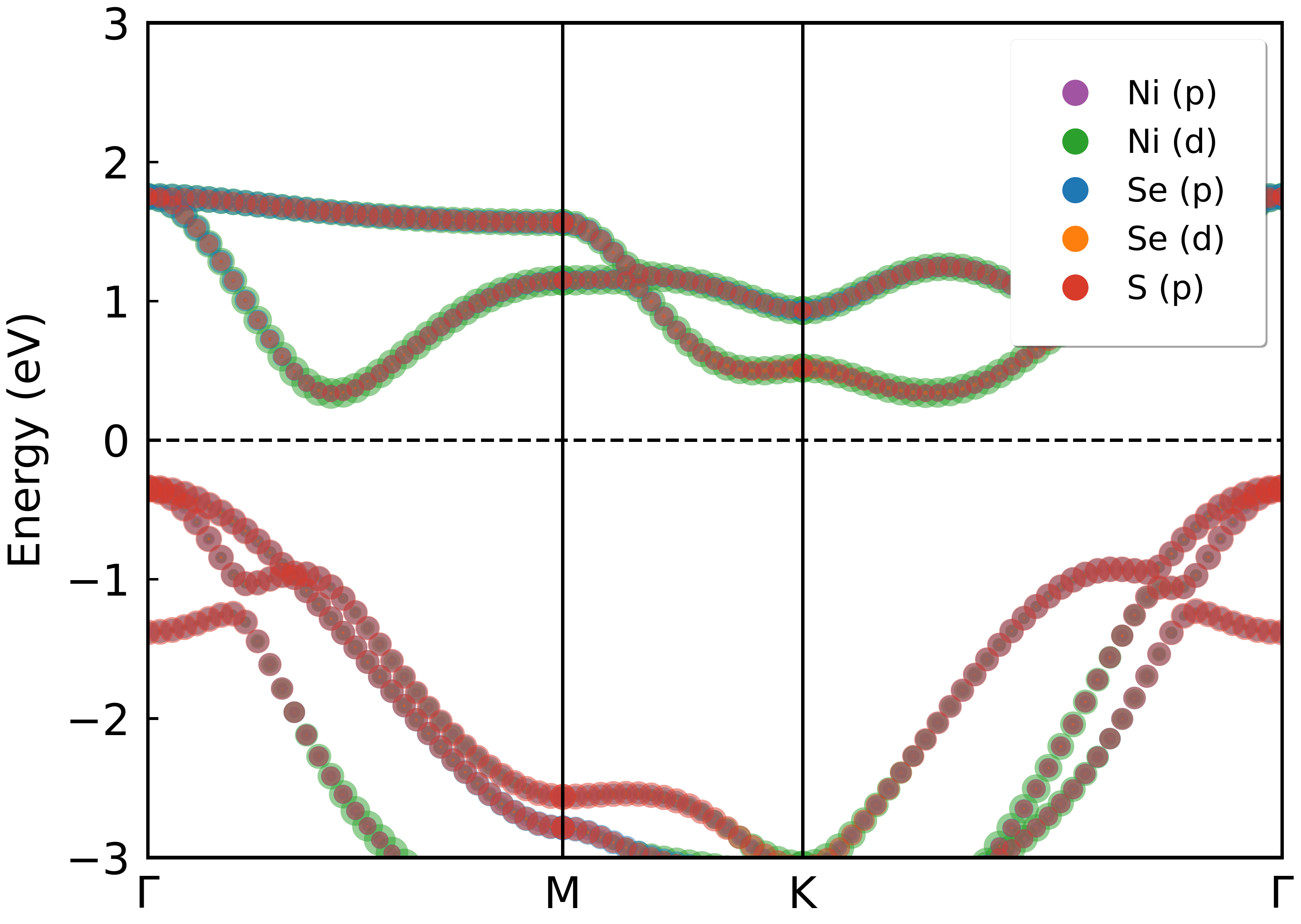}
		\end{subfigure}
		\caption{Orbital projected band structure for 1T-dichalcogenides: a) PdS$_2$, b) PtS$_2$, c) NiS$_2$, d) PdSe$_2$, e) PtSe$_2$, f) NiSe$_2$, g) PdSSe, h) PtSSe and i) NiSSe.}
		\label{fig:mx2bandstructures_1T}
	\end{figure}

\begin{figure}[!ht]
\centering
\begin{subfigure}[b]{0.32\columnwidth}
\subcaption[]{}
\includegraphics[width=\columnwidth,clip=true]{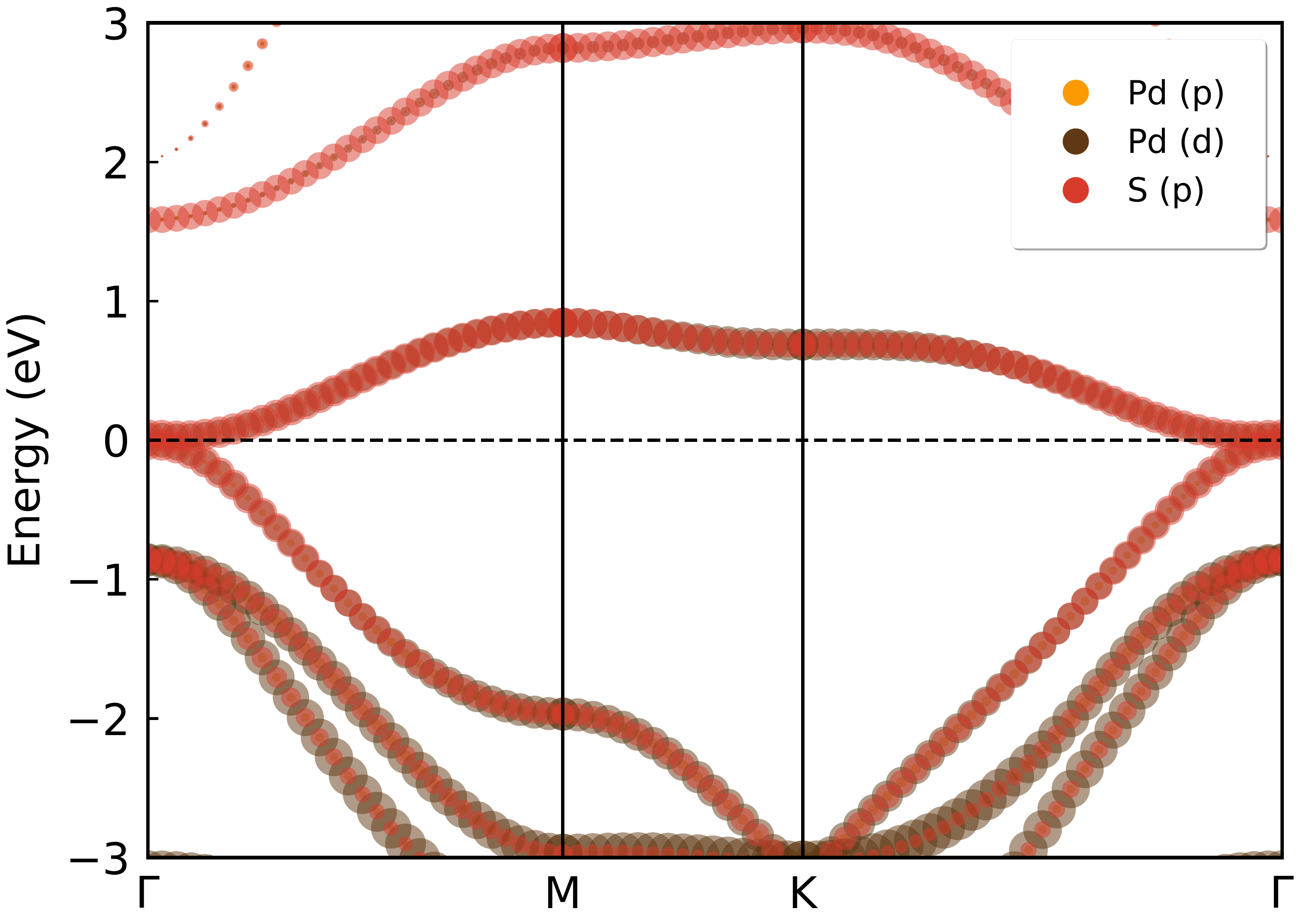}
\end{subfigure}\hfill
		\begin{subfigure}[b]{0.32\columnwidth}
			\subcaption[]{}
		\includegraphics[width=\columnwidth,clip=true]{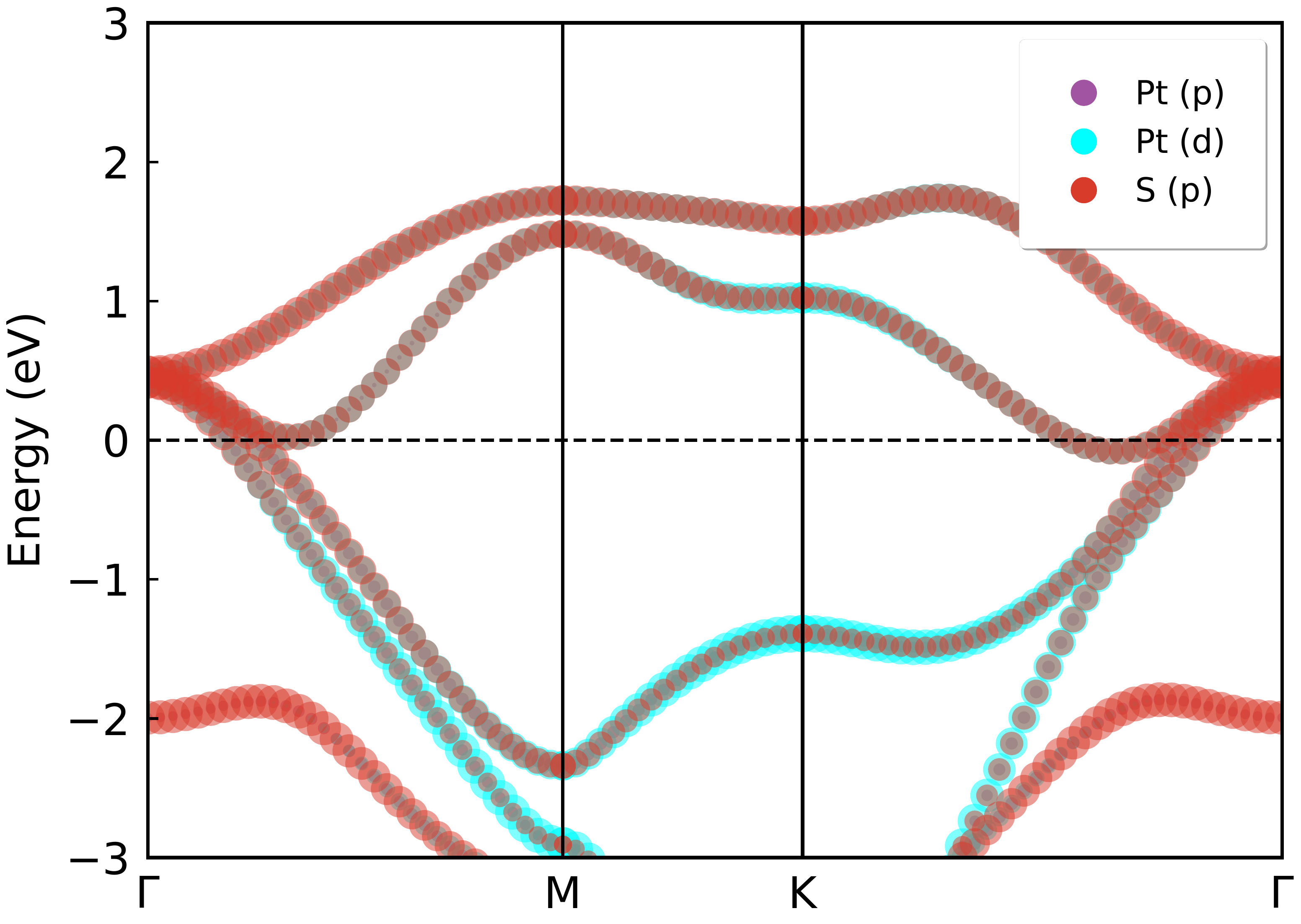}
		\end{subfigure}\hfill
		\begin{subfigure}[b]{0.32\columnwidth}
			\subcaption[]{}
		\includegraphics[width=\columnwidth,clip=true]{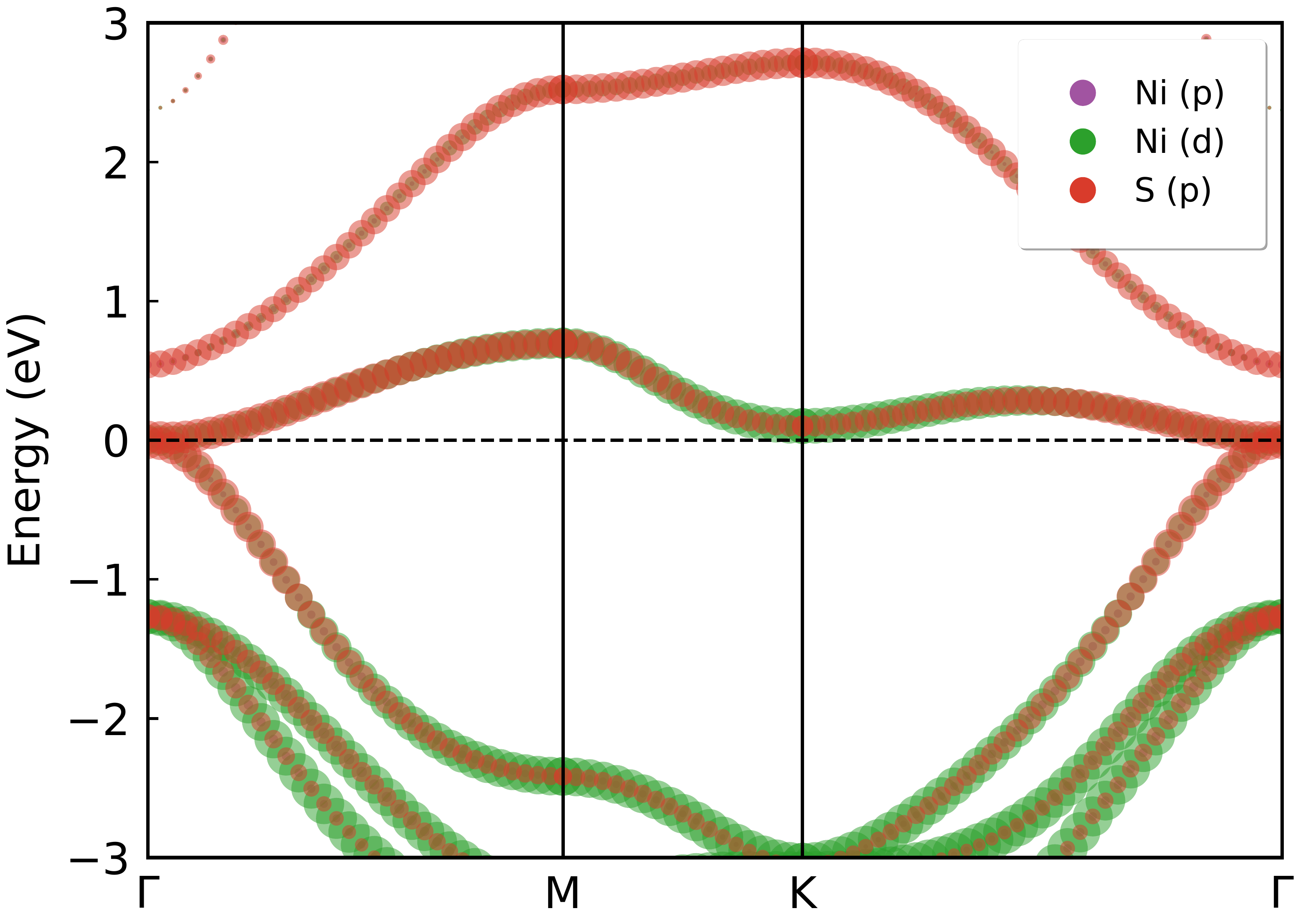}
		\end{subfigure}
		\\
		\begin{subfigure}[b]{0.32\columnwidth}
			\subcaption[]{}
			\includegraphics[width=\textwidth,clip=true]{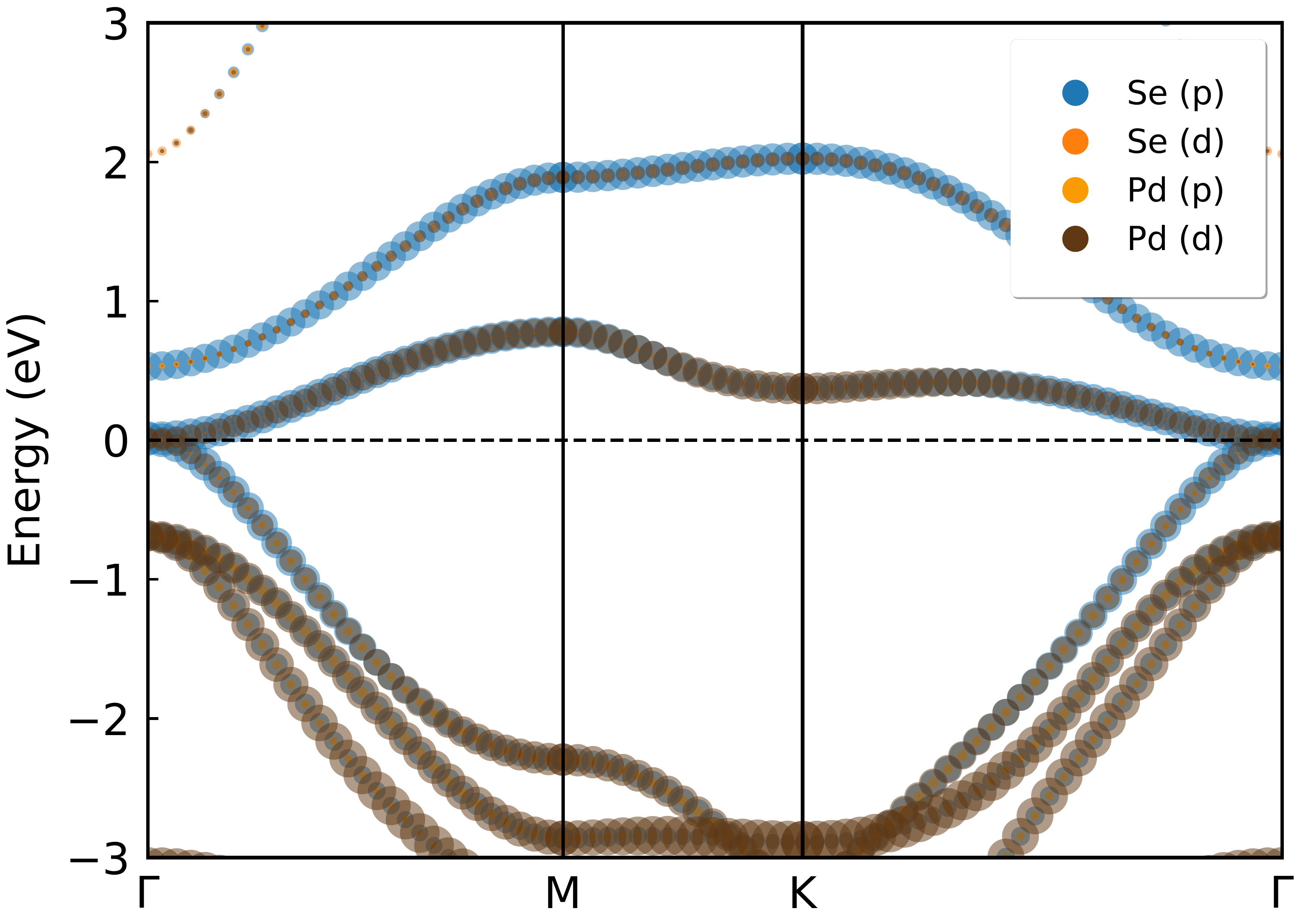}
		\end{subfigure}\hfill
		\begin{subfigure}[b]{0.32\columnwidth}
			\subcaption[]{}
			\includegraphics[width=\textwidth,clip=true]{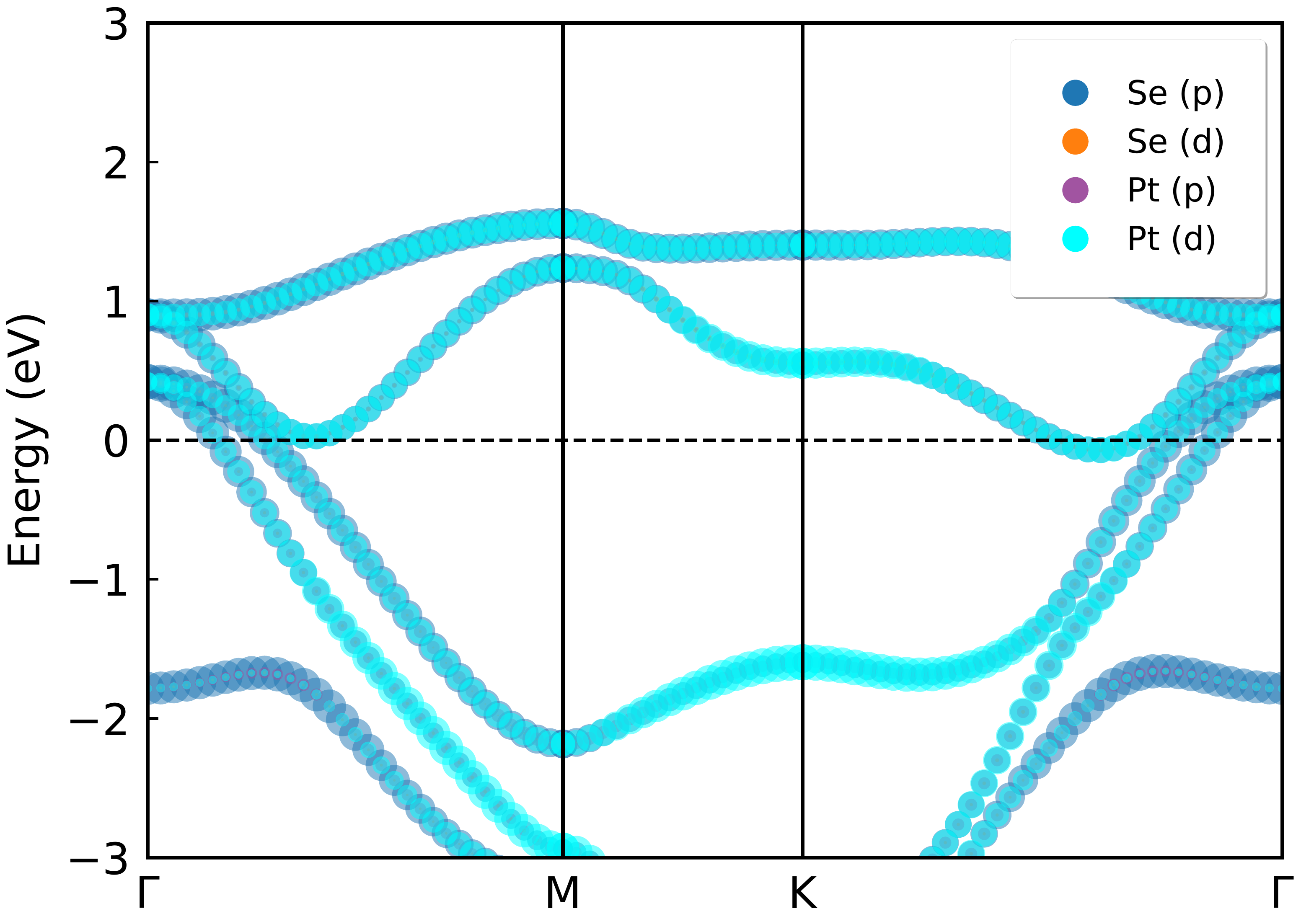}
		\end{subfigure}\hfill
		\begin{subfigure}[b]{0.32\columnwidth}
			\subcaption[]{}
			\includegraphics[width=\columnwidth,clip=true]{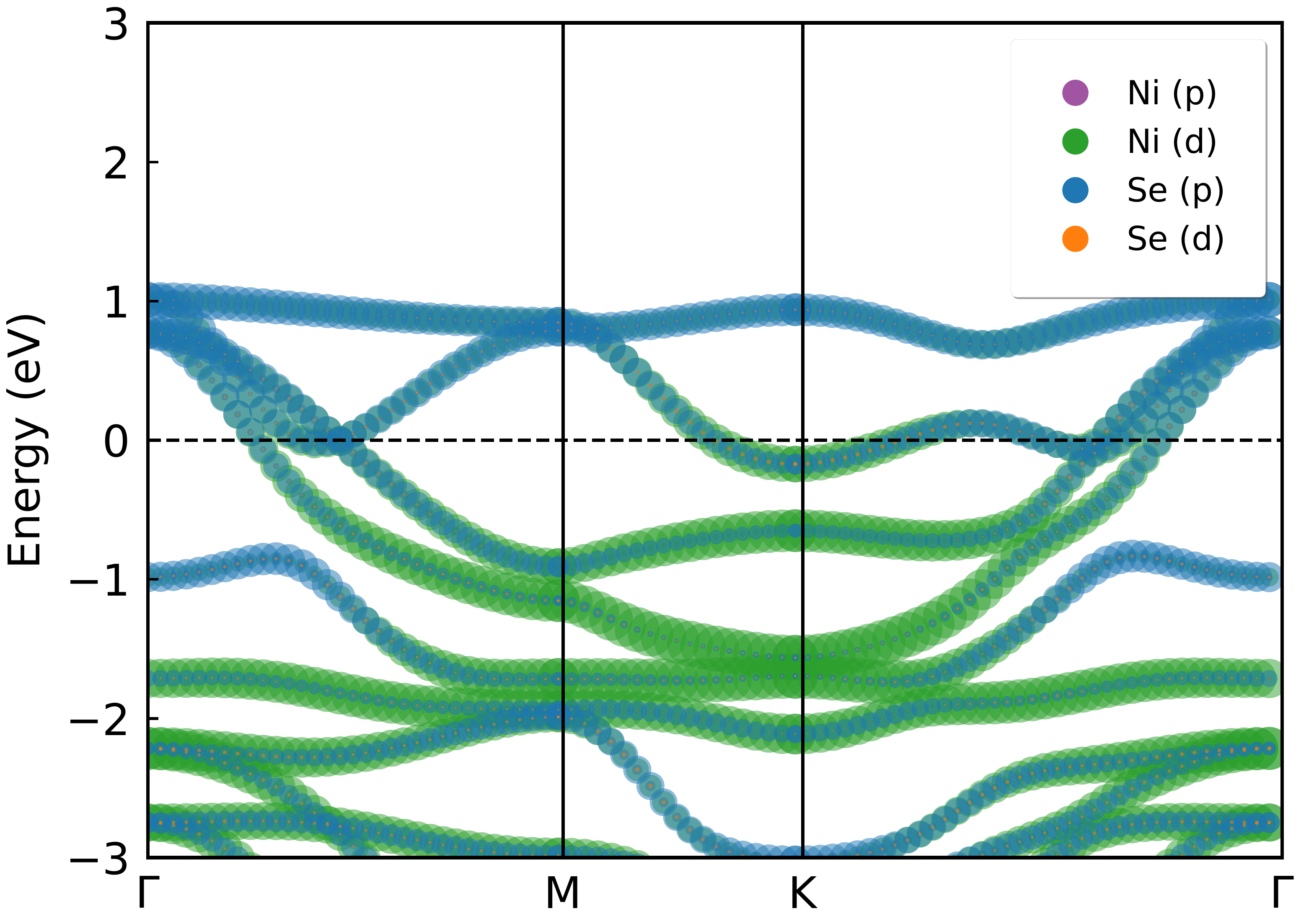}
		\end{subfigure}
		\\
		\begin{subfigure}[b]{0.32\columnwidth}
			\subcaption[]{}
			\includegraphics[width=\columnwidth,clip=true]{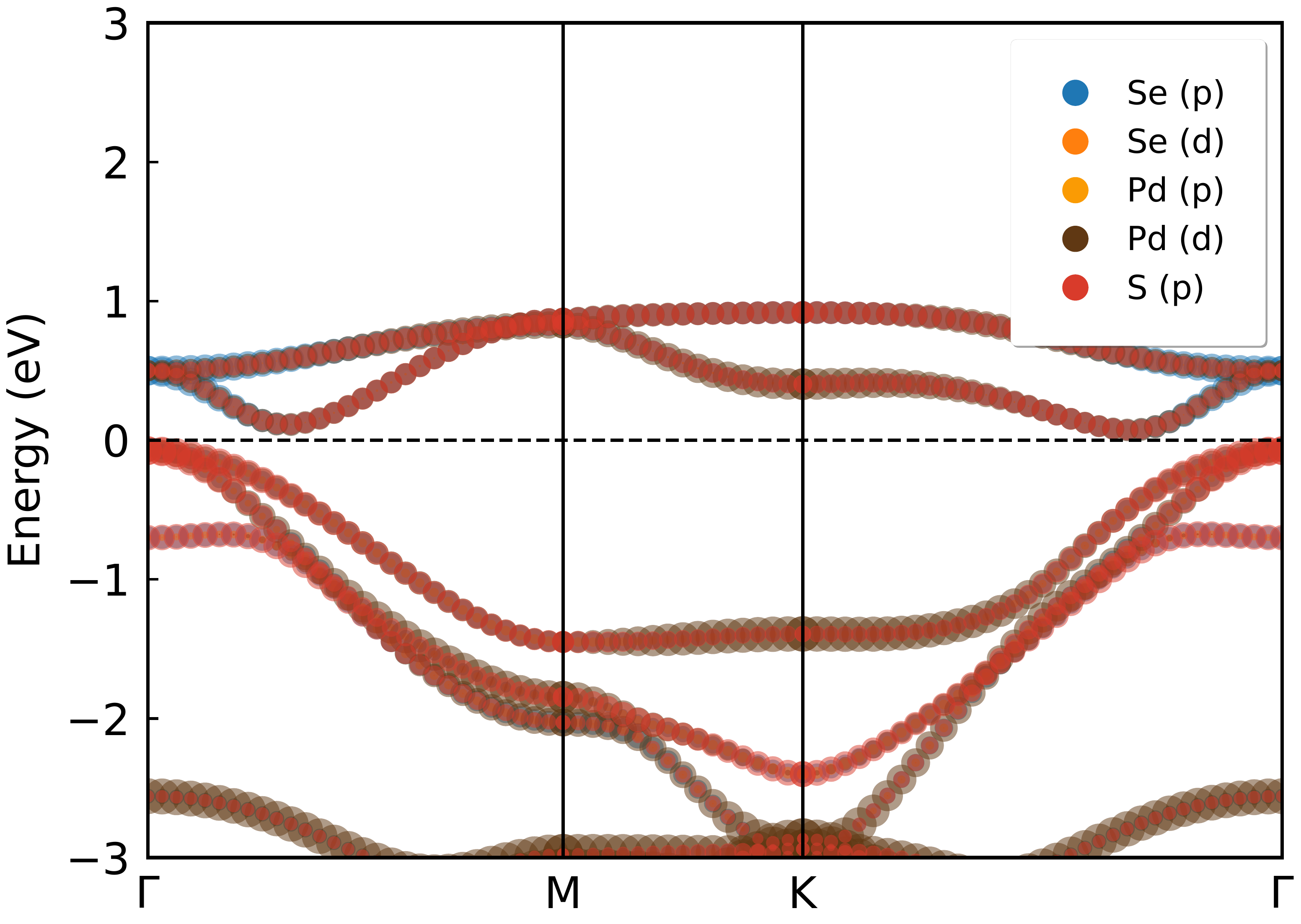}
		\end{subfigure}\hfill
		\begin{subfigure}[b]{0.32\columnwidth}
			\subcaption[]{}
			\includegraphics[width=\columnwidth,clip=true]{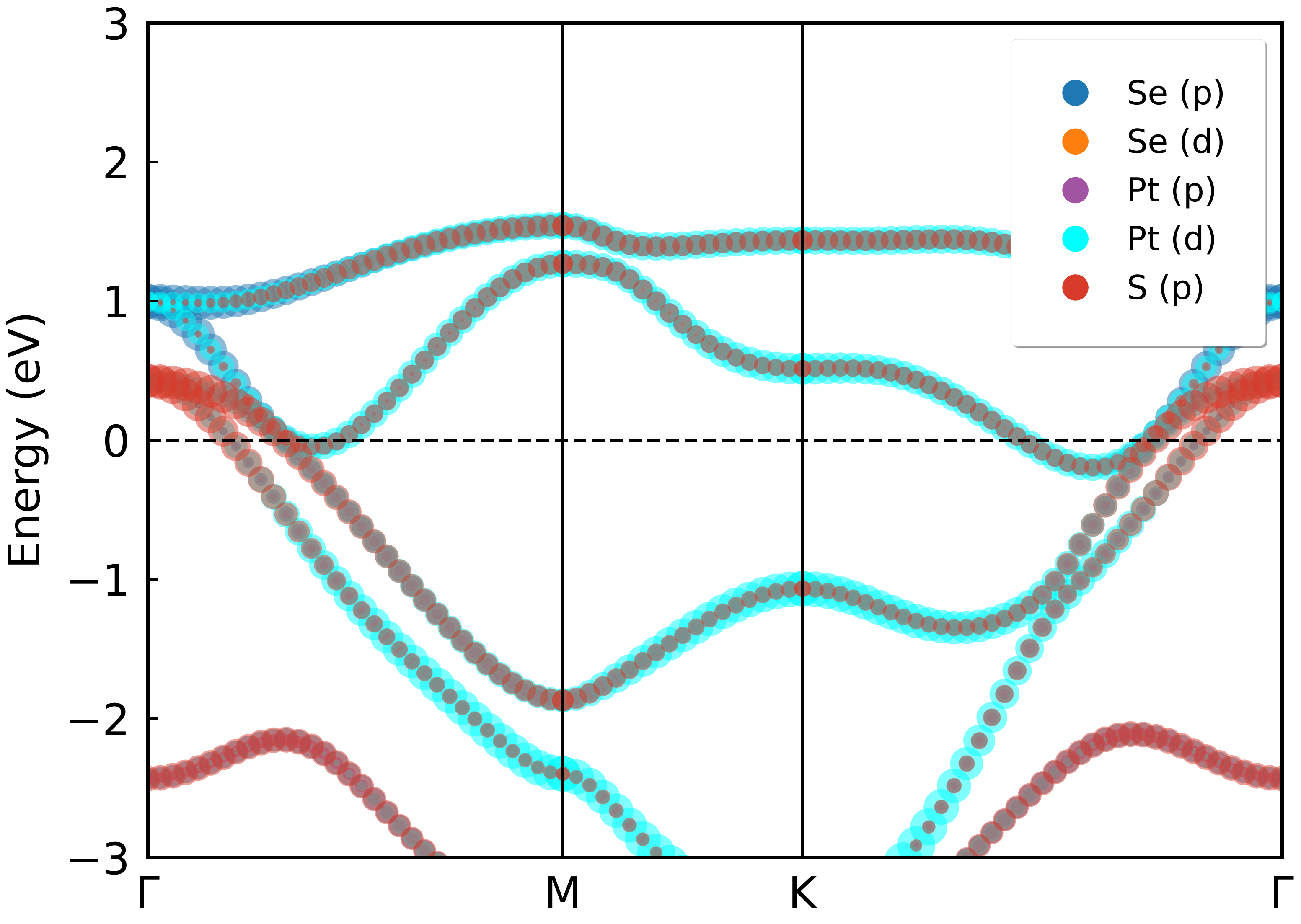}
		\end{subfigure}\hfill
		\begin{subfigure}[b]{0.32\columnwidth}
			\subcaption[]{}
		\includegraphics[width=\columnwidth,clip=true]{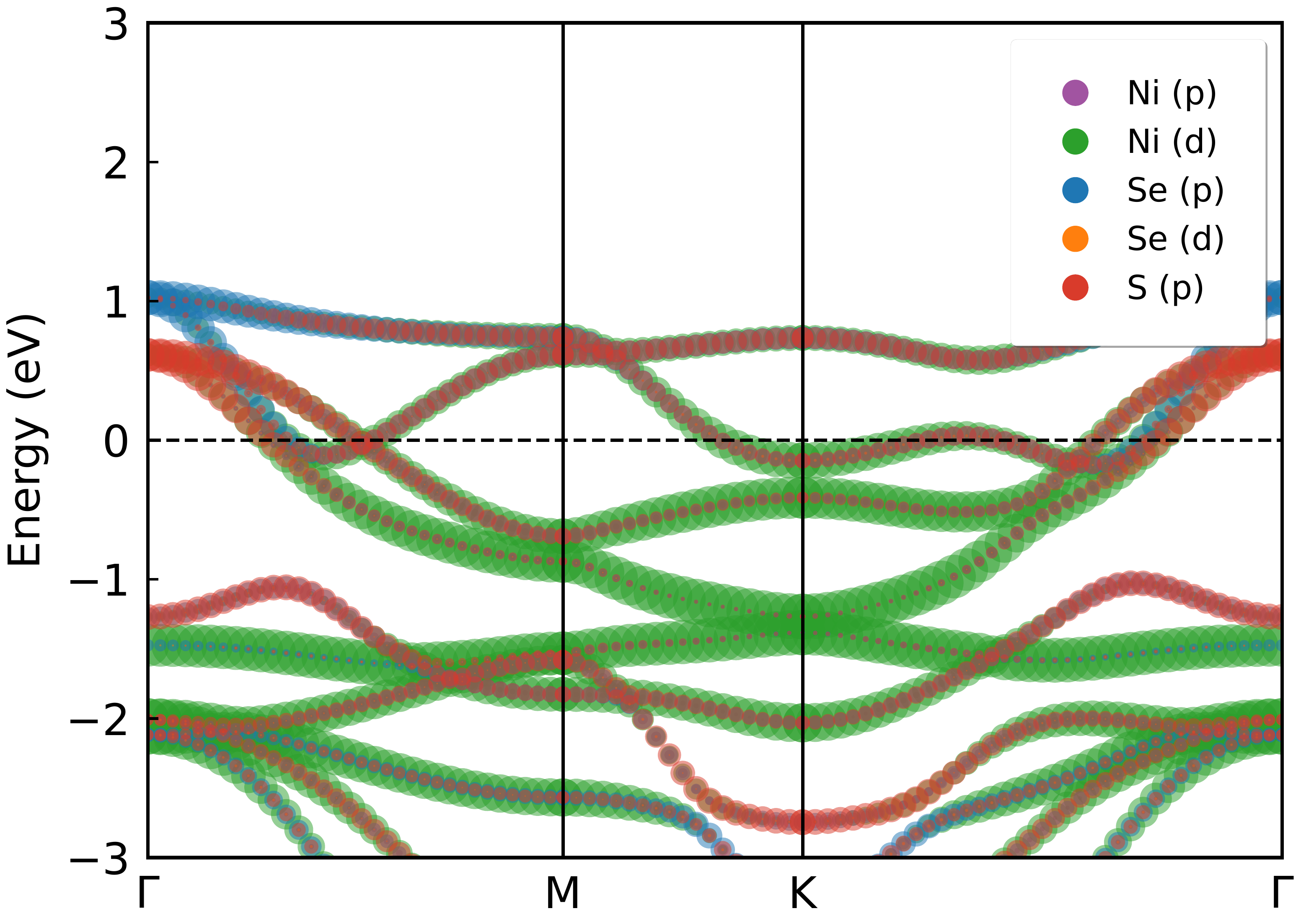}
		\end{subfigure}
		\caption{Orbital projected band structure for 2H-dichalcogenides: a) PdS$_2$, b) PtS$_2$, c) NiS$_2$, d) PdSe$_2$, e) PtSe$_2$, f) NiSe$_2$, g) PdSSe, h) PtSSe and i) NiSSe.}
		\label{fig:mx2bandstructures_2H}
	\end{figure}

\begin{table}
\caption{\label{tab:gaps}Electronic band gaps (eV) of pristine and Janus
transition-metal dichalcogenide monolayers. All gaps are indirect.}
\centering
\renewcommand{\arraystretch}{1.25}
\setlength{\tabcolsep}{10pt}

\begin{tabular}{cccccc}
\toprule
\multicolumn{6}{c}{\text{Sulfides}} \\
\midrule
\makecell{1T-PdS$_2$ \\ 1.81} &
\makecell{2H-PdS$_2$ \\ 0.06} &
\makecell{1T-PtS$_2$ \\ 2.57} &
\makecell{2H-PtS$_2$ \\ metallic} &
\makecell{1T-NiS$_2$ \\ 0.95} &
\makecell{2H-NiS$_2$ \\ 0.03} \\
\midrule
\multicolumn{6}{c}{\text{Selenides}} \\
\midrule
\makecell{1T-PdSe$_2$ \\ 1.04} &
\makecell{2H-PdSe$_2$ \\ semimetallic} &
\makecell{1T-PtSe$_2$ \\ 1.74} &
\makecell{2H-PtSe$_2$ \\ metallic} &
\makecell{1T-NiSe$_2$ \\ 0.60} &
\makecell{2H-NiSe$_2$ \\ metallic} \\
\midrule
\multicolumn{6}{c}{\text{Janus}} \\
\midrule
\makecell{1T-PdSSe \\ metallic} &
\makecell{2H-PdSSe \\ 0.15} &
\makecell{1T-PtSSe \\ 2.23} &
\makecell{2H-PtSSe \\ metallic} &
\makecell{1T-NiSSe \\ 0.68} &
\makecell{2H-NiSSe \\ metallic} \\
\bottomrule
\end{tabular}
\end{table}

The electronic band structures of the pristine MX$_2$ monolayers in the
1T and 2H polymorphs are presented in
Figs.~\ref{fig:mx2bandstructures_1T} and
\ref{fig:mx2bandstructures_2H}, respectively, with the corresponding
band-gap values summarized in Table~\ref{tab:gaps}. The calculations
reveal a strong dependence of the electronic structure on both the
structural polytype and the chemical composition.

For the 1T sulfides, PdS$_2$ and PtS$_2$ exhibit semiconducting behavior
with sizable band gaps. The valence-band maximum is primarily composed
of chalcogen $p$ states, while the conduction-band minimum also retains
a dominant chalcogen character with only minor contributions from metal
$d$ orbitals. This electronic structure reflects relatively weak
$d$–$p$ hybridization near the Fermi level and limited electronic
screening.

In contrast, 1T-NiS$_2$ displays a substantially reduced band gap, with
significant Ni 3$d$ contributions appearing near the valence-band edge.
The increased localization of Ni $d$ orbitals enhances the density of
states close to the Fermi level, signaling stronger electronic
polarizability. This trend becomes more pronounced upon replacing S
with Se. In PdSe$_2$ and PtSe$_2$, the higher energy of Se $p$ states
systematically raises the valence-band manifold and narrows the gap,
while in NiSe$_2$ multiple Ni-derived bands cross the Fermi level,
resulting in a metallic ground state.

The 2H polymorphs exhibit analogous chemical trends but with notable
differences arising from their trigonal-prismatic coordination. For
PdS$_2$ and PtS$_2$, the 2H phase remains semiconducting with band gaps
comparable to or slightly larger than those of the corresponding 1T
phases. However, the orbital composition near the band edges differs:
the reduced symmetry of the 2H phase leads to stronger crystal-field
splitting of the metal $d$ states, pushing them away from the Fermi
level and suppressing metallic behavior.

Across both polymorphs, two robust trends emerge. First, chalcogen
substitution from S to Se consistently reduces the band gap by raising
the valence-band edge, reflecting the increased polarizability and
weaker electronegativity of Se. Second, moving from Pd and Pt to Ni
increases the participation of metal $d$ states near the Fermi level,
promoting metallicity or near-metallicity. These trends are summarized
quantitatively in Table~\ref{tab:gaps} and are central to understanding
the adsorption behavior discussed in subsequent sections.

\subsection{Molecular hydrogen adsorption on pristine MX$_2$: site selectivity and chemical trends}

\begin{figure}[!ht]
		\centering
        \begin{subfigure}[b]{0.2\columnwidth}
			\subcaption[]{}
\includegraphics[width=\columnwidth,height=3cm,keepaspectratio]{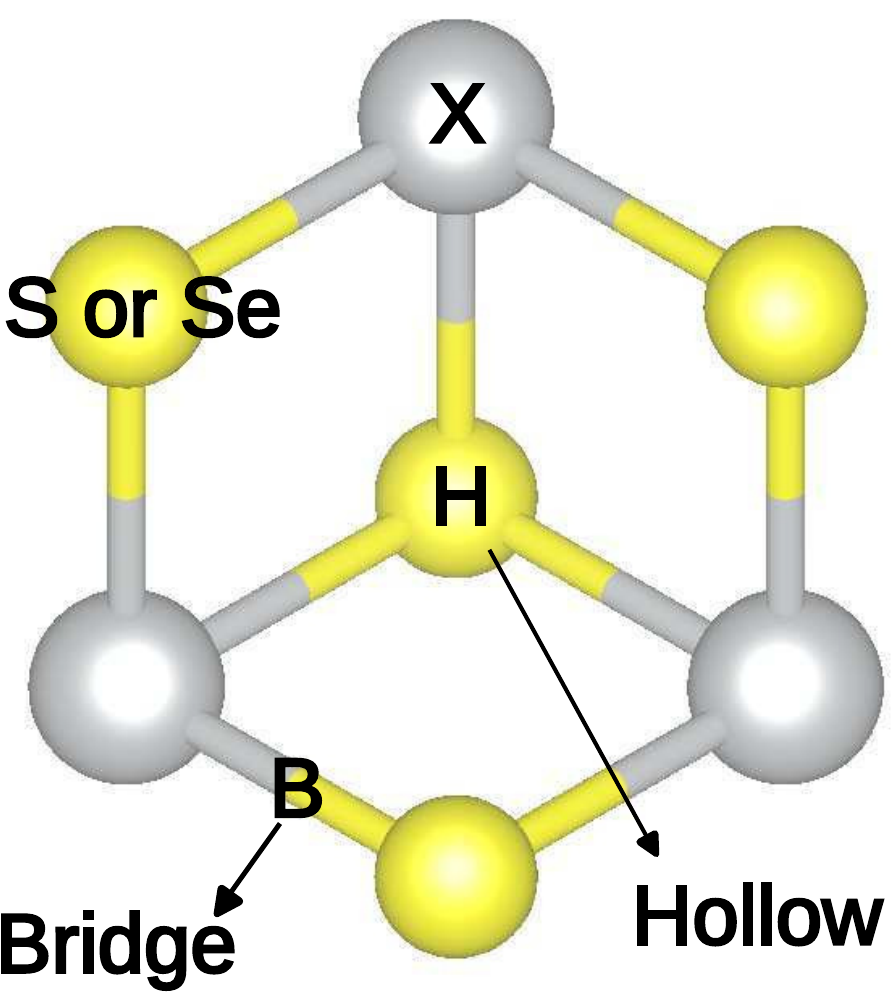}
    \end{subfigure}
\begin{subfigure}[b]{0.2\columnwidth}
			\subcaption[]{}
\includegraphics[width=\columnwidth,height=3cm,keepaspectratio]{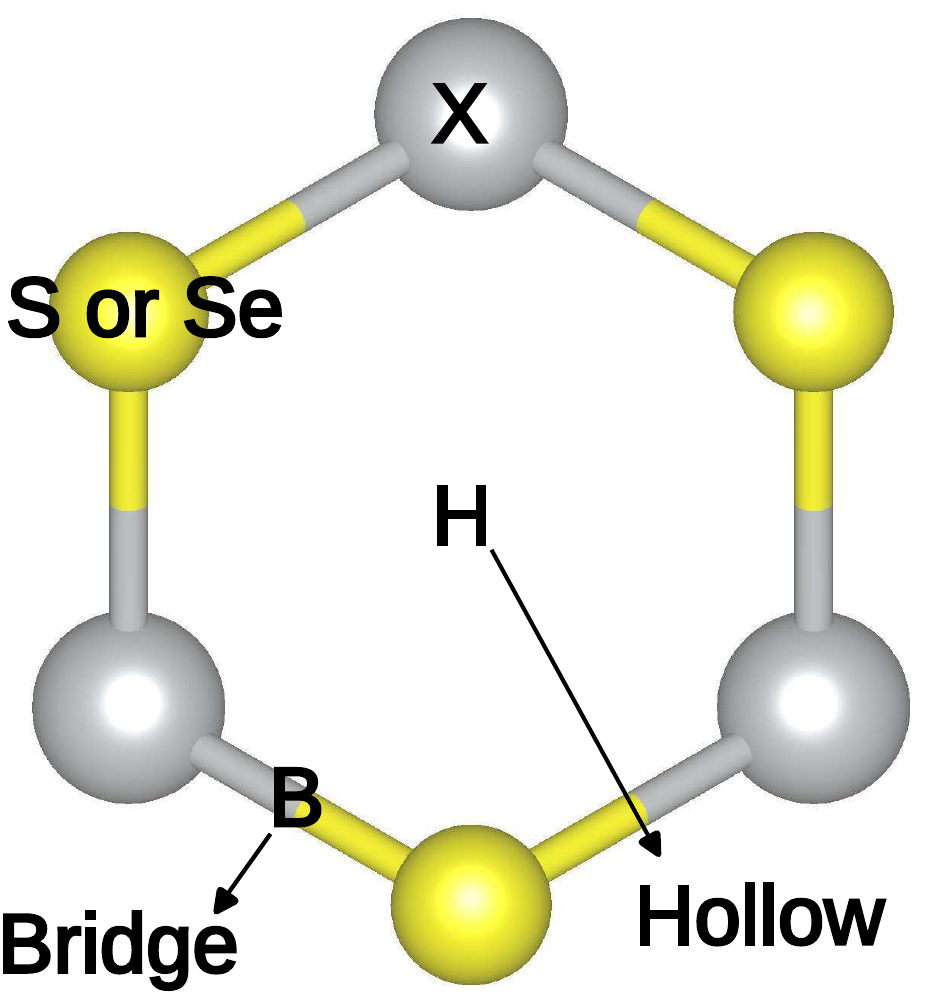}
\end{subfigure}\\
\vspace{1cm}
    \begin{subfigure}[b]{0.3\columnwidth}
			\subcaption[]{}
	\includegraphics[width=\columnwidth,height=3cm,clip=true,keepaspectratio]{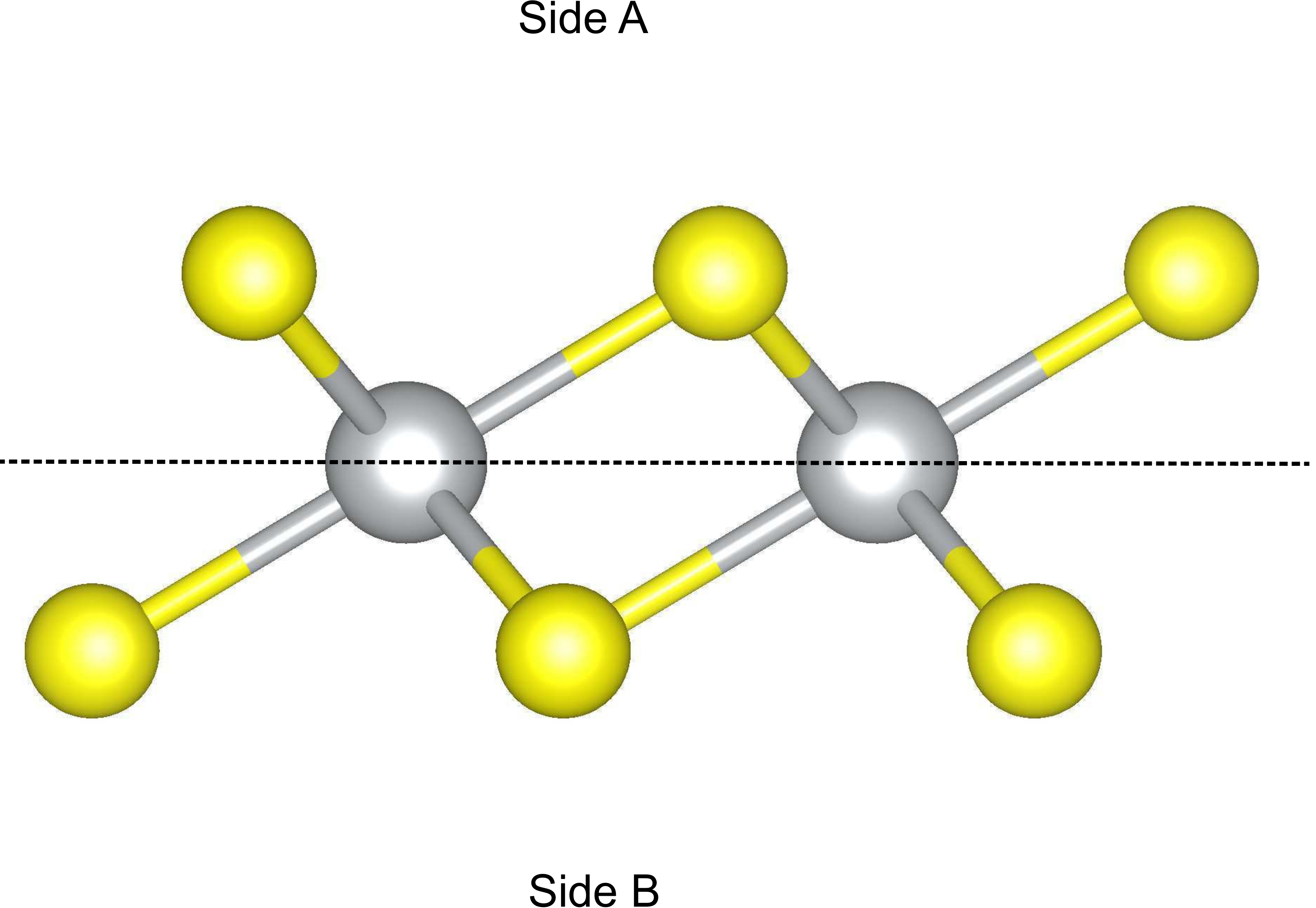}
    \end{subfigure}
     \begin{subfigure}[b]{0.3\columnwidth}
			\subcaption[]{}
	\includegraphics[width=\columnwidth,height=3cm,clip=true,keepaspectratio]{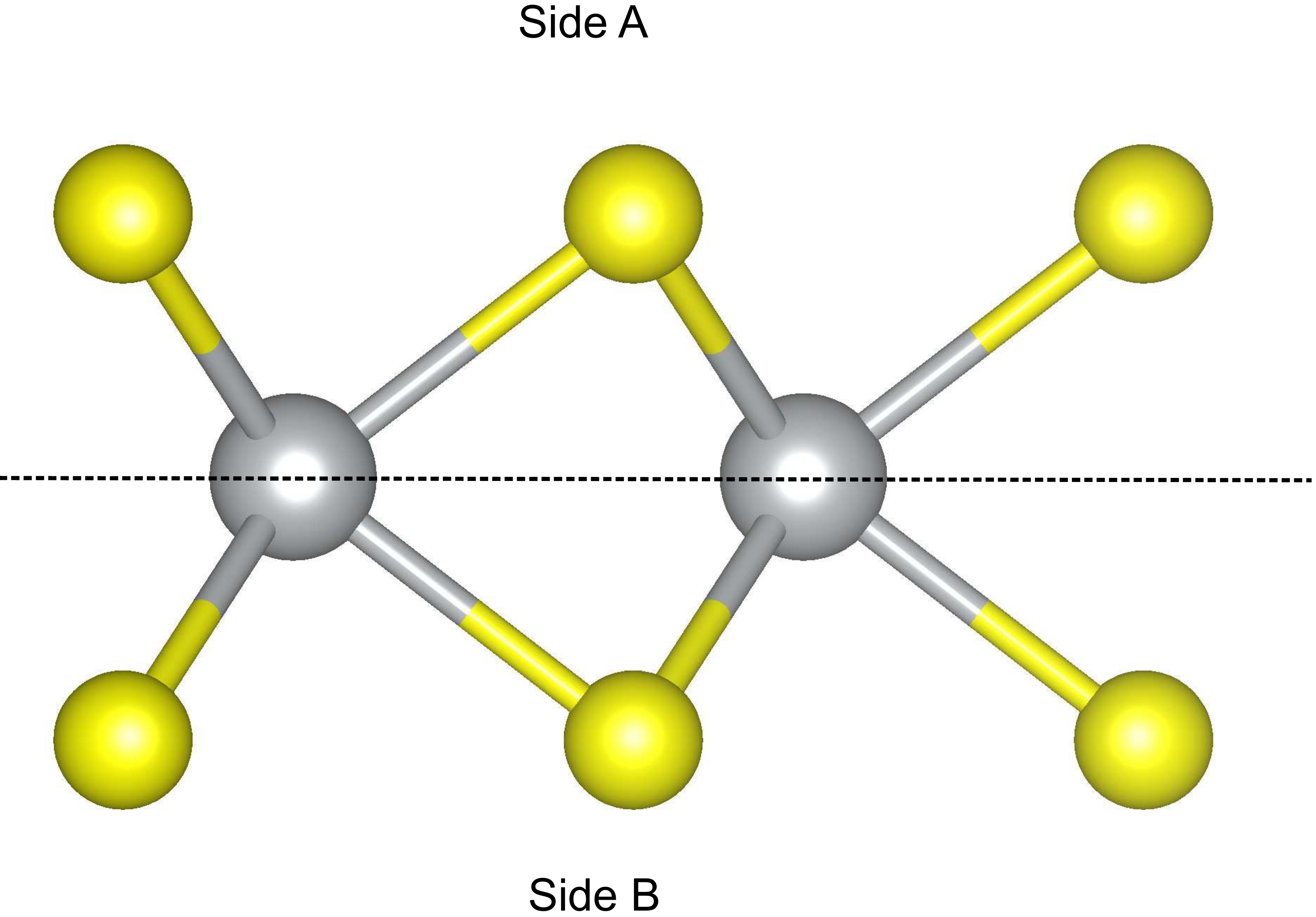}
    \end{subfigure}
\caption{Adsorption sites for H$_2$ adsorption on  MX$_2$ (M = Ni,Pt,Pd; X = S,Se) structures. a) top view of 1T-phase, b) top view of 2H-phase, c) side view of 1T-phase and d) side view of 2H-phase.}\label{fig:adsorption_position_MX2}
	\end{figure}

Recent studies have shown that reversible hydrogen storage in two-dimensional
materials requires adsorption energies within an intermediate physisorption
window, rather than maximal binding strength \cite{Chen2022PhysisorptionWindow}.

\begin{figure}[!ht]
		\centering
		\begin{subfigure}[b]{0.3\columnwidth}
			\subcaption[]{bridge (side A)}
	\includegraphics[width=\columnwidth,clip=true]{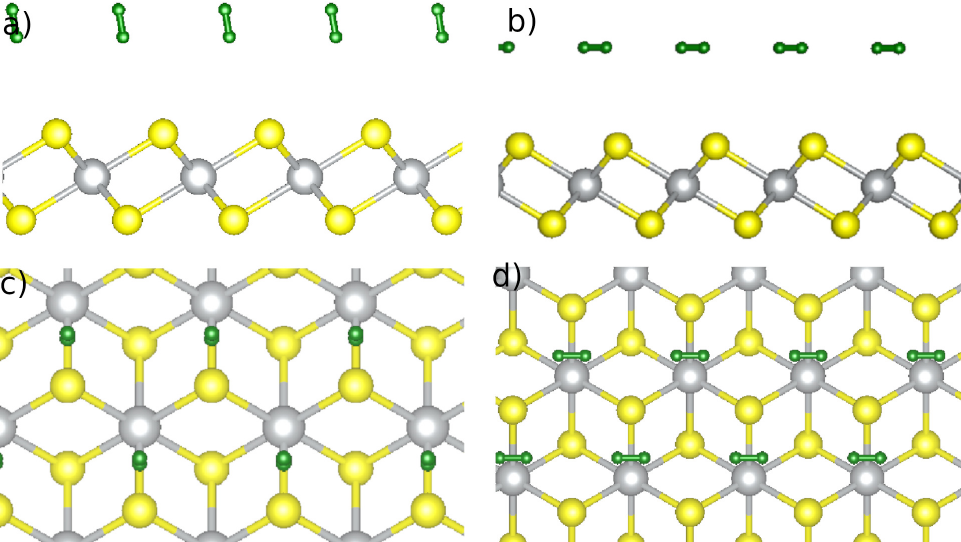}
		\end{subfigure}\hfill
		\begin{subfigure}[b]{0.3\columnwidth}
			\subcaption[]{bridge (side B)}
			\includegraphics[width=\columnwidth,clip=true]{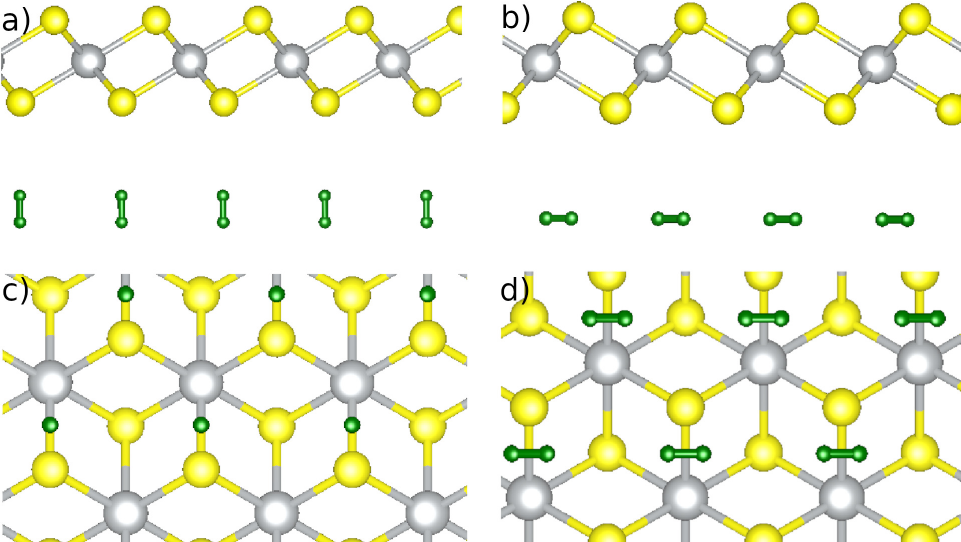}
		\end{subfigure}\hfill
		\begin{subfigure}[b]{0.3\columnwidth}
			\subcaption[]{on top-M (side A)}
			\includegraphics[width=\textwidth,clip=true]{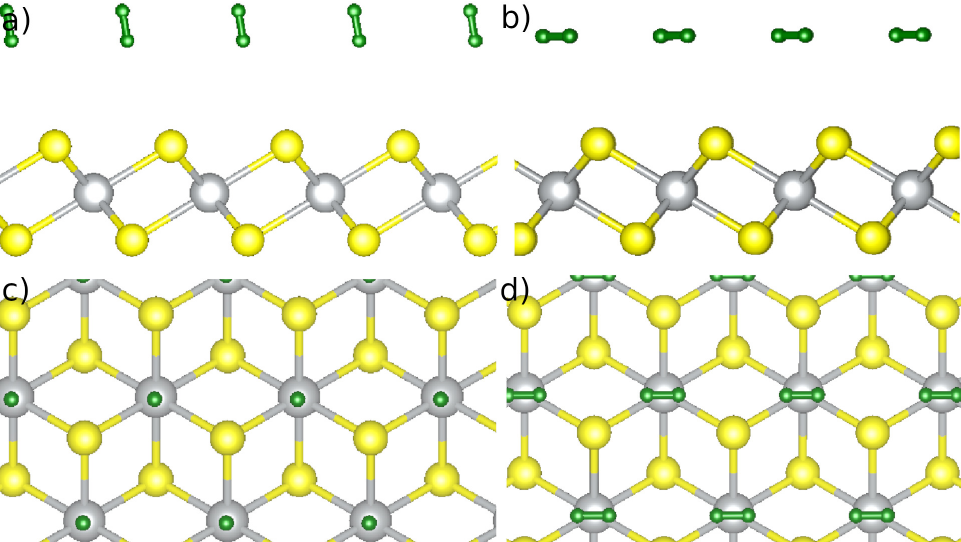}
		\end{subfigure}
		\\
		\begin{subfigure}[b]{0.3\columnwidth}
			\subcaption[]{on top-M (side B)}
			\includegraphics[width=\textwidth,clip=true]{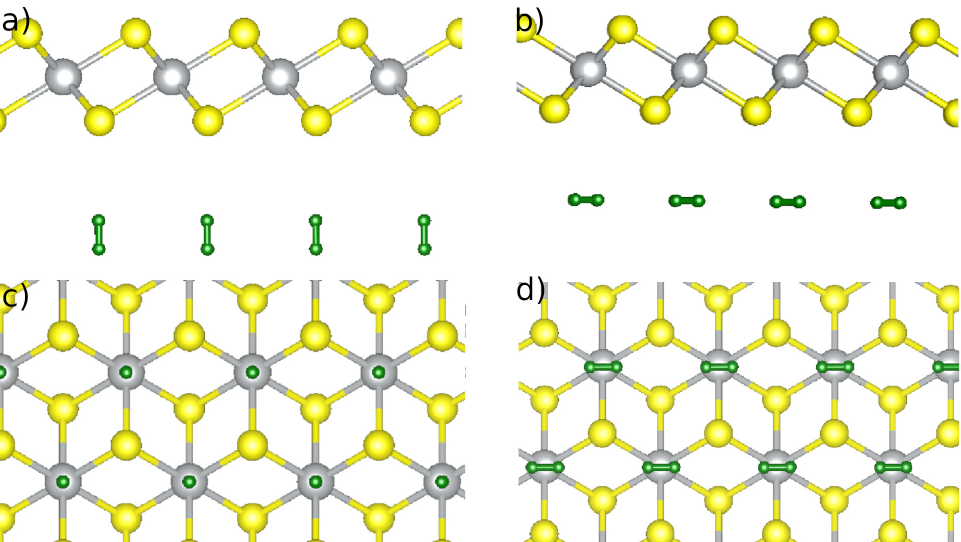}
		\end{subfigure}\hfill
		\begin{subfigure}[b]{0.3\columnwidth}
			\subcaption[]{ontop-S (side A)}
			\includegraphics[width=\textwidth,clip=true]{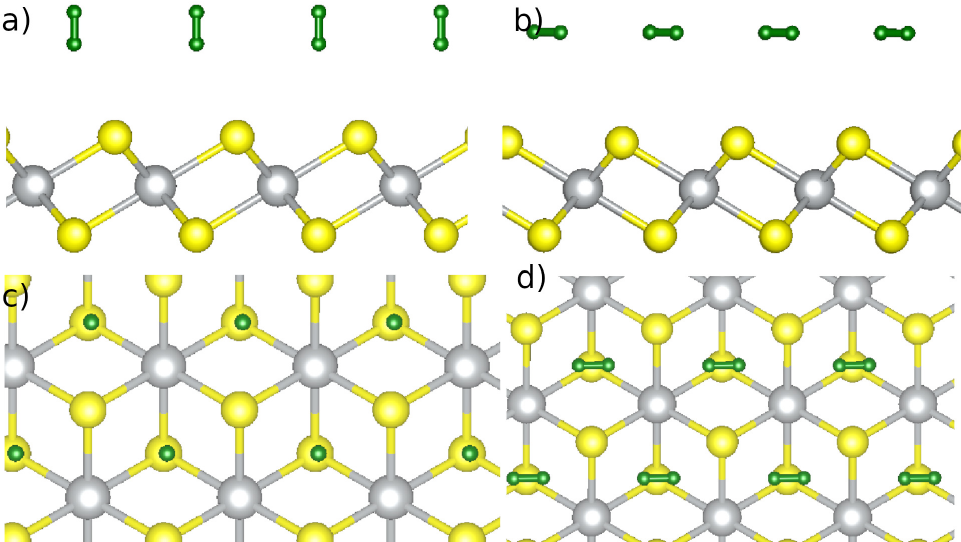}
		\end{subfigure}\hfill
		\begin{subfigure}[b]{0.3\columnwidth}
			\subcaption[]{on top-S (side B)}
			\includegraphics[width=\textwidth,clip=true]{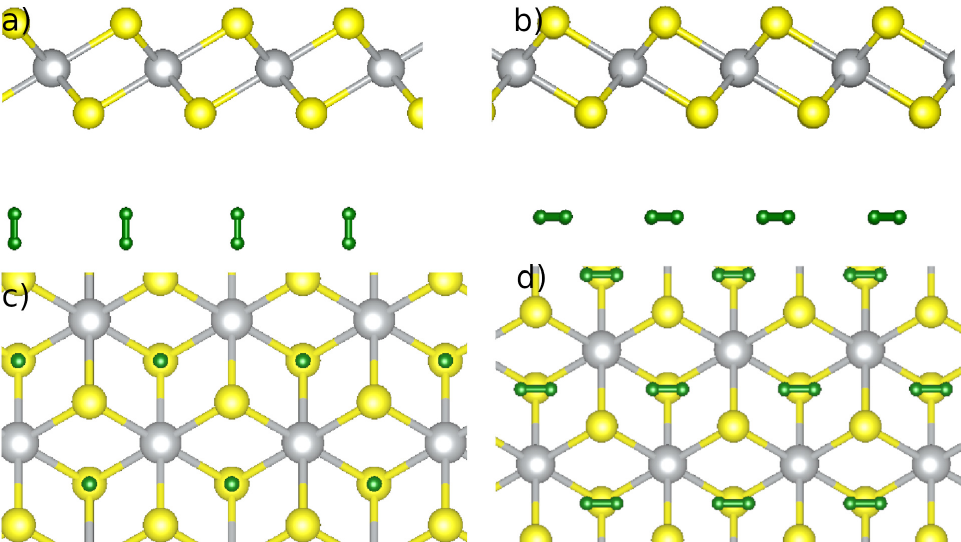}
		\end{subfigure}
		\caption{Relaxed geometries for a single hydrogen molecule adsorption on NiS$_2$: a) bridge (side A), b) bridge (side B), c) on top-Ni (side A), d) on top-Ni (side B), e) hollow (side A) and f) hollow (side B).}\label{fig:adsorption_MX2}
\end{figure}

From the perspective of hydrogen adsorption, the electronic structure
differences identified here have direct physical consequences. Systems
with a higher density of metal $d$ states near the Fermi level exhibit
enhanced electronic screening and surface polarizability, which
strengthen long-range dispersion interactions with molecular hydrogen.
Conversely, wide-gap semiconductors with chalcogen-dominated band edges
provide weaker screening and therefore weaker physisorption. The
interplay between structural polytype, chalcogen chemistry, and metal
$d$-orbital character thus establishes the electronic foundation for
the phase- and composition-dependent adsorption trends analyzed below.

\begin{table}
  \caption{Adsorption energies and optimized interatomic distances for molecular H$_2$ adsorbed on NiS$_2$. Values in parentheses correspond to the perpendicular orientation of H$_2$ with respect to the monolayer, while values without parentheses refer to the parallel configuration. Energies are given in (meV) and distances in (\AA).}
  \centering
\begin{tabular}{lccc}
\toprule
\multicolumn{4}{c}{NiS$_2$ -- 1T phase} \\
\midrule
Site & E$_{\rm ads}$ & d$_{\rm H-S}$ & d$_{\rm H-Ni}$ \\
\midrule
bridge (side A)  & -564(-557) & 2.86(3.24) & 3.90(4.14) \\
hollow (side A)  & -522(-548) & 2.98(3.22) & 3.92(4.27) \\
on top Ni (side A)   & -533(-544) & 2.98(3.37) & 3.38(3.83) \\
on top S (side A)       & -570(-564) & 2.92(3.30) & 4.51(4.87) \\
bridge (side B)  & -566(-557) & 3.15(3.14) & 3.95(4.14) \\
on top Ni (side B)   & -535(-535) & 5.75(3.05) & 4.26(3.70) \\
on top S (side B)       & -562(-543) & 2.95(3.16) & 3.40(4.34) \\
hollow (side B)  & -535(-536) & 4.35(3.12) & 4.38(4.61) \\
\midrule
\multicolumn{4}{c}{NiS$_2$ -- 2H phase} \\
\midrule
Site & E$_{\rm ads}$ & d$_{\rm H-S}$ & d$_{\rm H-M}$ \\
\midrule
bridge (side A)  & -0.64(-0.63) & 2.62(2.94) & 4.62(4.51) \\
hollow (side A)  & -0.65(-0.62) & 3.11(2.97) & 3.96(3.98) \\
on top Ni (side A)   & -0.66(-0.64) & 3.11(3.10) & 3.41(3.62) \\
on top S (side A)       & -0.65(-0.64) & 2.94(3.01) & 4.52(4.33) \\
bridge (side B)  & -0.64(-0.62) & 2.62(2.88) & 3.57(3.78) \\
on top Ni (side B)   & -0.65(-0.62) & 3.14(2.96) & 3.96(3.97) \\
on top S (side B)       & -0.65(-0.63) & 3.13(3.05) & 3.40(3.61) \\
hollow (side B)  & -0.65(-0.66) & 2.71(3.55) & 4.31(4.92) \\
\bottomrule
\end{tabular}
\label{tab:tabnis2}
\end{table}

\begin{table}
\centering
\caption{Adsorption energies $E_{\rm ads}$ (meV) and optimized interatomic distances $d$ (\AA) for molecular H$_2$ adsorbed on PdSe$_2$ in the 1T and 2H phases at different adsorption sites. Values in parentheses correspond to the perpendicular orientation of H$_2$ with respect to the monolayer, while values without parentheses refer to the parallel configuration. A dash (--) indicates that H$_2$ does not lead to stable structures.}
\begin{tabular}{lccc}
\toprule
\multicolumn{4}{c}{PdSe$_2$ -- 1T phase} \\
\hline
Site & $E_{\rm ads}$ & $d_{\rm H\!-\!Se}$ & $d_{\rm H\!-\!Pd}$  \\
\hline
bridge (side A)  & 27(51)      & 3.10(3.23) & 4.16(4.27)  \\
hollow (side A)  & 701(739)    & 2.99(3.32) & 4.05(5.10)  \\
on top-Pd (side A)   & 718(725)    & 3.02(3.53) & 5.05(4.05)  \\
hollow (side A)  & 32(735)     & 2.88(3.57) & 4.71(4.77)  \\
bridge (side B)  & --          & --         & --          \\
on top Pd (side B)   & --          & --         & --          \\
on top-Se (side B)      & --          & --         & --          \\
hollow (side B)  & --          & --         & --          \\
\hline
\multicolumn{4}{c}{PdSe$_2$ -- 2H phase} \\
\hline
Site & $E_{\rm ads}$ & $d_{\rm H\!-\!Se}$ & $d_{\rm H\!-\!Pd}$  \\
\hline
bridge (side A)  & --          & --         & --          \\
hollow (side A)  & -578(-540)  & 3.36(3.00) & 4.25(4.12)  \\
on top Pd (side A)   & -561(-554)  & 3.03(3.21) & 3.39(3.71)  \\
on top Se (side A)  & --          & --         & --          \\
bridge (side B)  & --          & --         & --          \\
on-top Pd (side B)   & -550(-550)  & 2.95(3.36) & 5.52(3.96)  \\
on top Se (side B)      & -585(-585)  & 3.28(3.17) & 3.52(4.77)  \\
hollow (side B)  & --          & --         & -- \\   
\bottomrule
\end{tabular}
\label{tab:pdse2}
\end{table}

\begin{table}
  \centering
\caption{Adsorption energies $E_{\rm ads}$ (meV) and optimized interatomic distances $d$ (\AA) for molecular H$_2$ adsorbed on NiSe$_2$ in the 1T and 2H phases at different adsorption sites. Values in parentheses correspond to the perpendicular orientation of H$_2$ with respect to the monolayer, while values without parentheses refer to the parallel configuration. A dash (--) indicates that H$_2$ does not lead to stable structures.}
\begin{tabular}{lccc}
\toprule
\multicolumn{4}{c}{NiSe$_2$ -- 1T phase} \\
\hline
Site & $E_{\rm ads}$ & $d_{\rm H\!-\!Se}$ & $d_{\rm H\!-\!Ni}$   \\
\hline
bridge (side A)   & -549(960)   & 3.05(3.19) & 4.09(4.12)  \\
hollow (side A)   & -576(928)   & 3.33(3.13) & 4.38(4.21)   \\
on top Ni (side A)    & -573(921)   & 3.03(3.28) & 3.47(4.09)  \\
on top Se (side B)   & -580(946)   & 2.90(3.20) & 4.62(4.76)  \\
bridge (side B)   & -558(938)   & 3.38(3.25) & 4.24(4.32)  \\
on top Ni (side B)    & -568(1052)  & 3.52(3.15) & 4.35(4.73)  \\
on top Se (side B)       & -576(975)   & 3.38(3.13) & 3.94(4.36)   \\
hollow (side B)   & -553(978)   & 2.87(3.06) & 4.59(3.79)   \\
\hline
\multicolumn{4}{c}{NiSe$_2$ -- 2H phase} \\
\hline
Site & $E_{\rm ads}$ & $d_{\rm H\!-\!Se}$ & $d_{\rm H\!-\!Ni}$  \\
\hline
bridge (side A)   & --            & --         & --          \\
hollow (side A)   & -567(-127)    & 3.01(3.43) & 3.99(4.69)  \\
on top Ni (side A)    & -572(-92)     & 2.96(3.06) & 3.41(3.77)  \\
on top Se (side B)   & --            & --         & --          \\
bridge (side B)   & --            & --         & --          \\
on top Ni (side B)    & -557(-1332)   & 3.20(3.21) & 5.79(3.98)   \\
on top Se (side B)   & -572(-1340)   & 3.20(3.36) & 3.73(4.98)   \\
hollow (side B)   & --            & --         & --           \\
\bottomrule
\end{tabular}
\label{tab:tabnise2}
\end{table}

The adsorption of a single H$_2$ molecule on pristine MX$_2$ monolayers
was investigated to establish the intrinsic interaction strength,
preferred adsorption sites, and the role of surface asymmetry prior to
considering coverage effects. The high-symmetry adsorption sites
considered in this work are illustrated in
Fig.~\ref{fig:adsorption_position_MX2}, and representative relaxed
adsorption geometries are shown in Fig.~\ref{fig:adsorption_MX2}. The
corresponding adsorption energies and key structural parameters are
reported in Tables~\ref{tab:tabnis2}--\ref{tab:tabnise2}.

Across all pristine systems and for both polymorphs, hydrogen adsorption
remains strictly molecular. In every relaxed configuration, the H--H
bond length remains close to its gas-phase value and no spontaneous
dissociation is observed. This result holds irrespective of transition-
metal species, chalcogen chemistry, adsorption site, or surface side.
The absence of dissociation confirms that hydrogen adsorption on
pristine MX$_2$ monolayers is governed by physisorption rather than
chemisorption, even in cases where the adsorption energy is relatively
large in magnitude.

Despite the universal molecular character, pronounced differences in
adsorption strength and site selectivity are observed. Adsorption at
sites located above chalcogen atoms or along metal--chalcogen bonds is
generally favored over hollow positions. This preference reflects the
dominant role of local polarization effects: proximity to chalcogen
atoms enhances the induced dipole interaction between the H$_2$
molecule and the surface, whereas hollow sites distribute the
interaction over multiple atoms and therefore weaken the net
van der Waals coupling. The relatively small energy differences between
bridge and on-top sites further indicate that adsorption occurs on a
shallow potential-energy surface, consistent with physisorption.

A robust chemical trend emerges when comparing different transition-
metal species. For a given chalcogen and structural polytype, the
adsorption strength follows the ordering Ni > Pd > Pt, as summarized in Tables~\ref{tab:tabnis2}--\ref{tab:tabnise2}. This trend correlates directly with the electronic structure calculations discussed above. Ni-based
systems exhibit a higher density of localized $d$ states near the Fermi level, which enhances surface polarizability and strengthens long-range dispersion interactions with molecular hydrogen. In contrast, the more delocalized $d$ states of Pd and Pt contribute less effectively
to screening, resulting in weaker adsorption.

The influence of the structural polytype is superimposed on this
chemical trend. For most compositions, the 1T phase exhibits
substantially stronger adsorption than the corresponding 2H phase. This enhancement is particularly striking in NiS$_2$, where the 1T polymorph supports uniformly strong molecular adsorption across all sites, while the 2H phase is effectively inert, with adsorption energies close to zero. The contrast between the two phases reflects the metallic character of the 1T phase, which provides enhanced electronic screening, and the semiconducting nature of the 2H phase, which suppresses surface
polarization.

Notable exceptions further highlight the interplay between electronic structure and chemistry. In PdSe$_2$, hydrogen adsorption is unfavorable in the 1T phase but becomes moderately stable in the 2H polymorph. This
behavior can be traced to the chalcogen-dominated band edges and the reduced contribution of Pd $d$ states near the Fermi level, which limits screening in the 1T phase despite its octahedral coordination. Conversely, NiSe$_2$ exhibits strong adsorption in both polymorphs,
consistent with its metallic or near-metallic electronic structure.

Finally, small but systematic differences are observed between the two
inequivalent surfaces (side A and side B) of the pristine
layers. These differences arise from the absence of inversion symmetry
in the slab geometry and manifest as modest variations in adsorption
energy and adsorption height. However, they do not alter the
qualitative trends with respect to metal species or structural
polytype.

Thus, the single-molecule adsorption results establish three key
points that guide the subsequent analysis. First, hydrogen adsorption
on pristine MX$_2$ monolayers is universally molecular and governed by
physisorption. Second, adsorption strength is controlled primarily by
transition-metal identity and electronic screening rather than by
specific adsorption sites. Third, the structural polytype plays a
decisive role, with metallic 1T phases generally enhancing adsorption
relative to their 2H counterparts. These insights provide the baseline
for understanding coverage-dependent behavior and finite-temperature
stability, which are examined next.

\subsection{Coverage-dependent adsorption and hydrogen--hydrogen interactions}

\begin{figure}[!ht]
\centering
\begin{subfigure}[b]{0.2\columnwidth}
\subcaption[]{}
\includegraphics[width=\columnwidth,clip=true]{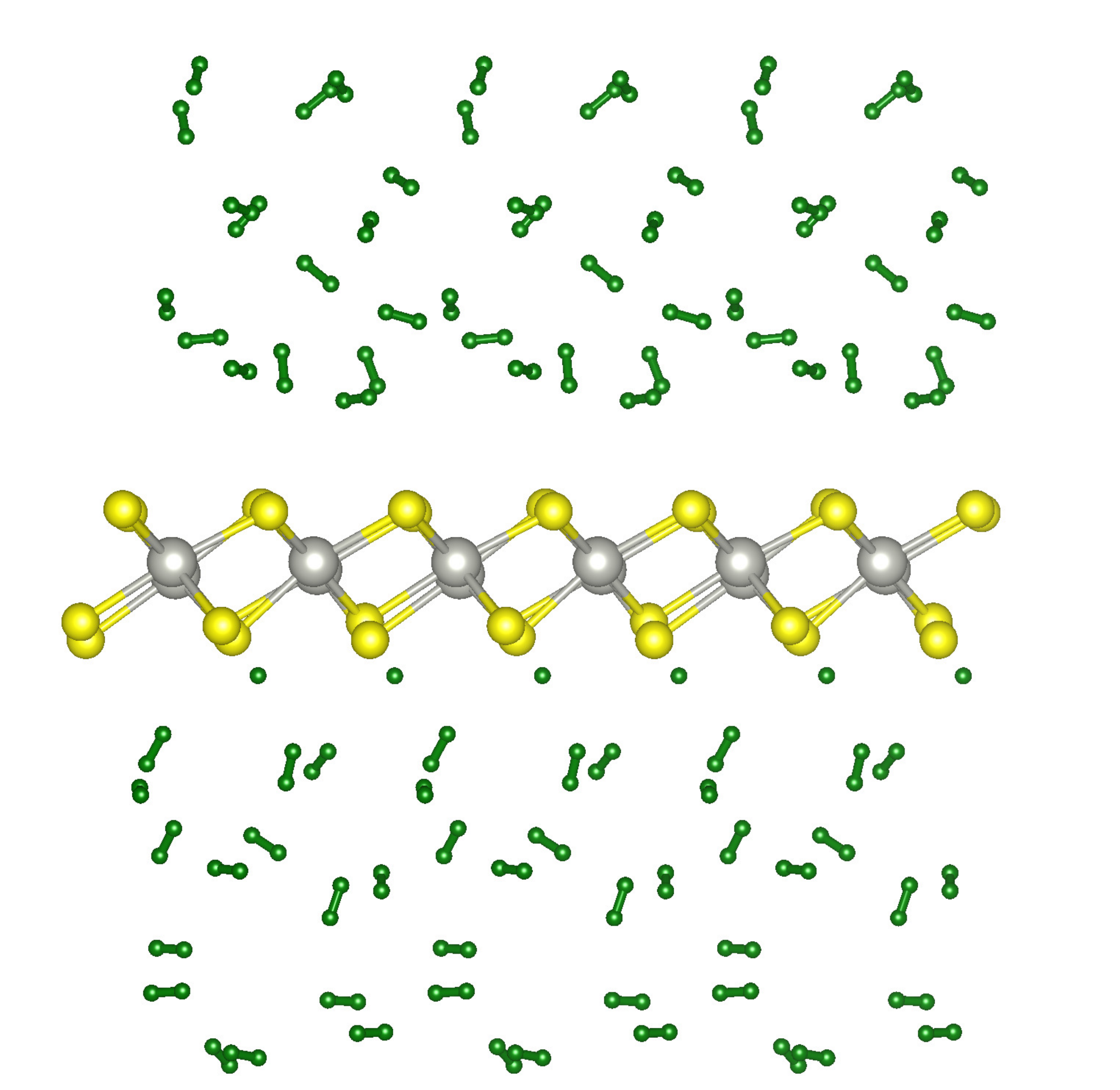}
\end{subfigure}
\begin{subfigure}[b]{0.2\columnwidth}
\subcaption[]{}
\includegraphics[width=\columnwidth,clip=true]{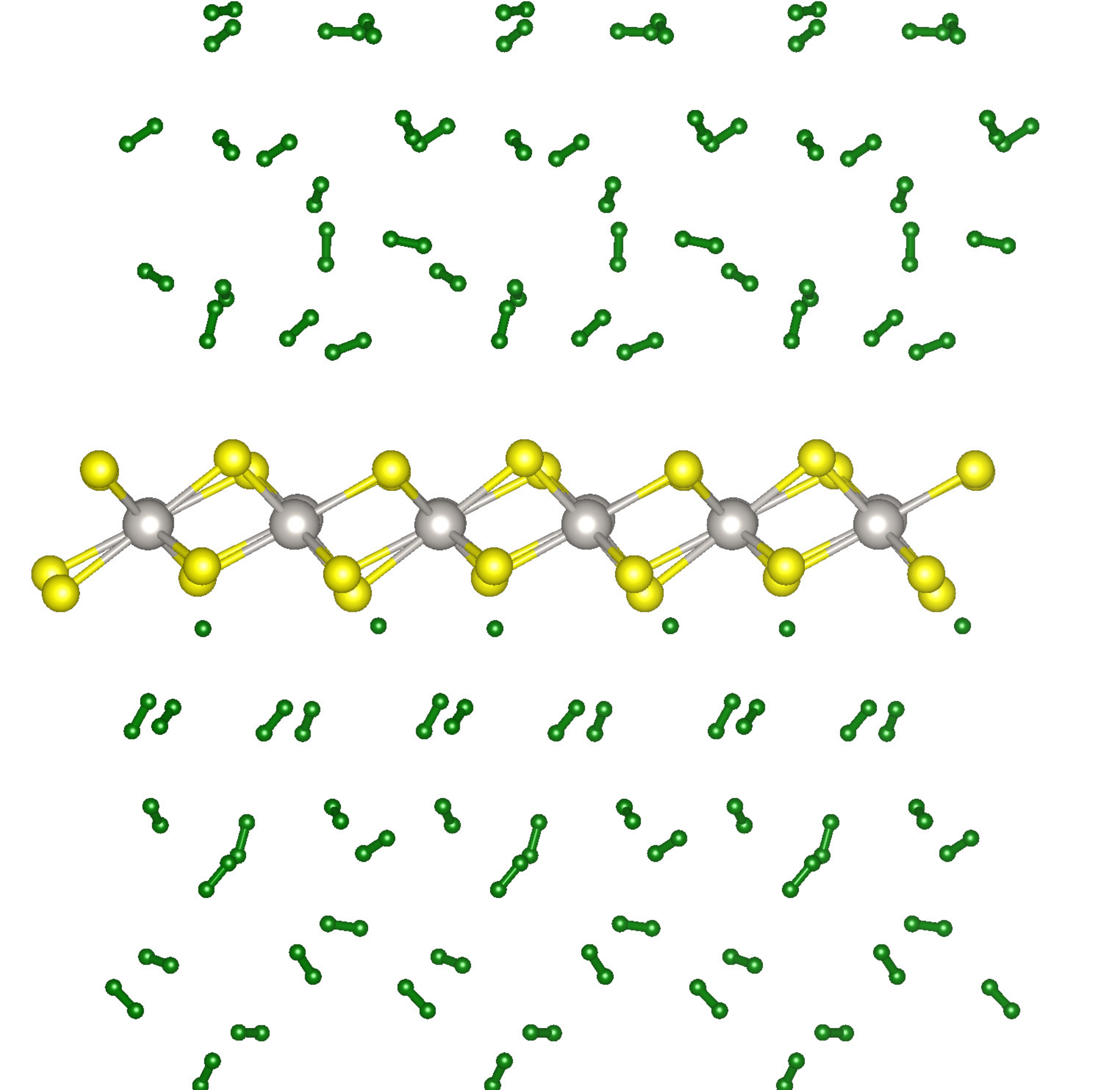}
\end{subfigure}
\begin{subfigure}[b]{0.2\columnwidth}
\subcaption[]{}
\includegraphics[width=\columnwidth,clip=true]{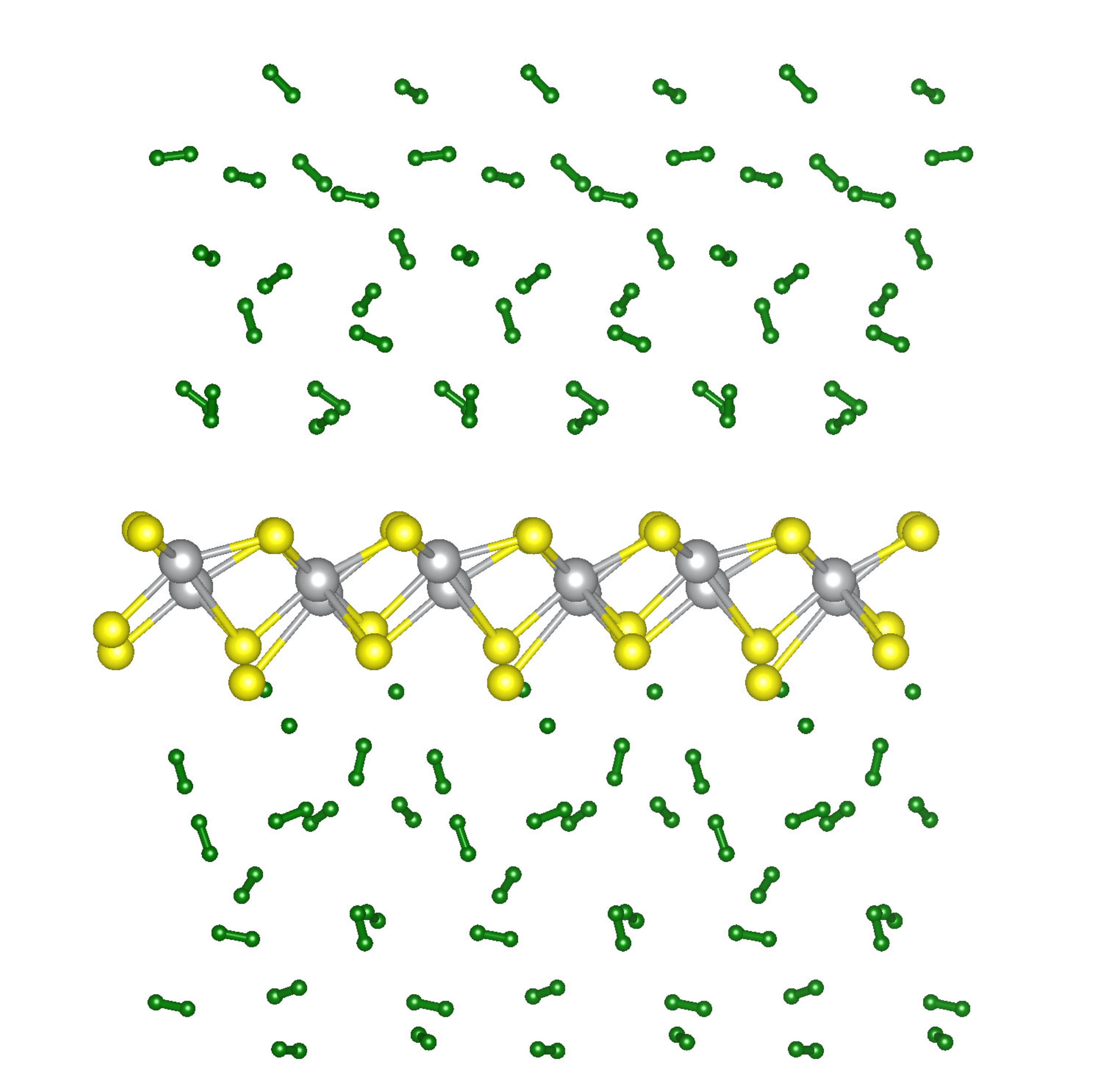}
\end{subfigure}
\begin{subfigure}[b]{0.2\columnwidth}
\subcaption[]{}
\includegraphics[width=\columnwidth,clip=true]{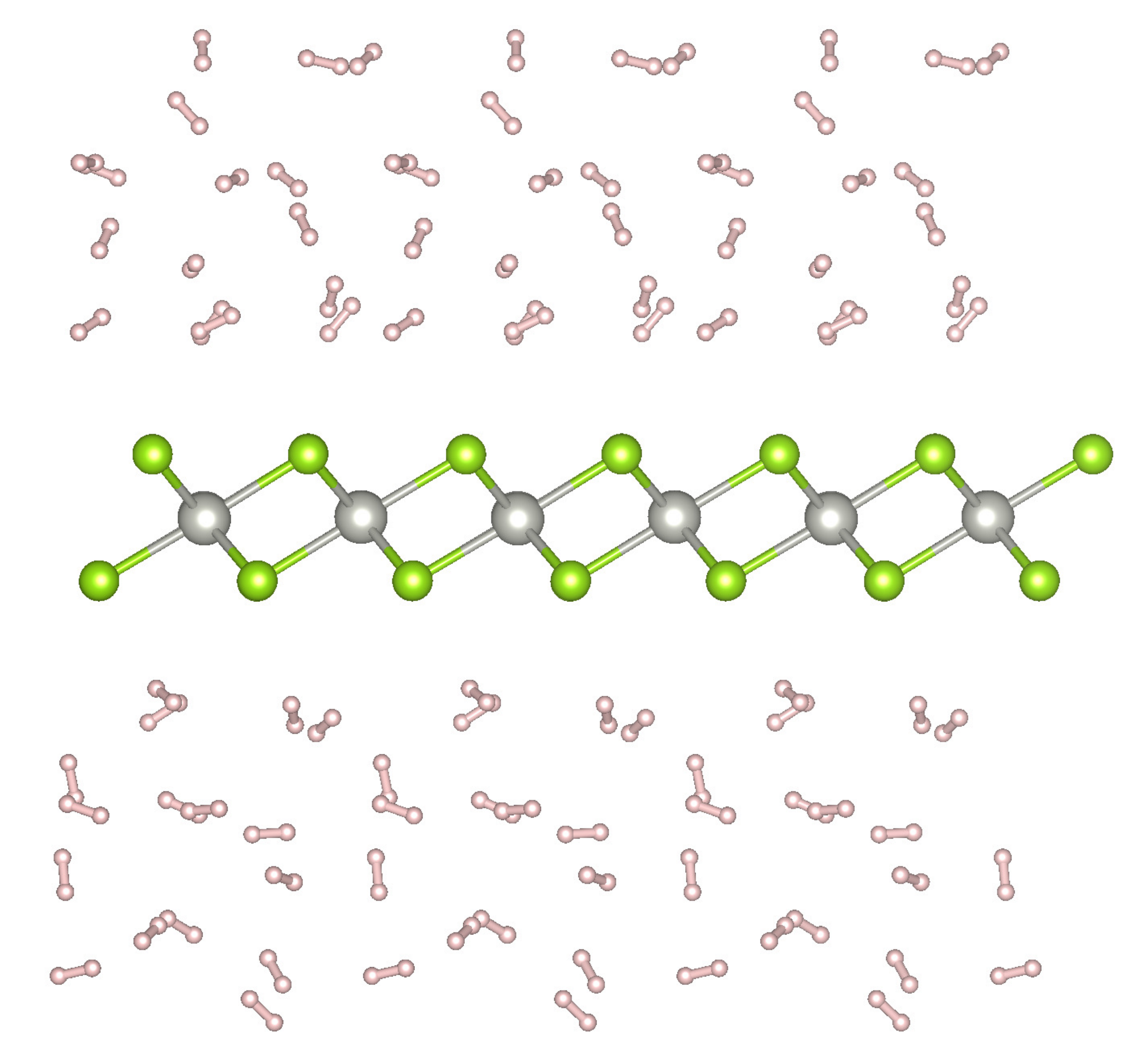}
\end{subfigure}\\
\begin{subfigure}[b]{0.2\columnwidth}
\subcaption[]{}
\includegraphics[width=\columnwidth,clip=true]{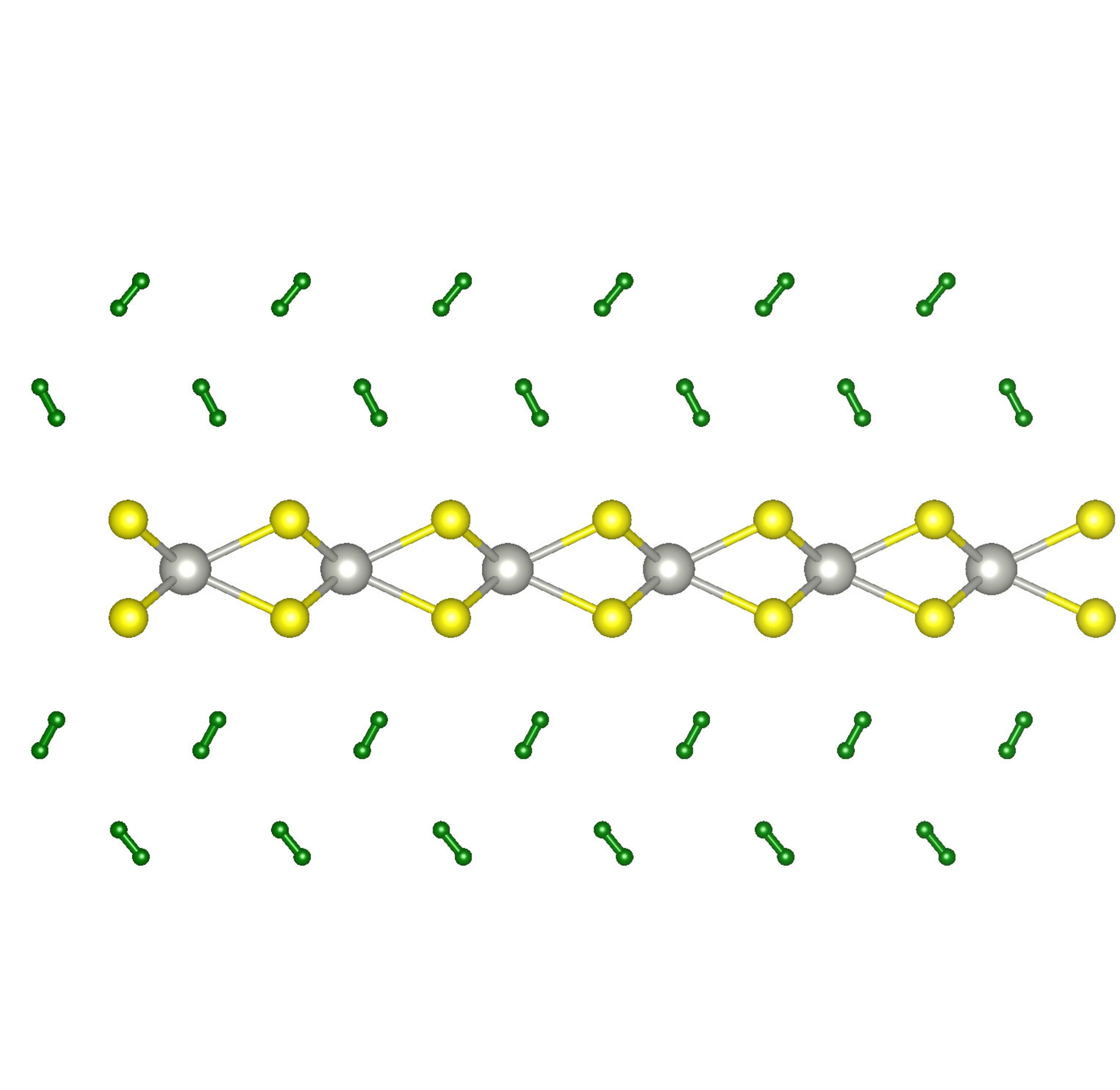}
\end{subfigure}
\begin{subfigure}[b]{0.2\columnwidth}
\subcaption[]{}
\includegraphics[width=\columnwidth,clip=true]{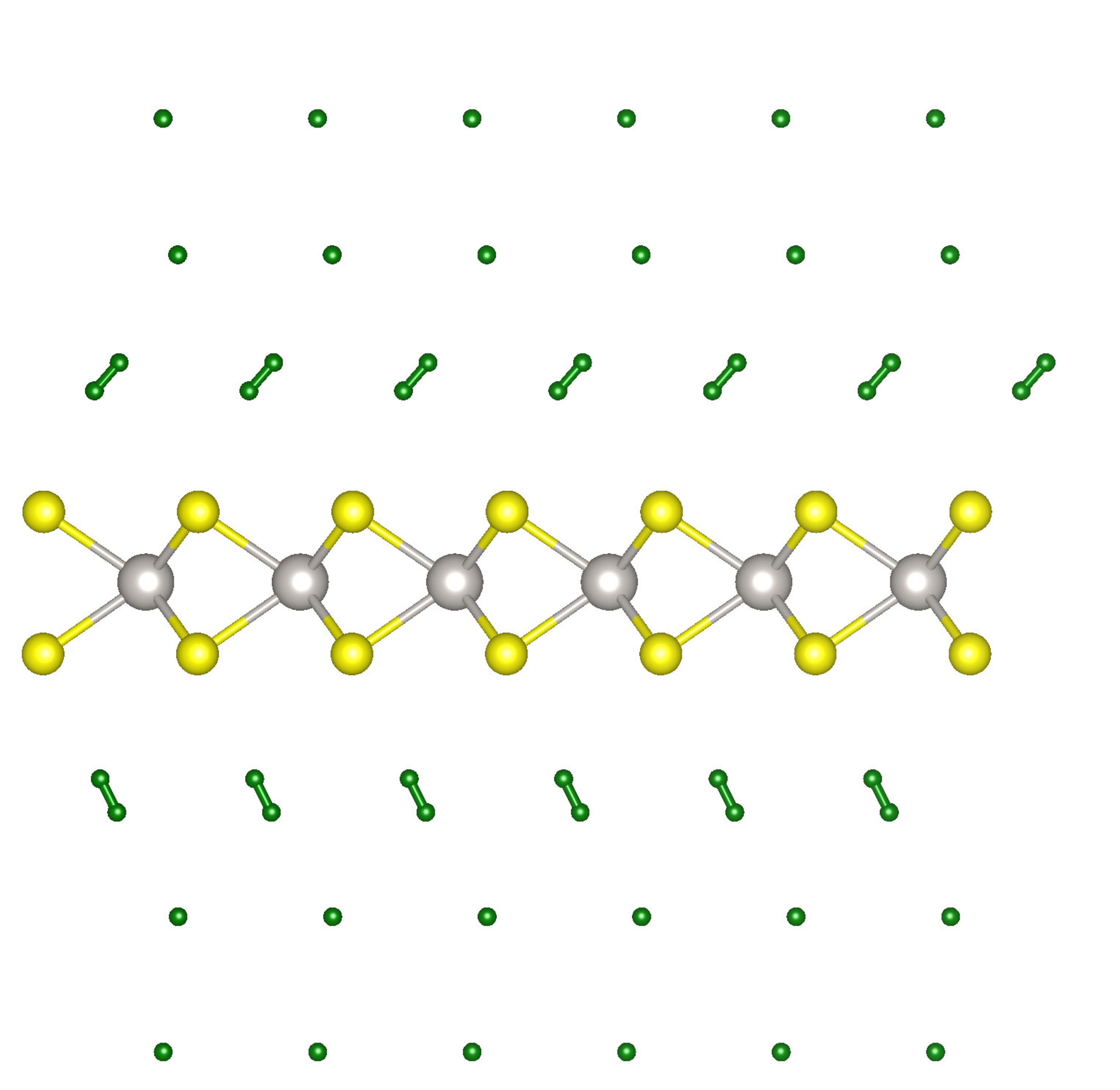}
\end{subfigure}
\begin{subfigure}[b]{0.2\columnwidth}
\subcaption[]{}
\includegraphics[width=\columnwidth,clip=true]{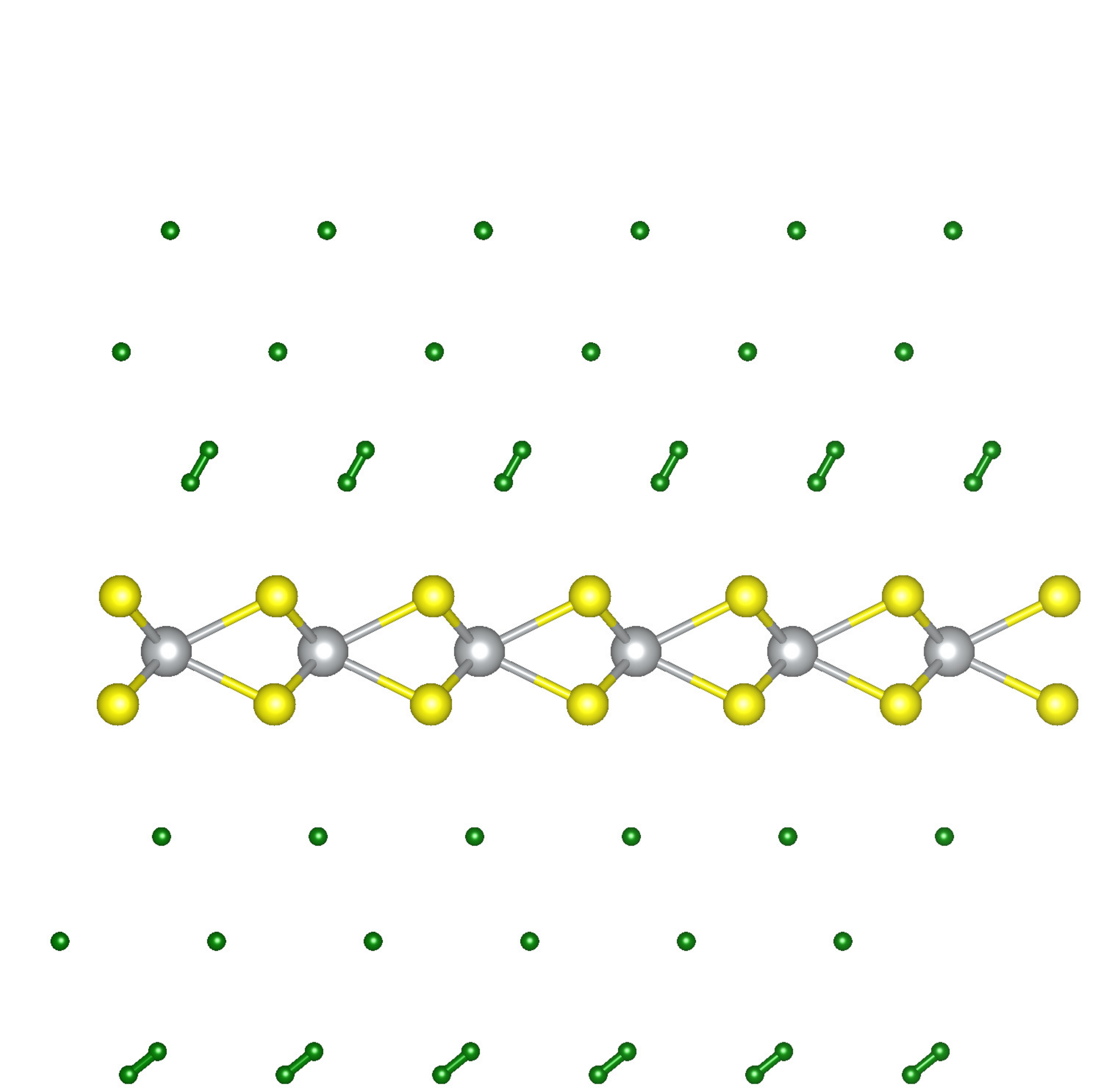}
\end{subfigure}
\begin{subfigure}[b]{0.2\columnwidth}
\subcaption[]{}
\includegraphics[width=\columnwidth,clip=true]{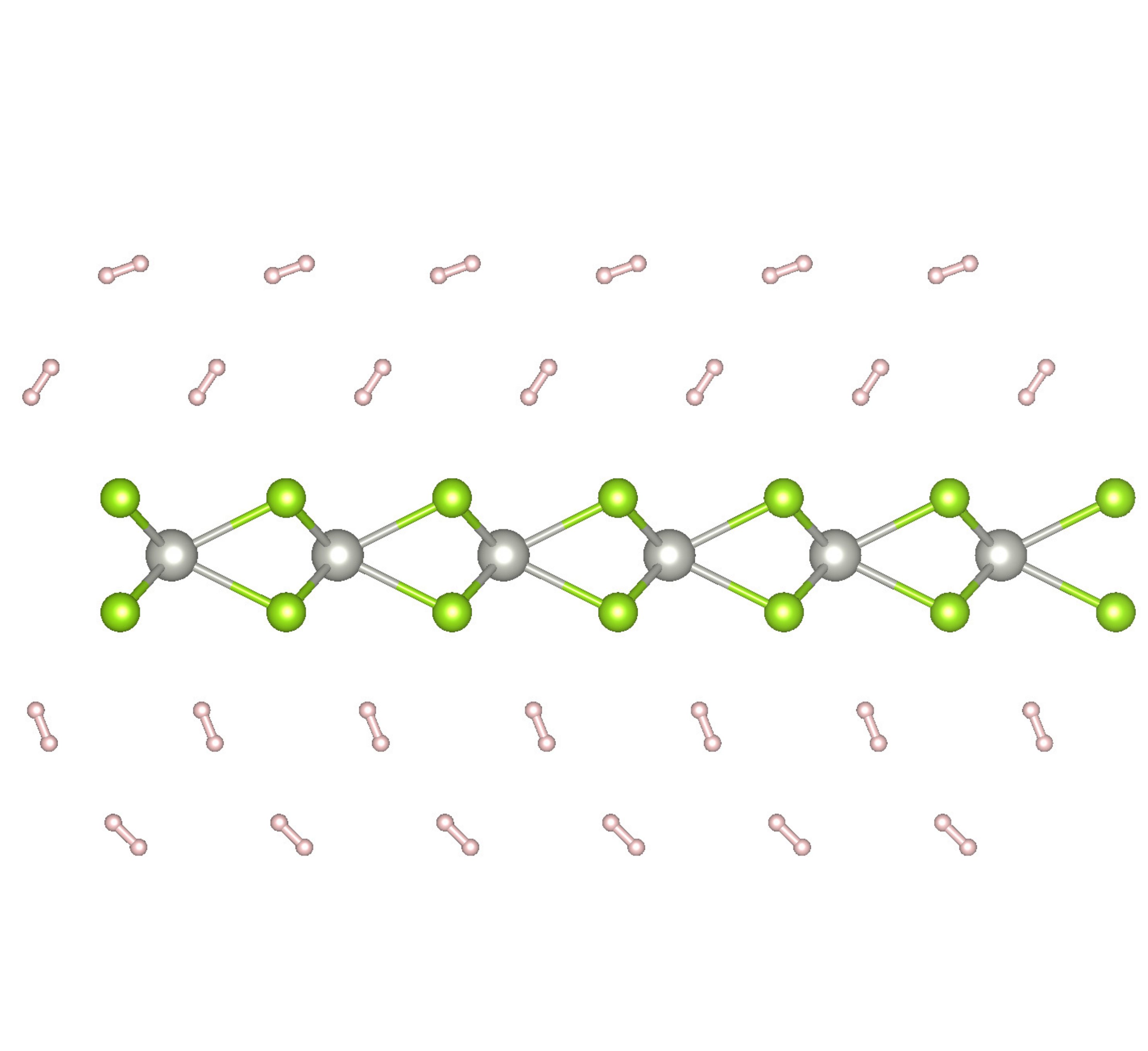}
\end{subfigure}
\caption{Relaxed geometries at T = 0K of 32 H$_2$ molecules adsorbed on a) 1T-PdS$_2$, b) 1T-PtS$_2$, c) 1T-NiS$_2$, d) 2H-PdSe$_2$, e) 2H-PdS$_2$, f) 2H-PtS$_2$, g) 2H-NiS$_2$ and h) 2H-PdSe$_2$ monolayers.}
\label{fig:h2_all_relaxed}
\end{figure}

\begin{figure}[!ht]
\centering
\begin{subfigure}[b]{0.5\columnwidth}
\subcaption[]{}
\includegraphics[width=\columnwidth,clip]{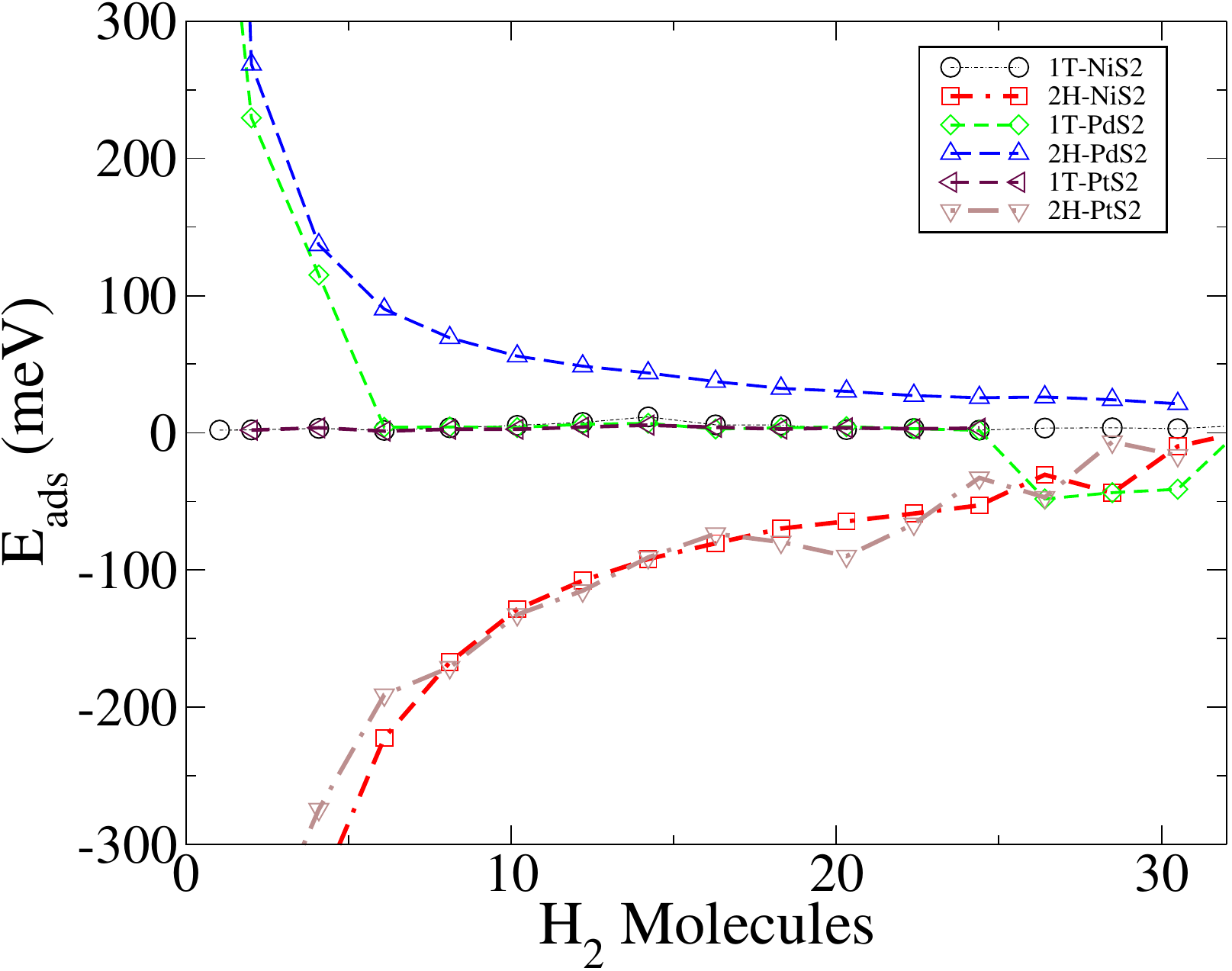}
\end{subfigure}\hfill
\begin{subfigure}[b]{0.5\columnwidth}
\subcaption[]{}
\includegraphics[width=\columnwidth,clip]{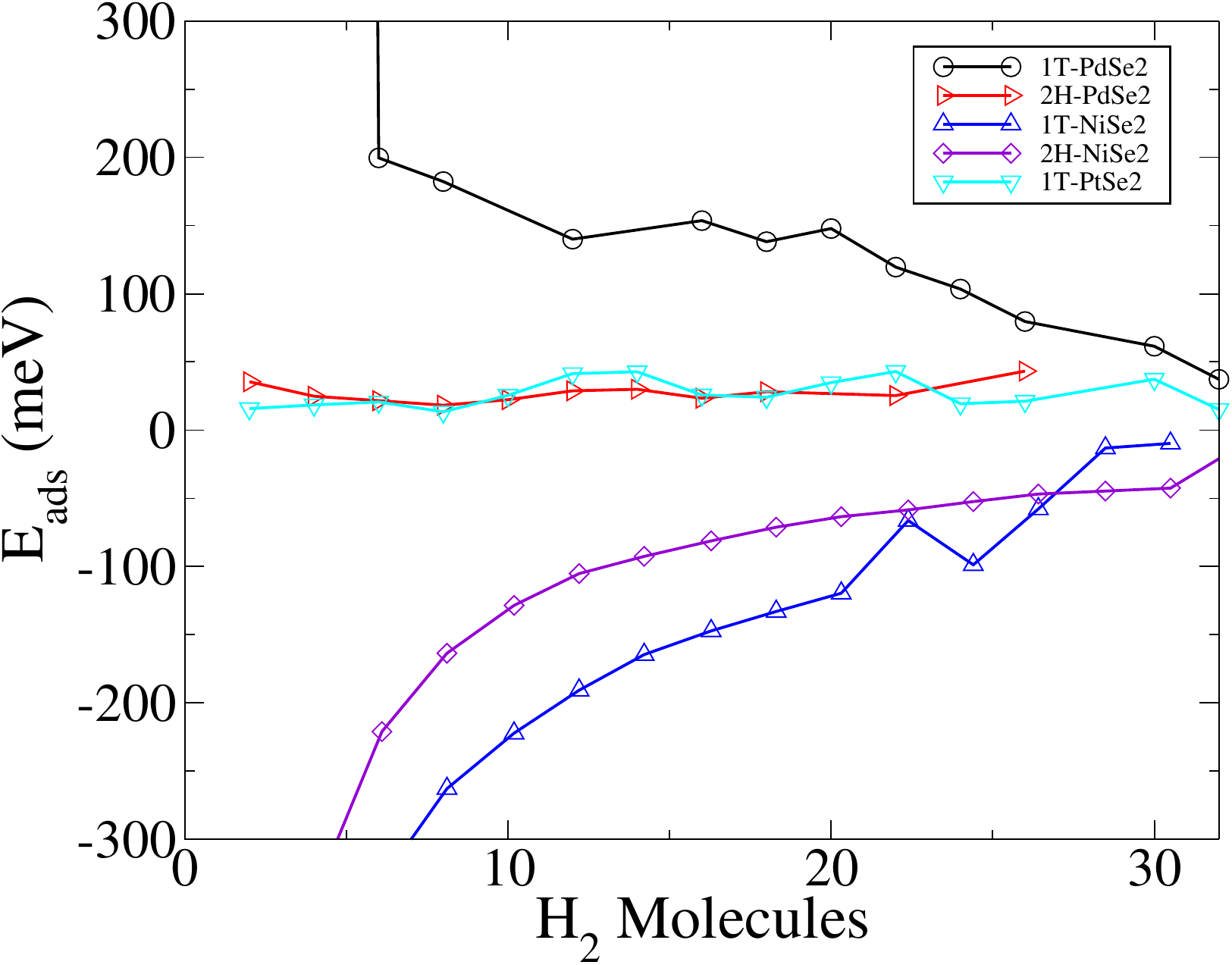}
\end{subfigure}
\caption{Adsorption energy of H$_2$ for a) MS$_2$ and b) MSe$_2$ (M= Pd,Pt and Ni) as a function of the number of H$_2$ molecules.}\label{fig:eadspermoleculesnojanus}
\end{figure}

While single-molecule adsorption establishes the intrinsic interaction
strength between hydrogen and the surface, practical hydrogen storage
necessarily involves finite and often high coverages. To assess the
evolution of adsorption behavior beyond the dilute limit, we therefore
investigate hydrogen adsorption as a function of H$_2$ loading, with up
to 32 molecules adsorbed per supercell. Representative relaxed
configurations at different coverages are shown in
Fig.~\ref{fig:h2_all_relaxed}, and the corresponding adsorption energy
per H$_2$ molecule is plotted as a function of coverage in
Fig.~\ref{fig:eadspermoleculesnojanus}. Structural and geometric metrics
relevant to multilayer formation are summarized in Table~\ref{tab:TMD_thickness},
while high-coverage adsorption data are reported in Table~\ref{tab:TMD_Janus_thickness}.

At low coverage, the adsorption energies closely follow the trends
identified for single-molecule adsorption in Sec.~3.3. In particular,
1T-NiS$_2$ exhibits the most negative adsorption energies, reflecting
strong surface polarizability and efficient screening. As the number of
adsorbed H$_2$ molecules increases, the adsorption energy per molecule
becomes progressively less negative for all systems. This monotonic
reduction reflects the gradual saturation of the most favorable
adsorption sites and the increasing importance of repulsive
hydrogen--hydrogen interactions within the adsorbed layer.

The competition between molecule--surface attraction and
molecule--molecule repulsion leads to markedly different coverage
behavior in the 1T and 2H polymorphs. In the 1T phase, the stronger
individual adsorption energies favor initial uptake at low coverage,
but the relatively compact surface geometry limits the available space
for additional hydrogen molecules. As a result, H$_2$--H$_2$ repulsion
sets in at comparatively lower coverage, leading to a faster reduction
of the average adsorption energy as the surface becomes crowded.

In contrast, the 2H polymorphs exhibit weaker adsorption at low
coverage but provide a more open trigonal-prismatic coordination
environment. This geometric openness increases the accessible adsorption
volume and allows hydrogen molecules to distribute more uniformly above
the surface. Consequently, although the adsorption energies per
molecule are smaller in magnitude, the onset of strong
hydrogen--hydrogen repulsion is delayed, enabling more efficient
multilayer adsorption at higher coverage.

This behavior is particularly evident when comparing sulfide and
selenide systems. Heavier Se-based compounds generally exhibit stronger
binding at low coverage due to enhanced polarizability, but their larger
mass reduces the achievable gravimetric hydrogen density. In contrast,
lighter sulfides can attain higher gravimetric capacities despite
weaker binding, underscoring that adsorption strength alone is not a
reliable indicator of storage performance.

The data in Table~\ref{tab:TMD_Janus_thickness} further illustrate this
decoupling between adsorption energy and storage capacity. Systems with
the strongest binding do not necessarily exhibit the highest hydrogen
uptake, as excessive binding strength amplifies intermolecular
repulsion and reduces packing efficiency. Optimal storage performance
instead arises from a balance between moderate adsorption energies and
sufficient geometric space to accommodate multiple hydrogen layers.

Overall, the coverage-dependent analysis reveals that hydrogen storage
in pristine MX$_2$ monolayers is governed by a subtle interplay between
electronic screening, surface geometry, and intermolecular repulsion.
While metallic 1T phases maximize adsorption strength at low coverage,
the more open 2H polymorphs are better suited for accommodating high
hydrogen loadings. These competing effects highlight the importance of
considering coverage and hydrogen--hydrogen interactions explicitly
when evaluating storage performance and motivate the finite-temperature
analysis presented in the following section.

\subsection{Finite-temperature stability and dynamical behavior from AIMD}

\begin{table}
  \centering
\begin{tabular}{lcccc}
\toprule
phase & surface H$_2$ & \multicolumn{2}{c}{$\Delta$ (\AA)} & gravimetric density (\%) \\
\hline
& & DFT ($T=0$ K) & AIMD ($T=300$ K) &\\
\hline
1T-PdS$_2$     & Side A (B)      & $9.70 \, (9.53)$ & $17.19 \, (15.82)$ & 8.04\\
1T-PtS$_2$     & Side A (B)      & $10.09 \, (11.36)$ & $15.21 \, (14.33)$ & 5.45 \\
1T-NiS$_2$     & Side A (B)      & $10.41 \, (8.27)$  & $15.02 \, (16.84)$ & 10.8\\
\hline
1T-PdSe$_2$    & Side A (B)    & $9.43 \, (9.38)$ & $14.88 \, (22.00)$ & 5.71\\
1T-PtSe$_2$    & Side A (B)    & $10.28 \, (11.36)$ & $15.22 \, (15.13)$ & 4.33\\
1T-NiSe$_2$    & Side A (B)    & $7.26 \, (7.28)$ & $14.94 \, (20.09)$ & 6.87\\
\hline
2H-PdS$_2$     & Side A (B)      & $5.03 \, (5.03)$ & $13.00 \, (14.91)$ & 8.53\\
2H-PtS$_2$     & Side A (B)      & $7.67 \, (7.77)$ & $16.35 \, (12.77)$ & 5.79\\
2H-NiS$_2$     & Side A (B)      & $7.76 \, (8.23)$ & $12.31 \, (15.89)$ & 11.44\\
\hline
2H-PdSe$_2$    & Side A (B)    & $6.36 \, (6.24)$ & $14.32 \, (14.31)$ & 3.99\\
2H-PtSe$_2$    & Side A (B)    & $9.18 \, (9.64)$ & $13.75 \, (16.33)$ & 3.28\\
2H-NiSe$_2$    & Side A (B)     & $9.48 \, (9.57)$  & $17.93 \, (15.03)$ & 6.87\\
\bottomrule
\end{tabular}
\caption{H$_2$ layer thickness $\Delta$ for  DFT and AIMD calculations on side A (top layers) and B (bottom layers). Gravimetric density is also given.}
\label{tab:TMD_thickness}
\end{table}

\begin{figure}[!ht]
\centering
\begin{subfigure}[b]{0.45\columnwidth}
\subcaption[]{}
\includegraphics[width=\columnwidth,clip=true]{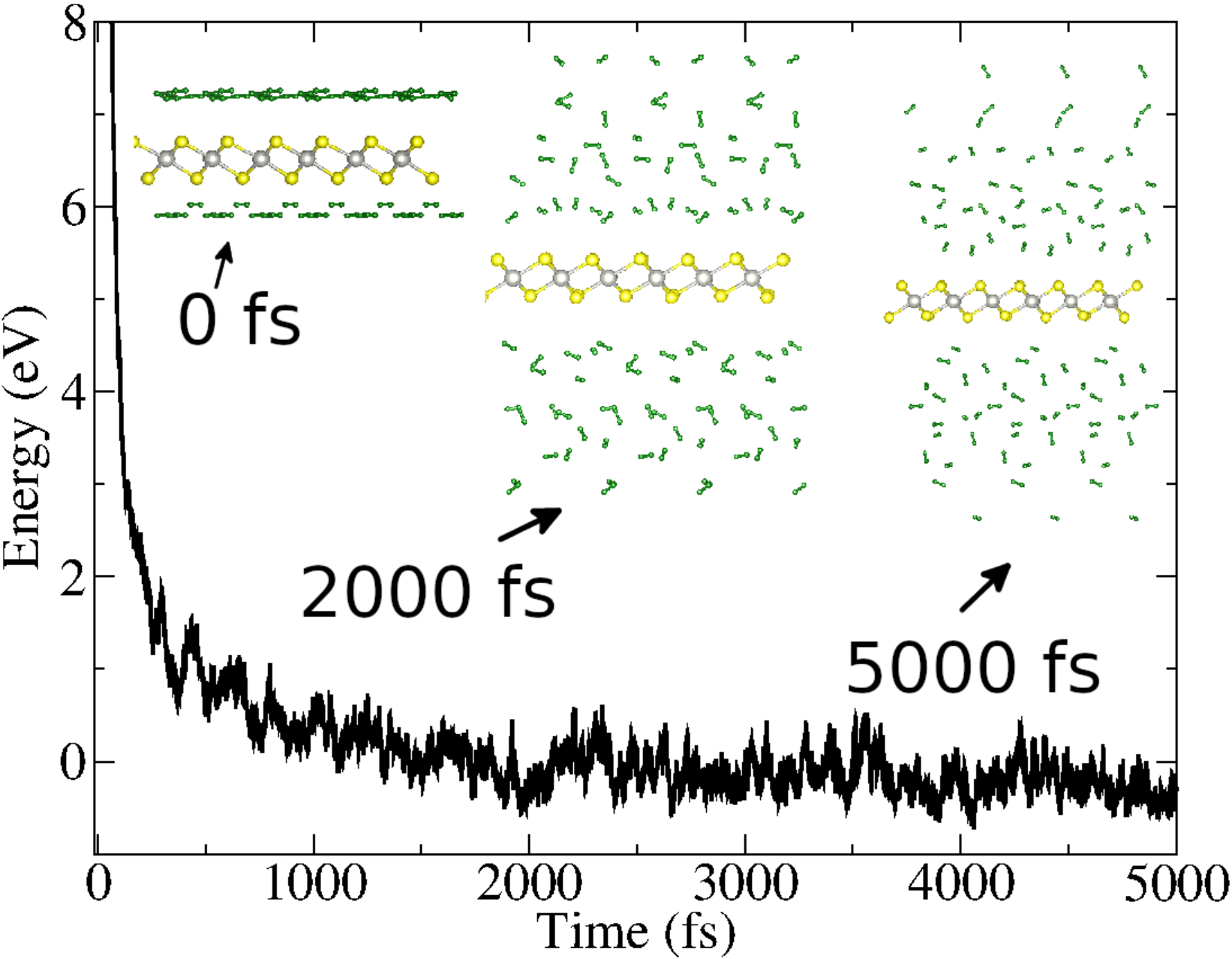}
\end{subfigure}\hfill
\begin{subfigure}[b]{0.45\columnwidth}
\subcaption[]{}
\includegraphics[width=\columnwidth,clip=true]{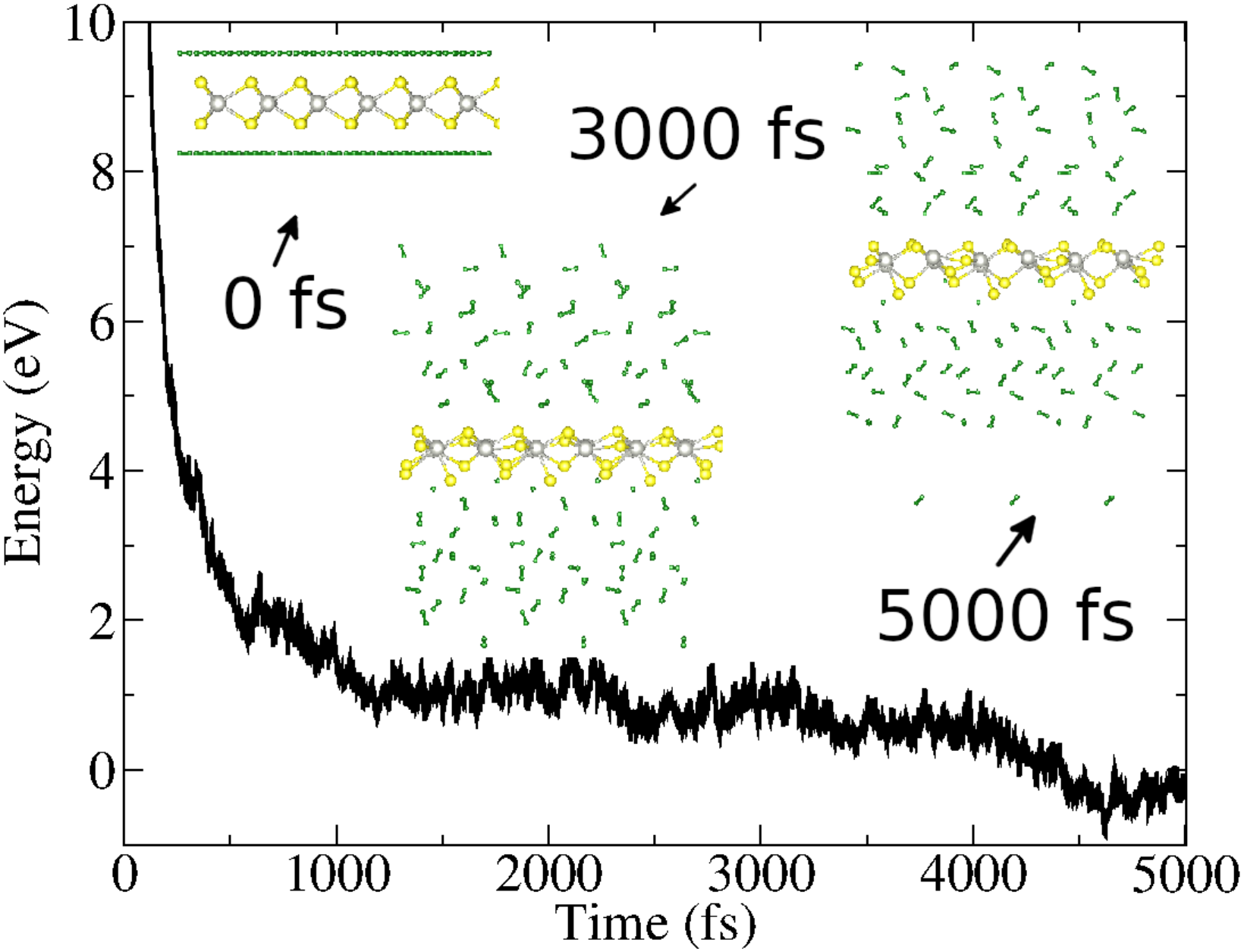}
\end{subfigure}
\\
\begin{subfigure}[b]{0.45\columnwidth}
\subcaption[]{}
\includegraphics[width=\columnwidth,clip=true]{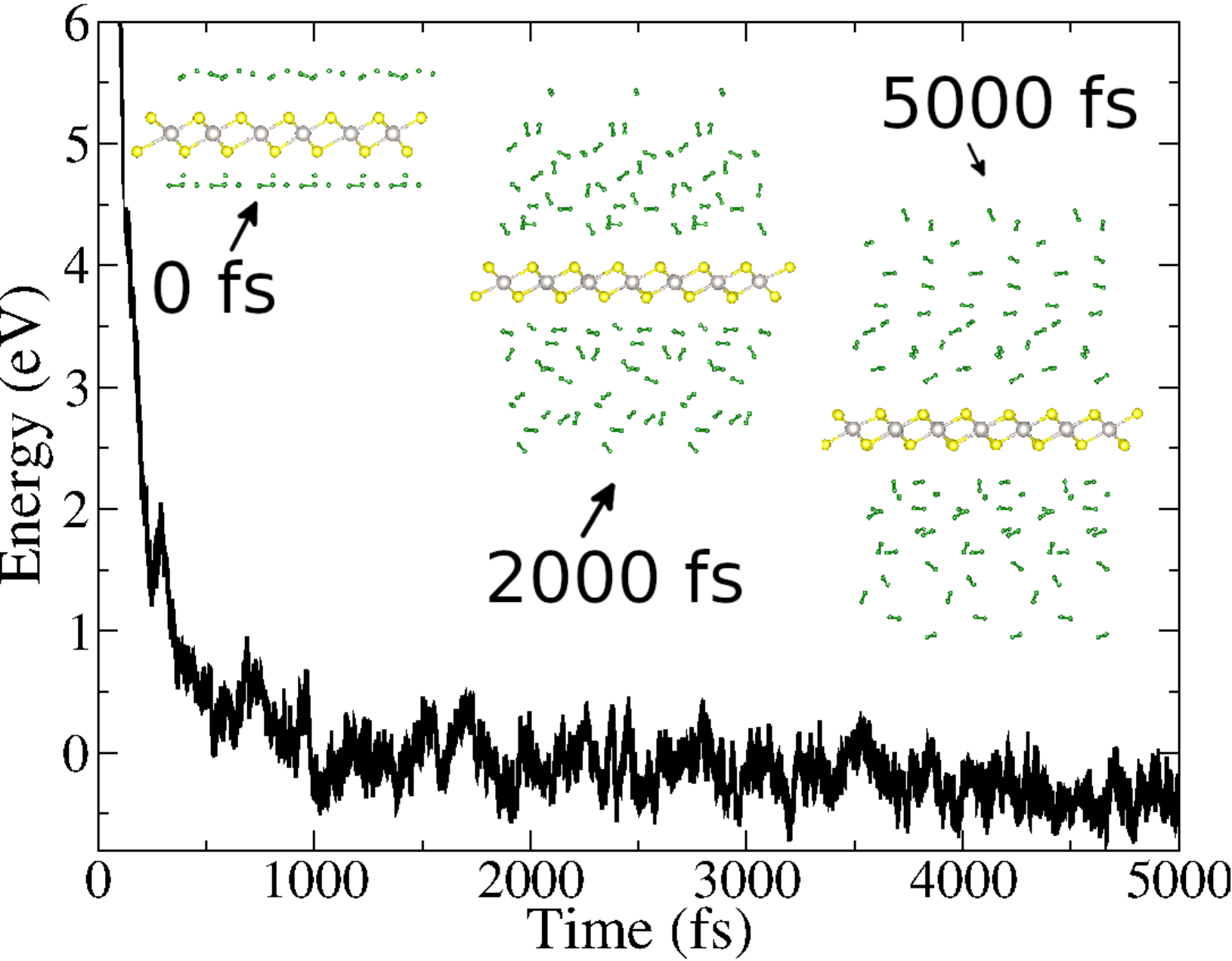}
\end{subfigure}
\hfill
\begin{subfigure}[b]{0.45\columnwidth}
\subcaption[]{}
\includegraphics[width=\columnwidth,clip=true]{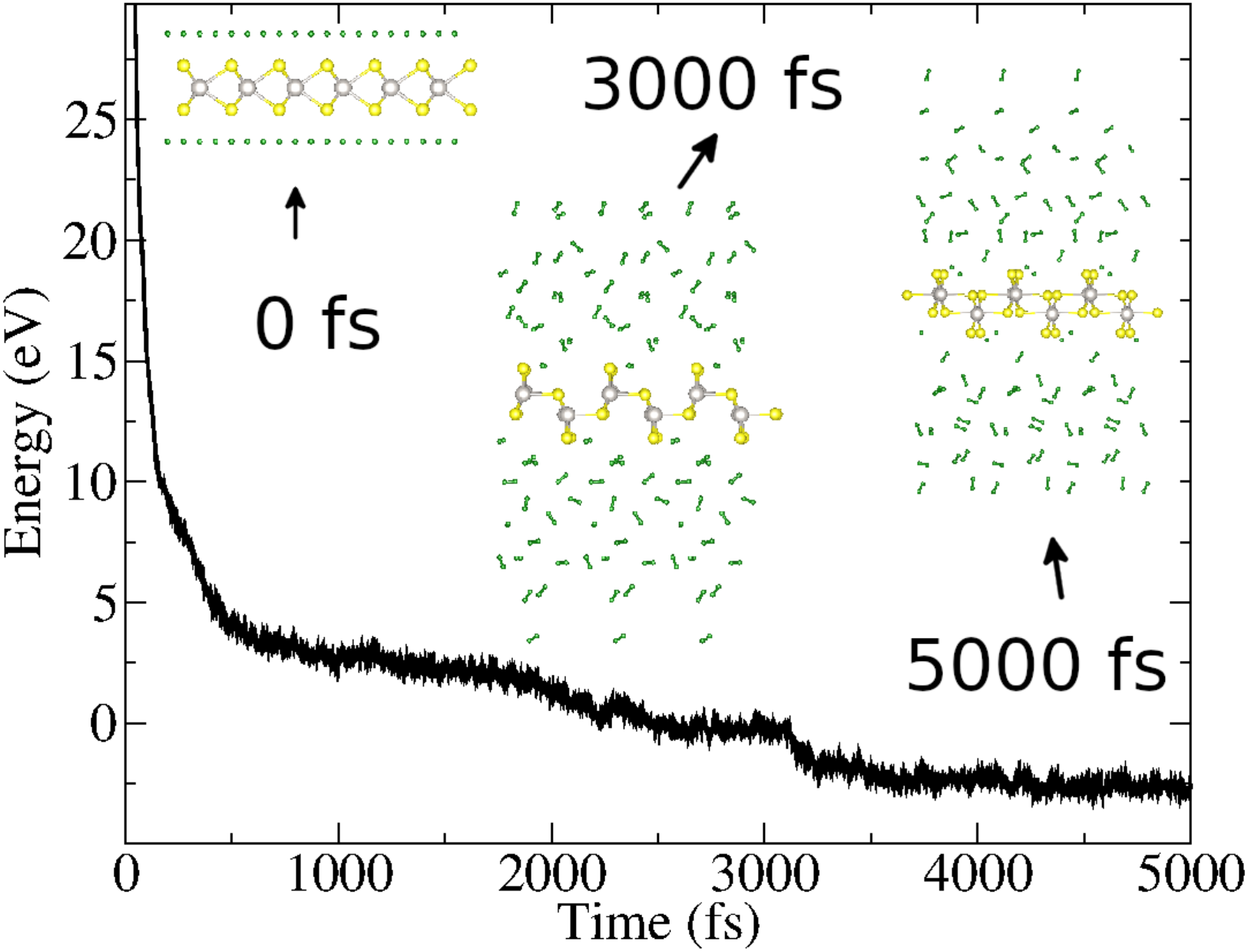}
\end{subfigure}
\caption{Energy as a function of time of AIMD simulations at \SI{300}{\kelvin} for 32 H$_2$ molecules concentration on a) 1T-PdS$_2$, b) 2H-PdS$_2$, c) 1T-PtS$_2$ and d) 2H-PtS$_2$. Snapshots at 0, 2000/3000 and \SI{5000}{\femto\second} simulation times are shown in the inset.}\label{fig:md_energy_PdPtS2}
\end{figure}

\begin{figure}[!ht]
\centering
\begin{subfigure}[b]{0.32\columnwidth}
		\subcaption[]{}
			\includegraphics[width=\columnwidth,clip=true]{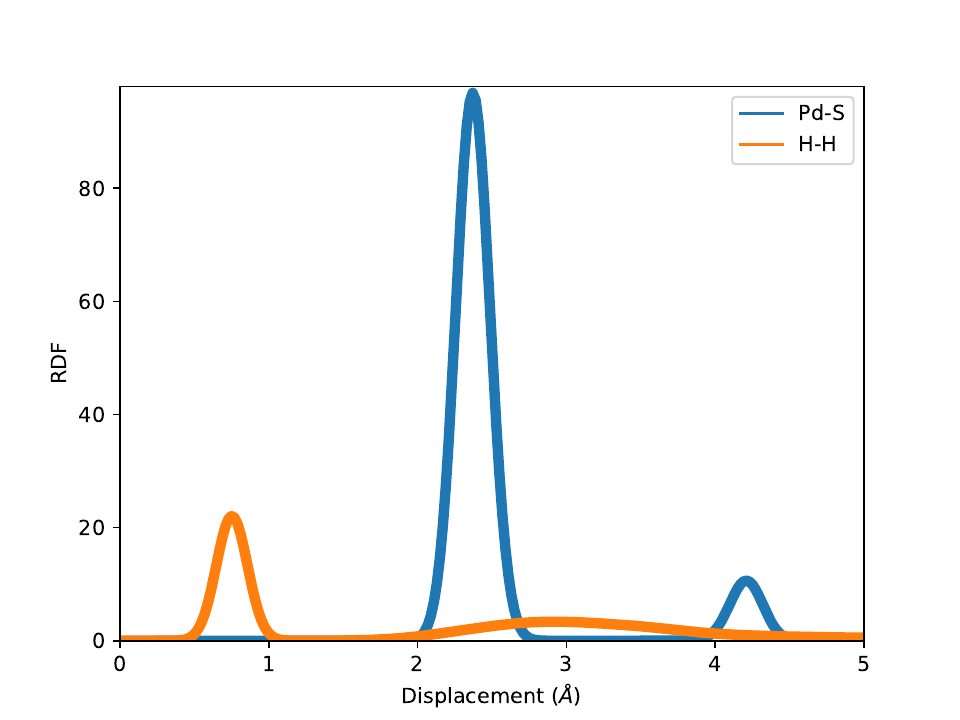}
		\end{subfigure}
		\begin{subfigure}[b]{0.32\columnwidth}
			\subcaption[]{}
			\includegraphics[width=\columnwidth,clip=true]{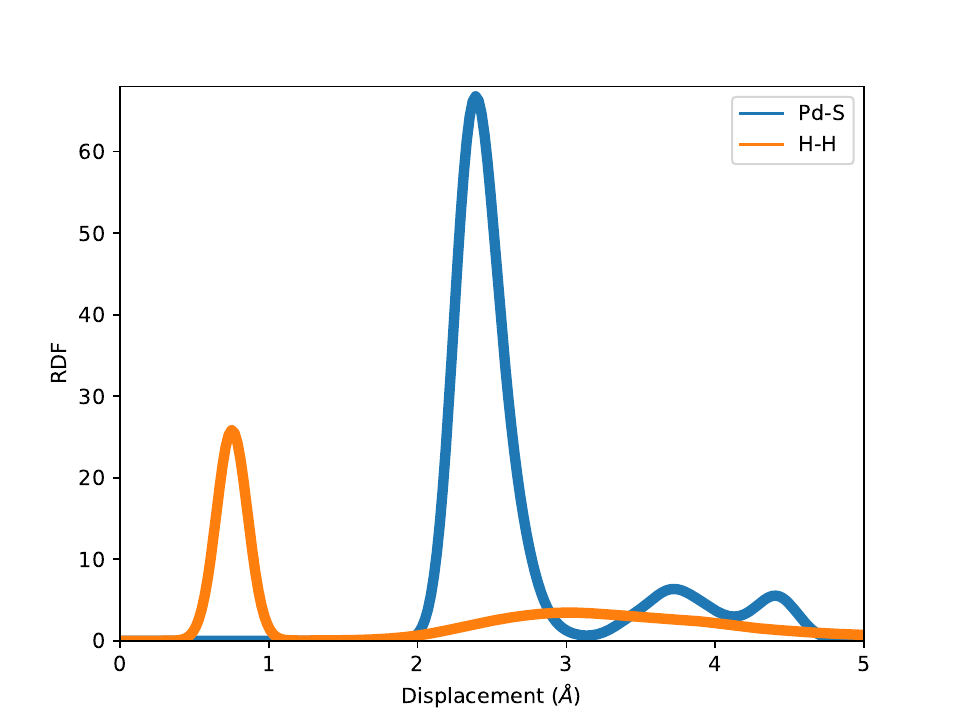}
		\end{subfigure}\\
		\begin{subfigure}[b]{0.32\columnwidth}
			\subcaption[]{}
			\includegraphics[width=\columnwidth,clip=true]{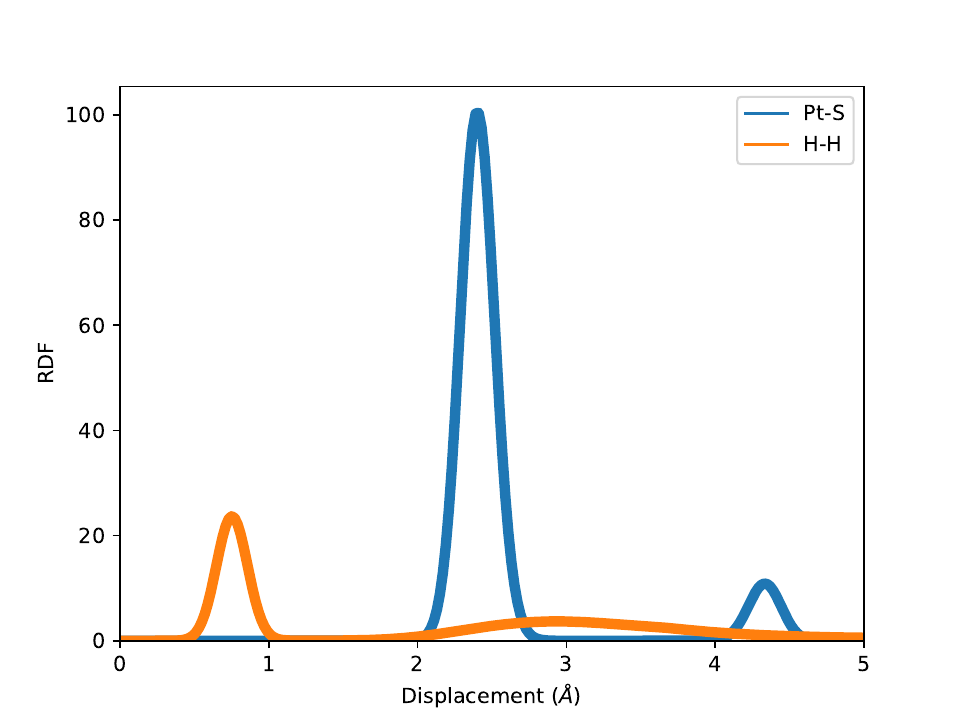}
		\end{subfigure}
		\begin{subfigure}[b]{0.32\columnwidth}
			\subcaption[]{}
			\includegraphics[width=\columnwidth,clip=true]{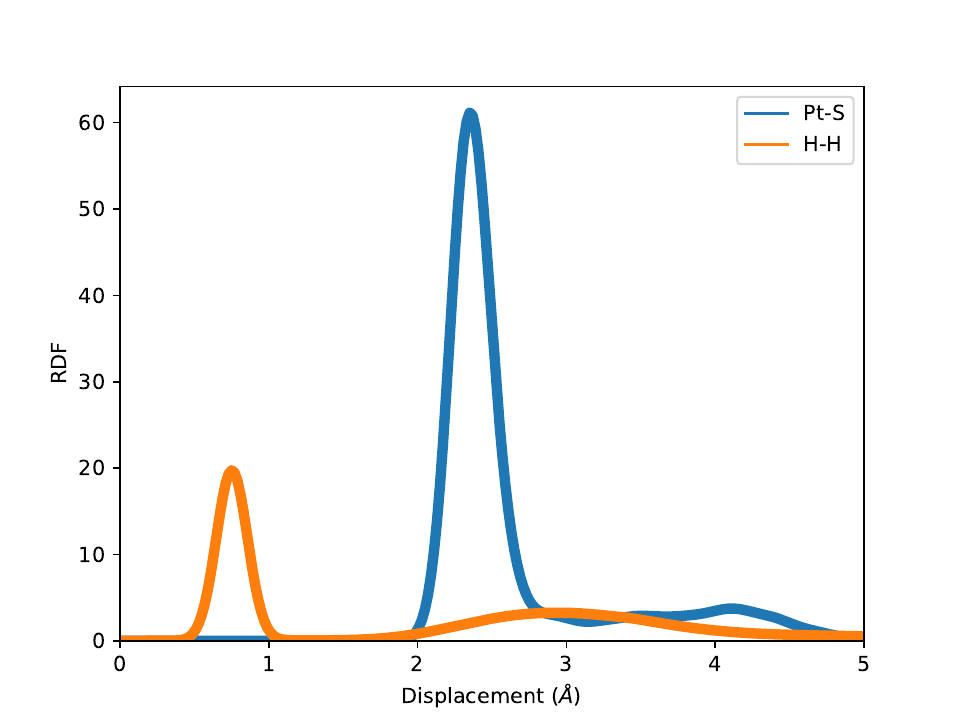}
		\end{subfigure}
\caption{RDF of AIMD simulations at \SI{300}{\kelvin} for 32 H$_2$ molecules on a) 1T-PdS$_2$, b) 2H-PdS$_2$, c) 1T-PtS$_2$ and d) 2H-PtS$_2$.}\label{fig:H2_1T_monoPds2_MD_rdf_PtS2}
\end{figure}

While static DFT calculations provide insight into adsorption energetics
and preferred configurations, they do not account for entropic effects,
molecular mobility, or thermal fluctuations that are critical for
assessing practical hydrogen storage. To address these aspects, we
performed \textit{ab initio} molecular dynamics (AIMD) simulations at
300 K for representative pristine MX$_2$ systems. The time evolution of
the total energy is shown in Fig.~\ref{fig:md_energy_PdPtS2}, and the
corresponding radial distribution functions (RDFs) are reported in
Fig.~\ref{fig:H2_1T_monoPds2_MD_rdf_PtS2}. 

In all investigated cases, the total energy exhibits a rapid initial
decrease during the first few hundred femtoseconds, followed by stable
fluctuations around a well-defined average value. This behavior
indicates efficient thermal equilibration and the absence of slow
structural relaxation or degradation processes. Importantly, no
systematic drift in the total energy is observed over the duration of
the simulations, confirming the thermal stability of both the
monolayers and the adsorbed hydrogen at room temperature.

Inspection of the AIMD trajectories reveals that hydrogen remains
strictly molecular throughout the simulations. The H--H bond length
fluctuates around its equilibrium value, and no dissociation events or
formation of surface hydrides are observed. This conclusion is further
supported by the RDFs shown in
Fig.~\ref{fig:H2_1T_monoPds2_MD_rdf_PtS2}, which display a pronounced
first peak corresponding to the intramolecular H--H distance, with no
additional features indicative of bond breaking or recombination. The
persistence of molecular hydrogen at finite temperature confirms that
adsorption remains physisorptive and reversible under ambient
conditions.

The AIMD results also provide insight into the dynamical confinement of
hydrogen near the surface. In metallic 1T phases, the RDFs exhibit
narrower peaks and reduced broadening compared to semiconducting 2H
phases, indicating stronger confinement of H$_2$ molecules in the
vicinity of the surface. This enhanced confinement originates from the
increased electronic screening and polarizability of the metallic
surface, which stabilizes hydrogen against thermal desorption. In
contrast, the broader RDFs observed for 2H phases reflect weaker
confinement and increased molecular mobility, consistent with their
smaller adsorption energies.

Despite these differences in confinement strength, all systems retain
their structural integrity during the simulations. No surface
reconstruction, layer buckling, or defect formation is observed, even
under relatively high hydrogen loading. The AIMD trajectories therefore
validate the static DFT predictions and demonstrate that the trends
identified at zero temperature persist under realistic thermal
conditions.

Taken together, the finite-temperature simulations establish three key
points. First, hydrogen adsorption on pristine MX$_2$ monolayers is
thermally stable and fully reversible at room temperature. Second, the
degree of dynamical confinement correlates with the electronic
character of the substrate, with metallic phases providing stronger
retention of hydrogen. Third, thermal effects do not induce qualitative
changes in adsorption mechanism or surface structure. These results
provide a critical link between static adsorption energetics and
practical operating conditions and set the stage for the analysis of
Janus systems, where additional asymmetry and polarity effects come into
play.

\begin{figure}[!ht]
		\centering
        \begin{subfigure}[b]{0.2\columnwidth}
			\subcaption[]{}
	\includegraphics[width=\columnwidth]{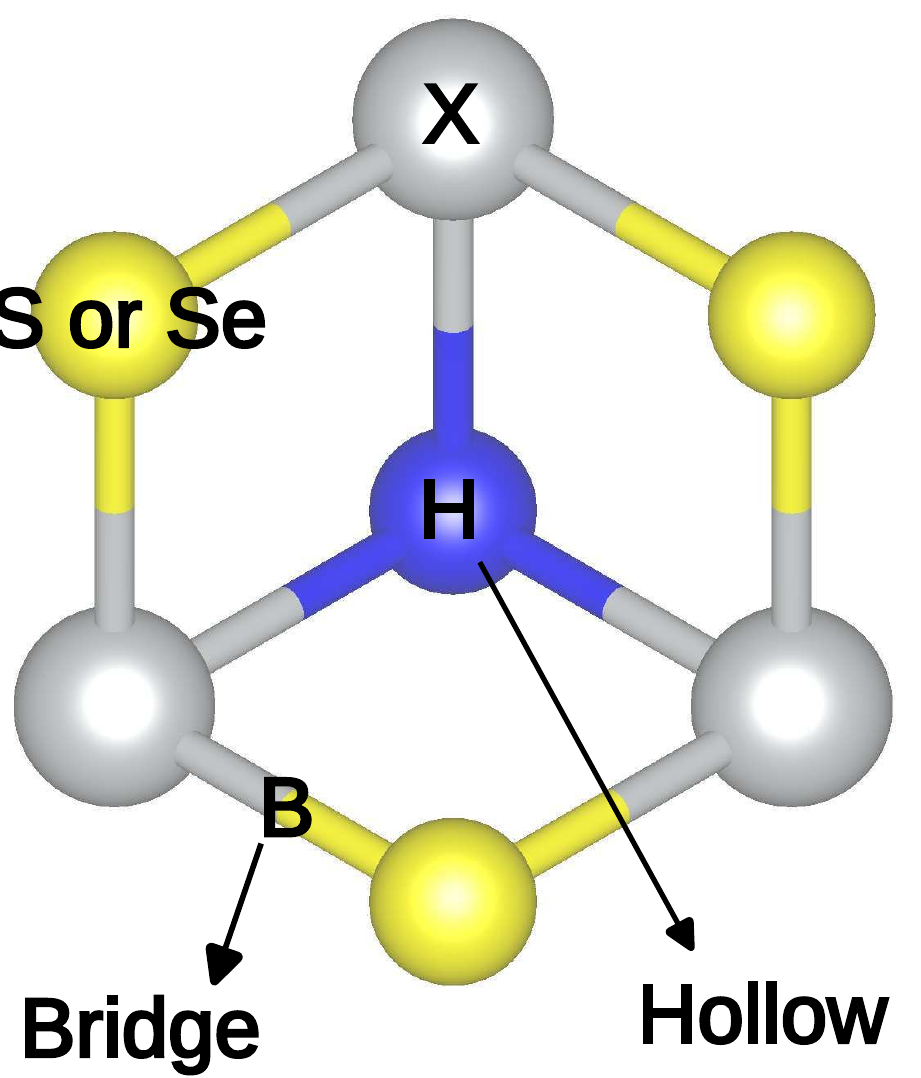}
    \end{subfigure}\hspace{0.5cm}
     \begin{subfigure}[b]{0.2\columnwidth}
			\subcaption[]{}
	\includegraphics[width=\columnwidth,clip]{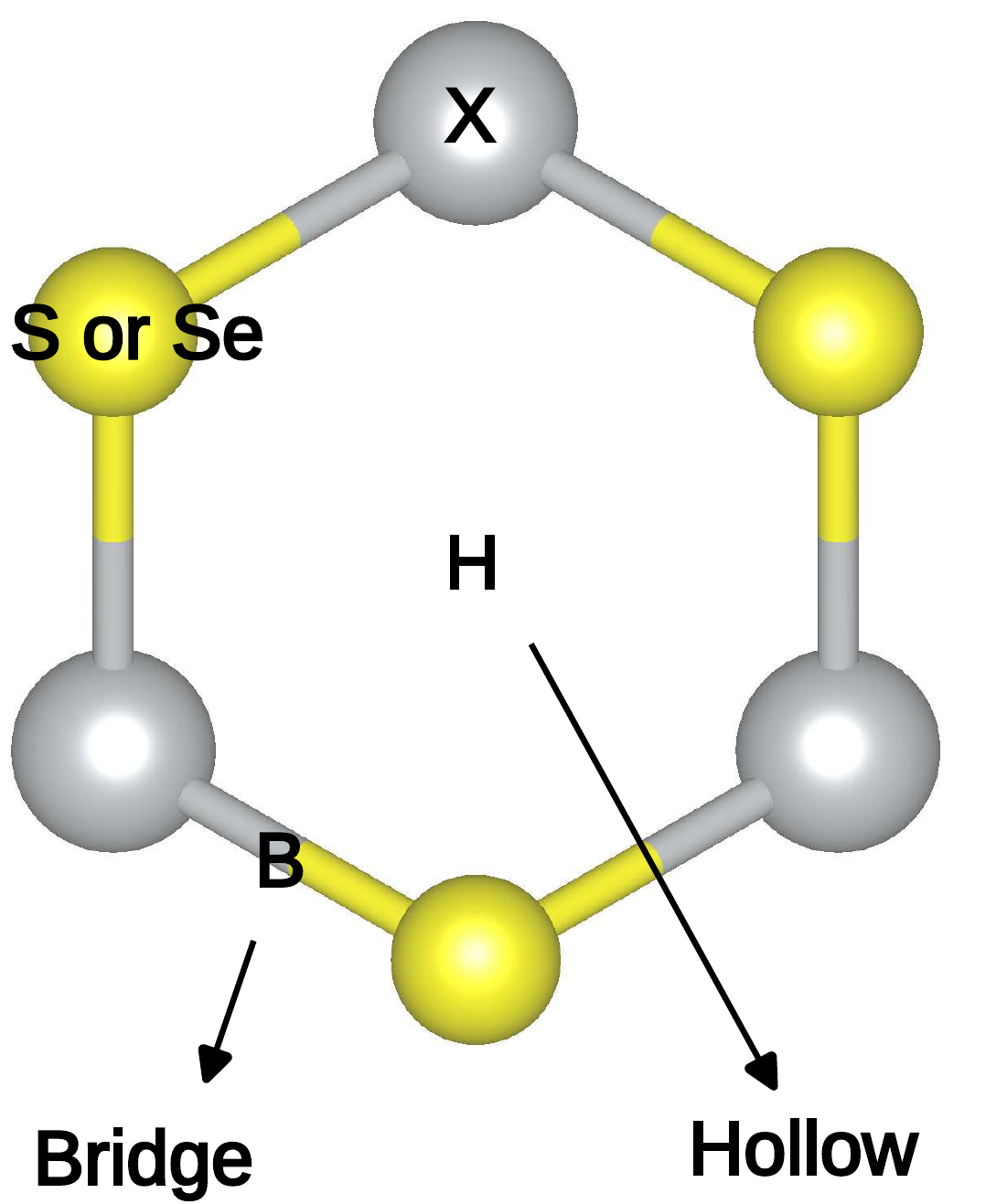}
     \end{subfigure}
     \\
    \begin{subfigure}[b]{0.32\columnwidth}
			\subcaption[]{}
	\includegraphics[width=\columnwidth,clip]{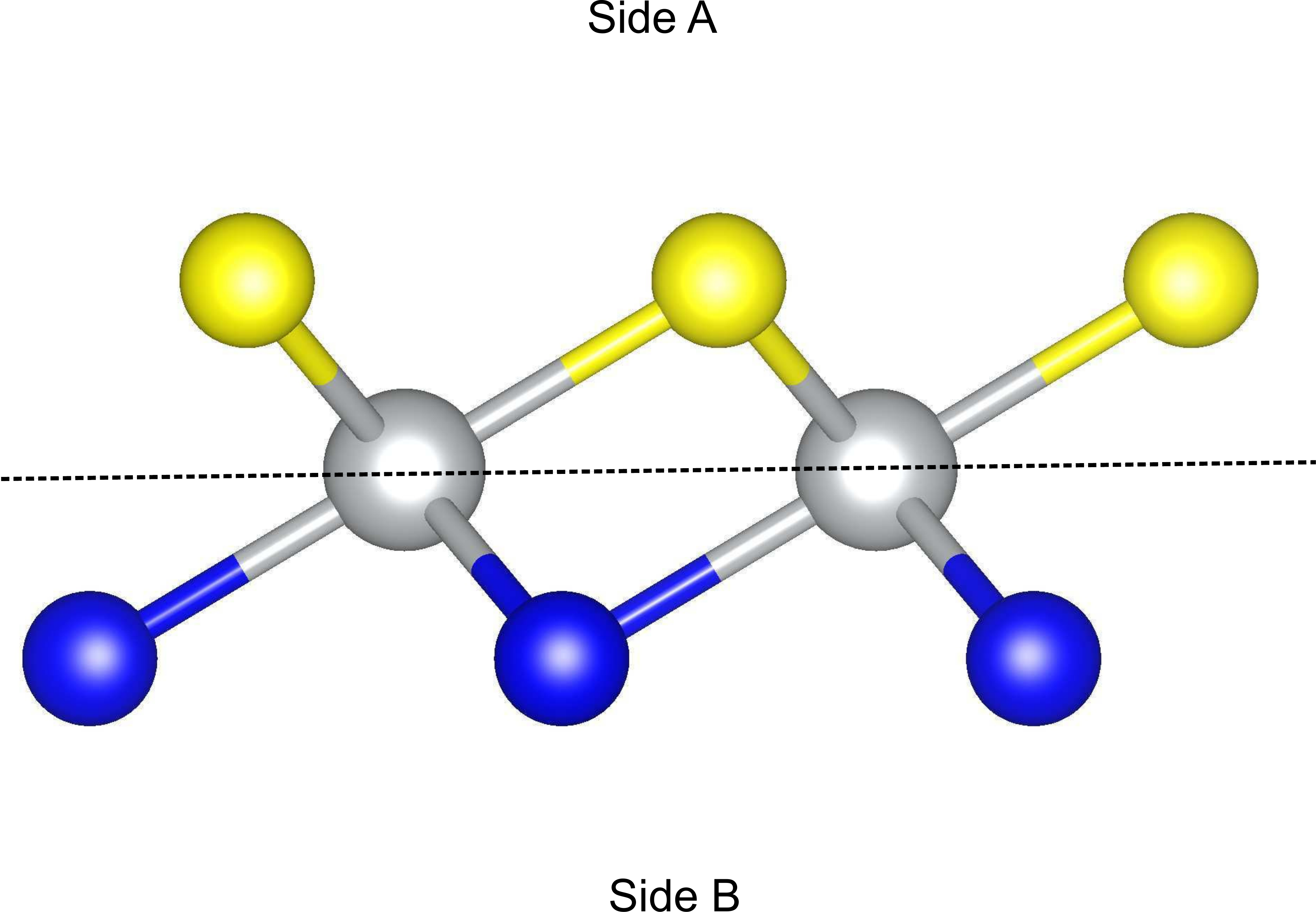}
    \end{subfigure}\hspace{0.5cm}
     \begin{subfigure}[b]{0.32\columnwidth}
			\subcaption[]{}
	\includegraphics[width=\columnwidth,clip]{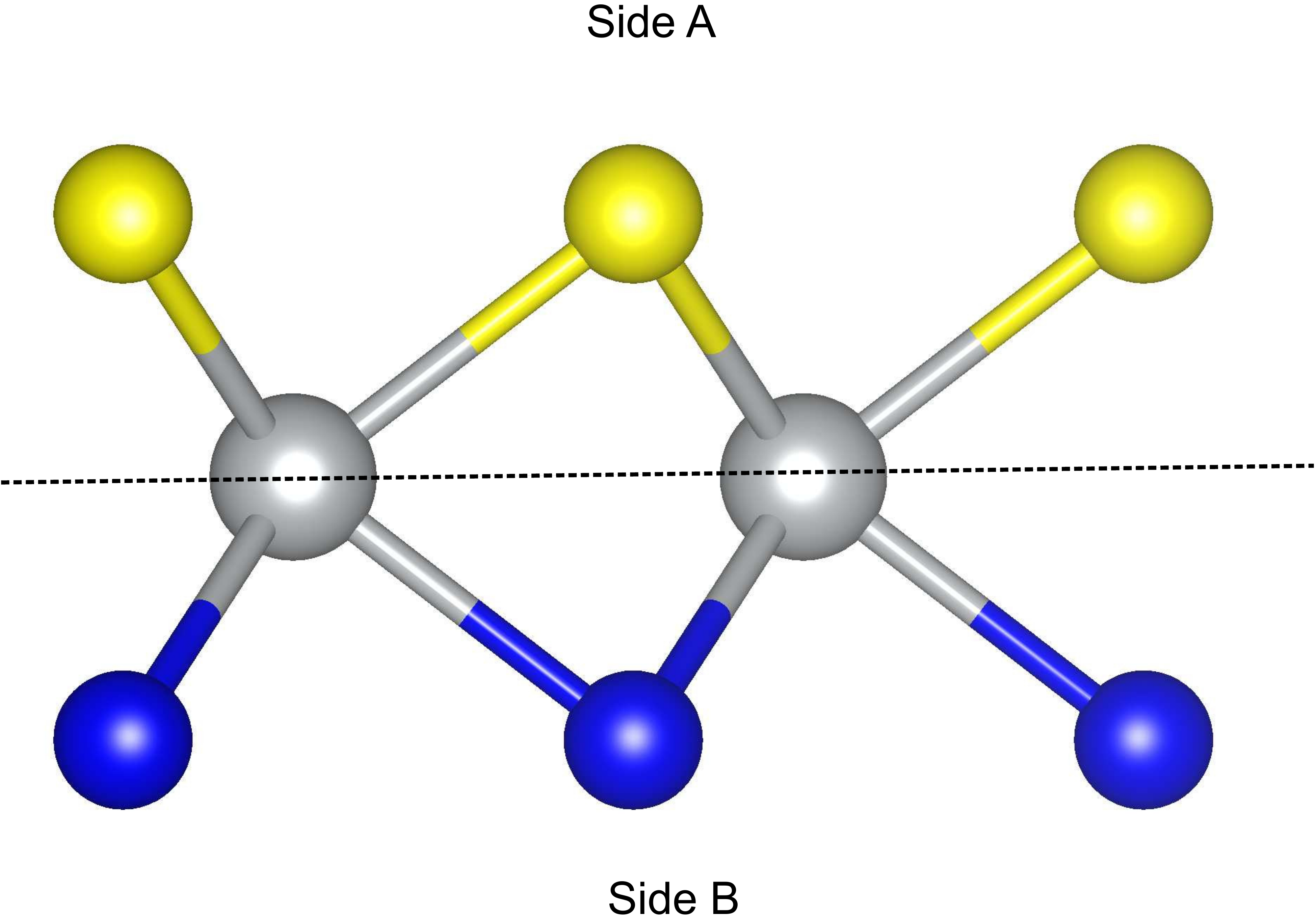}
    \end{subfigure}
		\caption{Adsorption sites for H$_2$ adsorption on  MSSe (M= Ni, Pt, and Pd) structures. a) top view of 1T-phase, b) top view of 2H-phase, c) side view of 1T-phase, d) side view of 2H-phase.}\label{fig:adsorption_position_MSSe}
	\end{figure}

\begin{figure}[!ht]
\centering
\begin{subfigure}[]{0.4\columnwidth}
\subcaption[]{S-side bridge}			
\includegraphics[width=\columnwidth,clip=true]{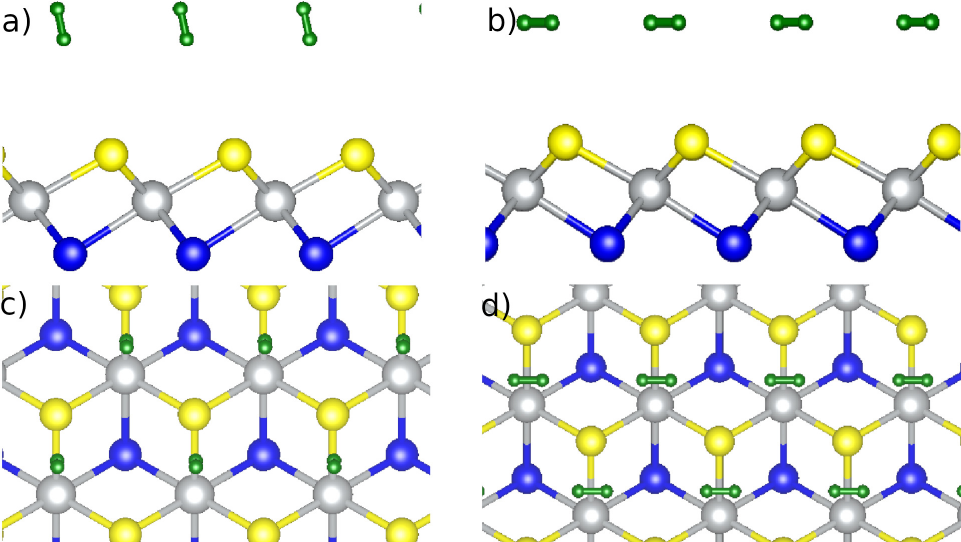}
		\end{subfigure}\hfill
		\begin{subfigure}[]{0.4\columnwidth}
			\subcaption[]{Se-side bridge}
			\includegraphics[width=\columnwidth,clip]{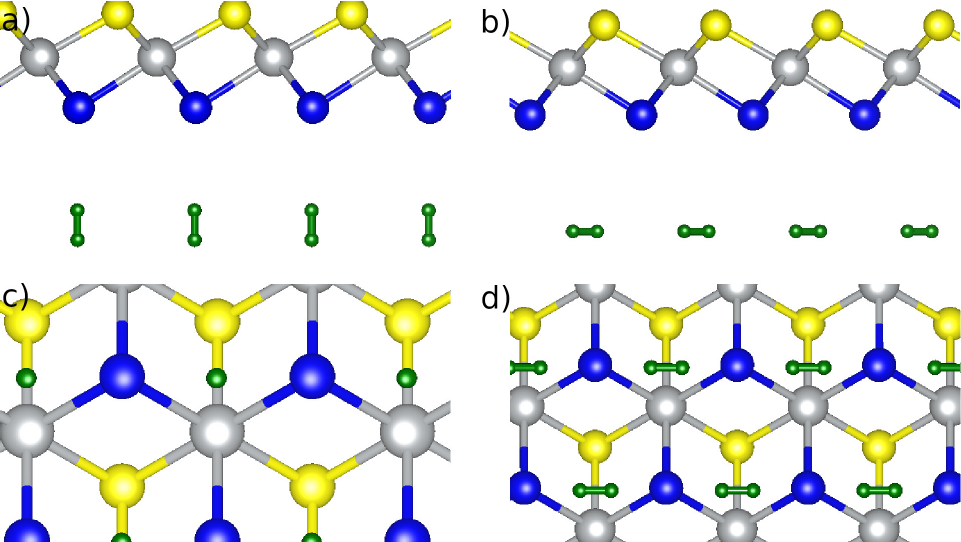}
		\end{subfigure}
		\\
		\begin{subfigure}[]{0.4\columnwidth}
			\subcaption[]{S-side hollow}
			\includegraphics[width=\columnwidth,clip=true]{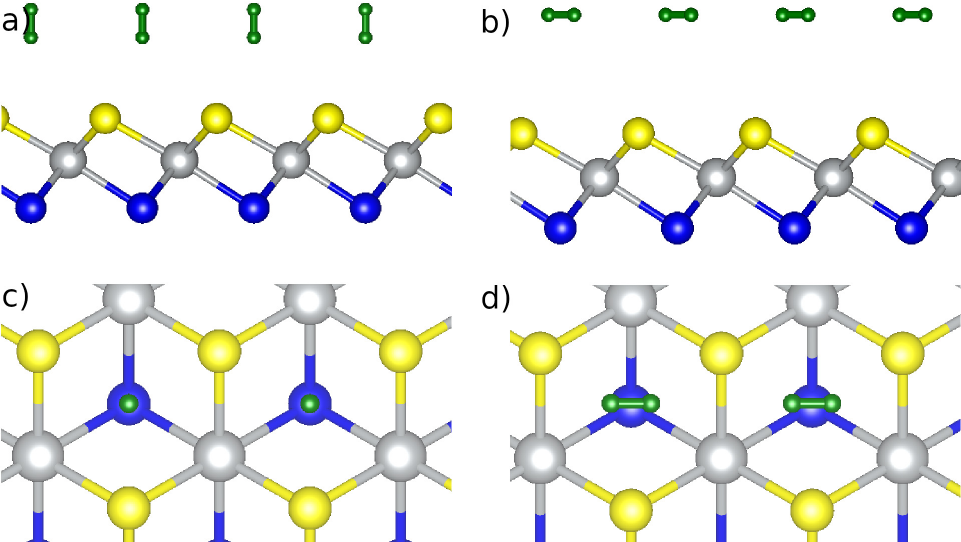}
		\end{subfigure}\hfill
		\begin{subfigure}[]{0.4\columnwidth}
			\subcaption[]{Se-side hollow}
			\includegraphics[width=\columnwidth,clip=true]{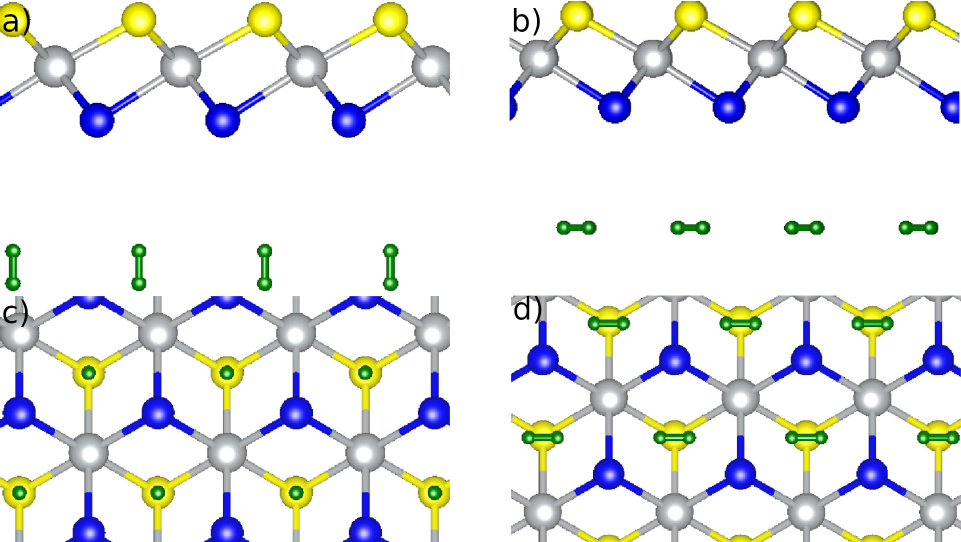}
		\end{subfigure}
		\\
		\begin{subfigure}[]{0.4\columnwidth}
			\subcaption[]{S-side on top-Ni}
			\includegraphics[width=\columnwidth,clip=true]{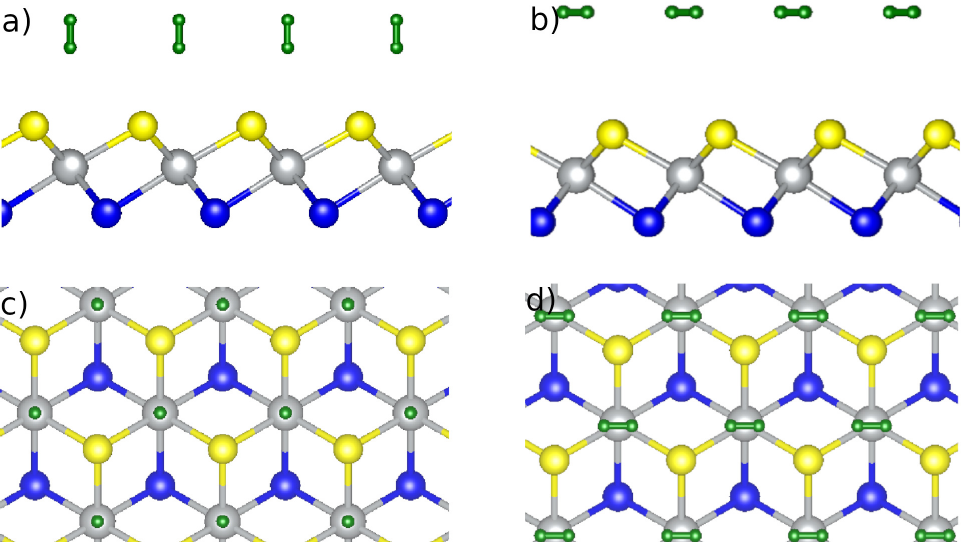}
		\end{subfigure}\hfill
		\begin{subfigure}[]{0.4\columnwidth}
			\subcaption[]{Se-side on top-Ni}
			\includegraphics[width=\columnwidth,clip=true]{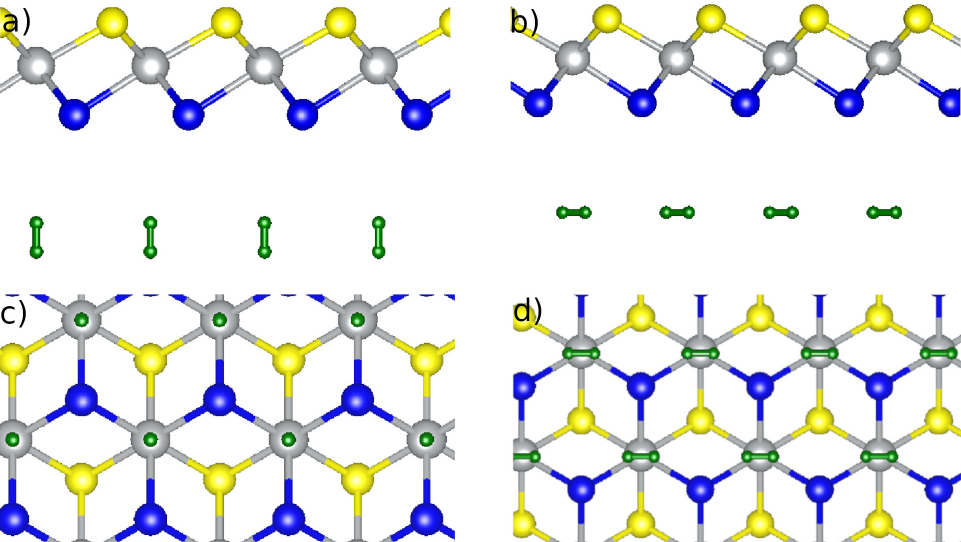}
		\end{subfigure}
		\\
		\begin{subfigure}[]{0.4\columnwidth}
			\subcaption[]{S-side on top S}
			\includegraphics[width=\columnwidth,clip=true]{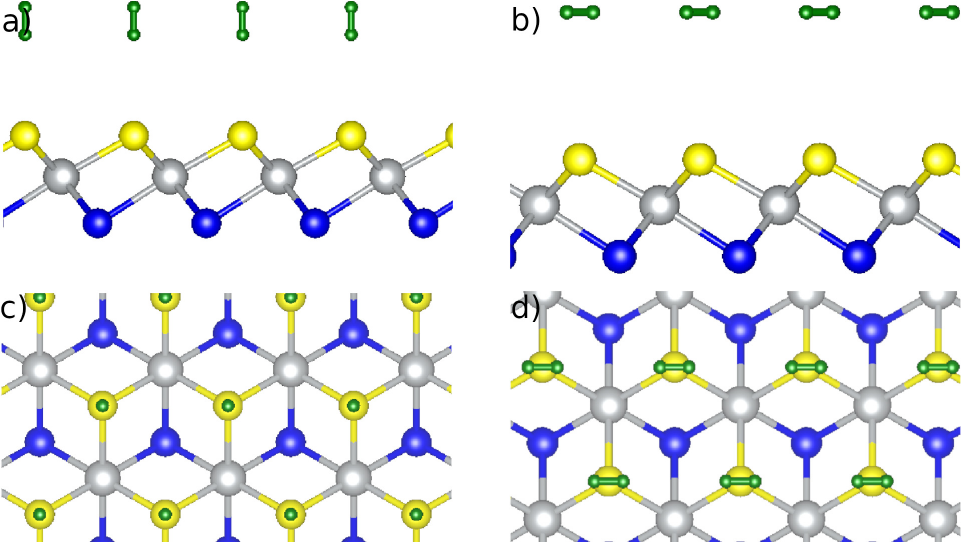}
		\end{subfigure}\hfill
		\begin{subfigure}[]{0.4\columnwidth}
			\subcaption[]{Se-side on top Se}
			\includegraphics[width=\columnwidth,clip=true]{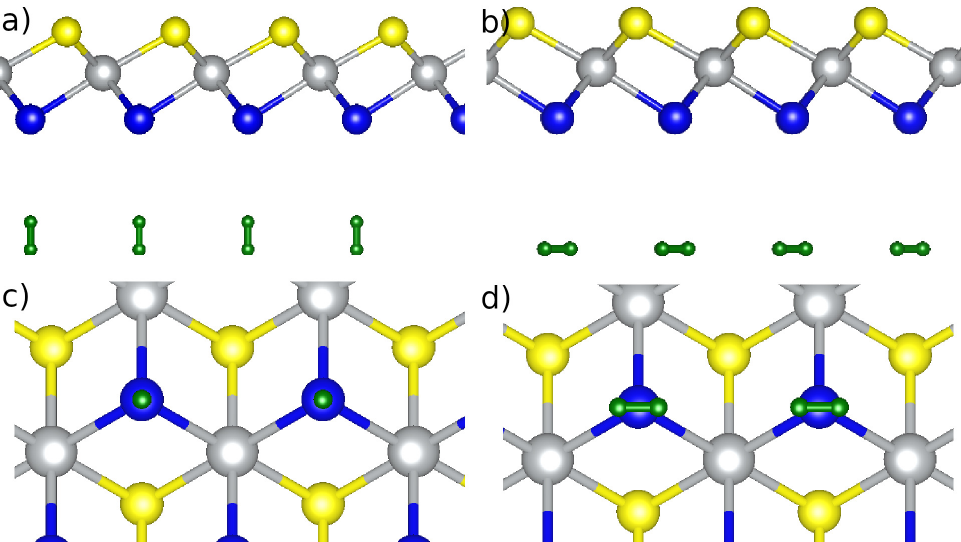}
		\end{subfigure}
		\caption{Relaxed geometry of a single H$_2$ molecule in the unit cell on NiSSe Janus structures. Clockwise: a) bridge on S-side, b) bridge on Se-side, c) hollow on S-side, d) hollow on Se-side, e) on top-Ni on S-side, f) on top-Ni on Se-side, g) on top-S on S-side and h) on top-S on Se-side.}\label{fig:relaxed_H2_MSSe}
	\end{figure}

\begin{table}[ht]
\centering
\caption{Adsorption energies $E_{\rm ads}$ (meV) and optimized interatomic distances (\AA) for molecular H$_2$ adsorbed on Janus PdSSe in the 1T and 2H phases at different adsorption sites. Values without parentheses correspond to H$_2$ oriented parallel to the monolayer, while values in parentheses refer to the perpendicular orientation. A dash (--) indicates that H$_2$ does not lead to stable structures.}
\begin{tabular}{lcccc}
\toprule
\multicolumn{5}{c}{PdSSe -- 1T phase} \\
\hline
Site & $E_{\rm ads}$ & $d_{\rm H\!-\!S}$ & $d_{\rm H\!-\!Se}$ & $d_{\rm H\!-\!Pd}$ \\
\hline
bridge (S-side)        & -242(-285) & 2.32(1.91) & --          & 3.86(3.71) \\
hollow (S-side)        & -200(-210) & 2.84(2.84) & --          & 4.15(4.15) \\
on top-Pd (S-side)           & -225(-207) & 2.92(3.32) & --          & 5.07(4.54) \\
on top-S (S-side)             & -251(-323) & 2.10(1.83) & --          & 4.28(3.52) \\
bridge (Se-side)     & -182(-213) & --          & 2.77(3.13) & 3.99(4.64) \\
on top-Pd (Se-side)          & -219(-207) & --          & 2.75(3.00) & 4.90(4.19) \\
on top-Se (Se-side)          & -230(-172) & --          & 3.38(3.17) & 4.84(4.65) \\
hollow (Se-side) & -213(-228) & --          & 2.68(3.30) & 4.83(5.19) \\
\hline
\multicolumn{5}{c}{PdSSe -- 2H phase} \\
\hline
site & $E_{\rm ads}$ & $d_{\rm H\!-\!S}$ & $d_{\rm H\!-\!Se}$ & $d_{\rm H\!-\!Pd}$ \\
\hline
bridge (S-side)        & 21(23) & 2.73(3.10) & --          & 3.71(4.10) \\
hollow   (S-side)     & 22(31) & 3.15(3.06) & --          & 4.02(4.02) \\
on top-Pd  (S-side)           & 27(39) & 2.97(2.97) & --          & 3.19(3.19) \\
on top-S (S-side)             & --     & --          & --          & --          \\
bridge (Se-side)     & 16(15) & --          & 2.97(3.06) & 3.95(4.10) \\
on top-Pd (Se-side)          & 21(21) & --          & 3.03(3.03) & 3.27       \\
on top-Se (Se-side)          & --     & --          & --          & --          \\
hollow (Se-side) & --     & --          & --          & 4.11       \\
\bottomrule
\end{tabular}
\label{tab:tabpdsse}
\end{table}

\begin{table}[ht]
\centering
\caption{Adsorption energies $E_{\rm ads}$ (meV) and optimized distances (\AA) for H$_2$ adsorbed on Janus PtSSe (1T and 2H phase). Values without parentheses correspond to H$_2$ oriented parallel to the monolayer while values in parentheses correspond to perpendicular orientation.}
\begin{tabular}{lcccc}
\toprule
\multicolumn{5}{c}{PtSSe -- 1T phase}\\
site & $E_{\rm ads}$ & $d_{\rm H\!-\!S}$ & $d_{\rm H\!-\!Se}$ & $d_{\rm H\!-\!Pt}$ \\
\hline
bridge (S-side)   & 49(46)  & 2.75(3.16) & --           & 3.61(4.04) \\
hollow (S-side)   & 22(4)   & 3.58(3.29) & --           & 4.61(4.49) \\
on top-Pt (S-side)    & 25(54)  & 3.35(3.17) & --           & 3.82(3.85) \\
on top-S (S-side)    & 19(33)  & 4.78(3.11) & --           & 4.77(4.65) \\
bridge (Se-side)  & 40(66)  & --         & 3.15(3.01)  & 4.77(4.13) \\
on top-Pt (Se-side)   & 30(82)  & --         & 3.33(3.02)  & 3.92(3.78) \\
on top-Se (Se-side)  & 22(120) & --         & 4.78(2.93)  & 3.63(4.07) \\
hollow (Se-side)  & 45(39)  & --         & 4.62(3.35)  & 2.78(5.00) \\
\midrule
\multicolumn{5}{c}{PtSSe -- 2H phase}\\
\midrule
site & $E_{\rm ads}$ & $d_{\rm H\!-\!S}$ & $d_{\rm H\!-\!Se}$ & $d_{\rm H\!-\!Pt}$ \\
\hline
bridge (S-side)   & 25(36) & 3.12(3.40) & --           & 4.21(4.28) \\
hollow (S-side)   & 31(25) & 3.32(3.58) & --           & 4.47(4.86) \\
on top-Pt (S-side)    & 27(51) & 3.38(3.17) & --           & 4.05(4.02) \\
on top-S (S-side)    & 35(51) & 2.79(3.13) & --           & 4.58(4.76) \\
bridge (Se-side)  & 33(35) & --         & 3.63(3.69)  & 4.49(4.79) \\
on top-Pt (Se-side)   & 42(42) & --         & 3.45(3.53)  & 4.47(5.01) \\
on top-Se (Se-side)  & 26(41) & --         & 3.11(3.11)  & 5.16(5.16) \\
hollow (Se-side)  & 25(33) & --         & 4.82(3.72)  & 4.70(5.01) \\
\bottomrule
\end{tabular}
\label{tab:tabptsse}
\end{table}

\begin{table}
  \centering
\caption{Adsorption energies $E_{\rm ads}$ (meV) and optimized interatomic distances (\AA) for molecular H$_2$ adsorbed on Janus NiSSe in the 1T and 2H phases at different adsorption sites. Values without parentheses correspond to H$_2$ oriented parallel to the monolayer, while values in parentheses refer to the perpendicular orientation.}
\begin{tabular}{lcccc}
\toprule
\multicolumn{5}{c}{NiSSe -- 1T phase} \\
\hline
site & $E_{\rm ads}$ & $d_{\rm H\!-\!S}$ & $d_{\rm H\!-\!Se}$ & $d_{\rm H\!-\!Ni}$ \\
\hline
bridge (S-side)   & 23(48) & 3.03(3.15) & --            & --           \\
hollow (S-side)   & 73(28) & 2.94(3.44) & --            & 3.84(4.60)  \\
on top-Ni (S-side)    & 68(17) & 2.93(3.73) & --            & 3.28(4.48)  \\
on top-S (S-side)    & 32(19) & 4.42(3.69) & --            & 4.37(5.09)  \\
bridge (Se-side)  & 39(33) & --         & 3.13(3.40)   & 4.02(4.05)  \\
on top-Ni (Se-side)   & 25(38) & --         & 3.46(3.56)   & 4.12(4.47)  \\
on top-Se (Se-side)  & 22(28) & --         & 3.69(3.55)   & 4.81(4.74)  \\
hollow (Se-side)  & 40(32) & --         & 2.82(3.35)   & 4.55(4.90)  \\
\midrule
\multicolumn{5}{c}{NiSSe -- 2H phase} \\
\hline
site & $E_{\rm ads}$ & $d_{\rm H\!-\!S}$ & $d_{\rm H\!-\!Se}$ & $d_{\rm H\!-\!Ni}$ \\
\hline
bridge (S-side)   & 24(71) & 3.63(2.87) & --            & 3.73(3.69)  \\
hollow (S-side)   & 60(45) & 2.86(3.05) & --            & 3.65(3.99)  \\
on top-Ni (S-side)    & 61(26) & 2.87(3.33) & --            & 3.05(3.92)  \\
on top-S (S-side)    & 32(50) & 2.65(2.96) & --            & 4.21(4.31)  \\
bridge (Se-side)  & 34(30) & --         & 2.86(3.42)   & 3.98(4.44)  \\
on top-Ni (Se-side)   & 46(18) & --         & 2.99(3.33)   & 3.51(4.19)  \\
on top-Se (Se-side)  & 54(37) & --         & 2.61(3.25)   & 4.42(4.86)  \\
hollow (Se-side)  & 52(35) & --         & 4.19(3.99)   & 3.12(5.24)  \\
\bottomrule
\end{tabular}
\label{tab:tabnisse}
\end{table}

\subsection{Janus dichalcogenides: polarity-driven adsorption asymmetry}

Janus MSSe monolayers introduce an additional degree of freedom absent
in pristine MX$_2$ systems: broken out-of-plane mirror symmetry and the
associated internal electrostatic dipole. The high-symmetry adsorption
sites on Janus surfaces are illustrated in
Fig.~\ref{fig:adsorption_position_MSSe}, and representative relaxed
adsorption geometries for molecular hydrogen are shown in
Fig.~\ref{fig:relaxed_H2_MSSe}. The corresponding adsorption energies and
structural parameters are summarized in
Tables~\ref{tab:tabpdsse}--\ref{tab:tabnisse}.

Across all Janus systems investigated, hydrogen adsorption remains
strictly molecular. As in pristine monolayers, the H--H bond length is
preserved and no dissociation events are observed for either surface
termination or structural polytype. This confirms that the introduction
of polarity modifies the strength of physisorption without altering its
fundamental nature.

A clear surface selectivity emerges as a direct consequence of Janus
asymmetry. For all metals and both polymorphs, adsorption on the
S-terminated side is systematically stronger than on the Se-terminated
side. This trend reflects the higher electronegativity and lower
polarizability of sulfur relative to selenium, which enhances local
electric fields and strengthens induced dipole interactions with the
H$_2$ molecule. The resulting asymmetry manifests as measurable
differences in adsorption energy and adsorption height between the two
surfaces, while preserving similar preferred adsorption sites.

The magnitude of the Janus-induced enhancement is, however, strongly
metal dependent. In PdSSe, the polarity-driven asymmetry leads to
moderate and consistently negative adsorption energies, particularly in
the 1T phase, placing this system within the optimal window for
reversible hydrogen storage. The balance between surface polarity and
moderate electronic screening provided by Pd $d$ states results in
stable molecular adsorption without excessive confinement.

In contrast, PtSSe exhibits either weak or overly strong adsorption,
depending on the polymorph and surface termination. In the 2H phase,
adsorption on the S-terminated side becomes excessively strong,
resulting in compressed hydrogen configurations that reduce mobility
and compromise reversibility. This behavior highlights that Janus
polarity alone does not guarantee favorable adsorption: when combined
with highly polarizable Pt $d$ states, the interaction can exceed the
ideal physisorption regime.

NiSSe represents an intermediate case. While Ni-based systems exhibit
strong electronic screening due to localized $d$ states, the Janus
asymmetry does not significantly improve adsorption relative to the
pristine NiSe$_2$ and NiS$_2$ counterparts. Instead, NiSSe displays
relatively uniform adsorption across both surfaces and polymorphs,
suggesting that the intrinsic metallicity of Ni-based monolayers
dominates over polarity effects. As a result, Janus functionalization
provides limited additional benefit for Ni systems.

These observations demonstrate that the impact of Janus asymmetry is
not universal but arises from a delicate interplay between surface
polarity, transition-metal electronic structure, and lattice geometry.
Janus functionalization is most effective in systems where adsorption is
initially too weak in the pristine material, as in Pd-based
dichalcogenides. When adsorption is already strong, as in Ni-based
systems, or becomes excessive, as in Pt-based Janus monolayers, polarity
offers little advantage and may even be detrimental.

First-principles studies on Janus dichalcogenides have similarly reported
enhanced molecular hydrogen adsorption driven by polarity-induced charge
redistribution, while preserving the molecular character of H$_2$
\cite{Zhao2021JanusH2,Wang2022JanusGas,Li2023TMDC_H2}.

Overall, the Janus results establish that polarity-driven symmetry
breaking provides a tunable but nontrivial route to modifying hydrogen
adsorption. Its effectiveness depends critically on matching the
strength of the induced electrostatic interaction to the intrinsic
electronic screening of the host material. These insights motivate the
analysis of high hydrogen loading and multilayer formation in Janus
systems, which is addressed in the following section.

\subsection{High hydrogen loading and multilayer formation in Janus systems}

\begin{figure}[!ht]
\centering
\begin{subfigure}{0.45\columnwidth}
\subcaption[]{}
\includegraphics[width=\columnwidth,clip]{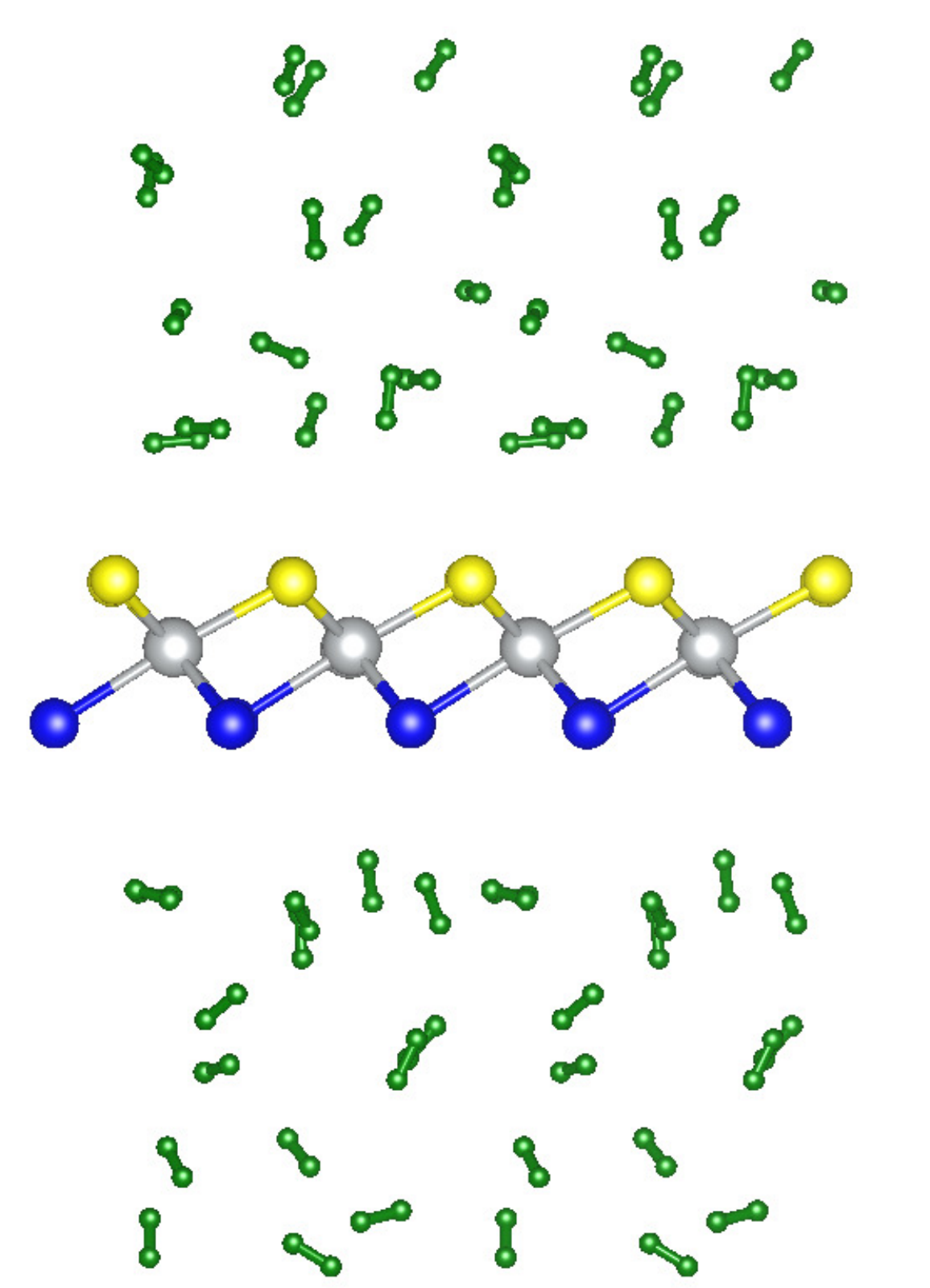}
\end{subfigure}
\hfill
\begin{subfigure}{0.45\columnwidth}
\subcaption[]{}
\includegraphics[width=\columnwidth,clip]{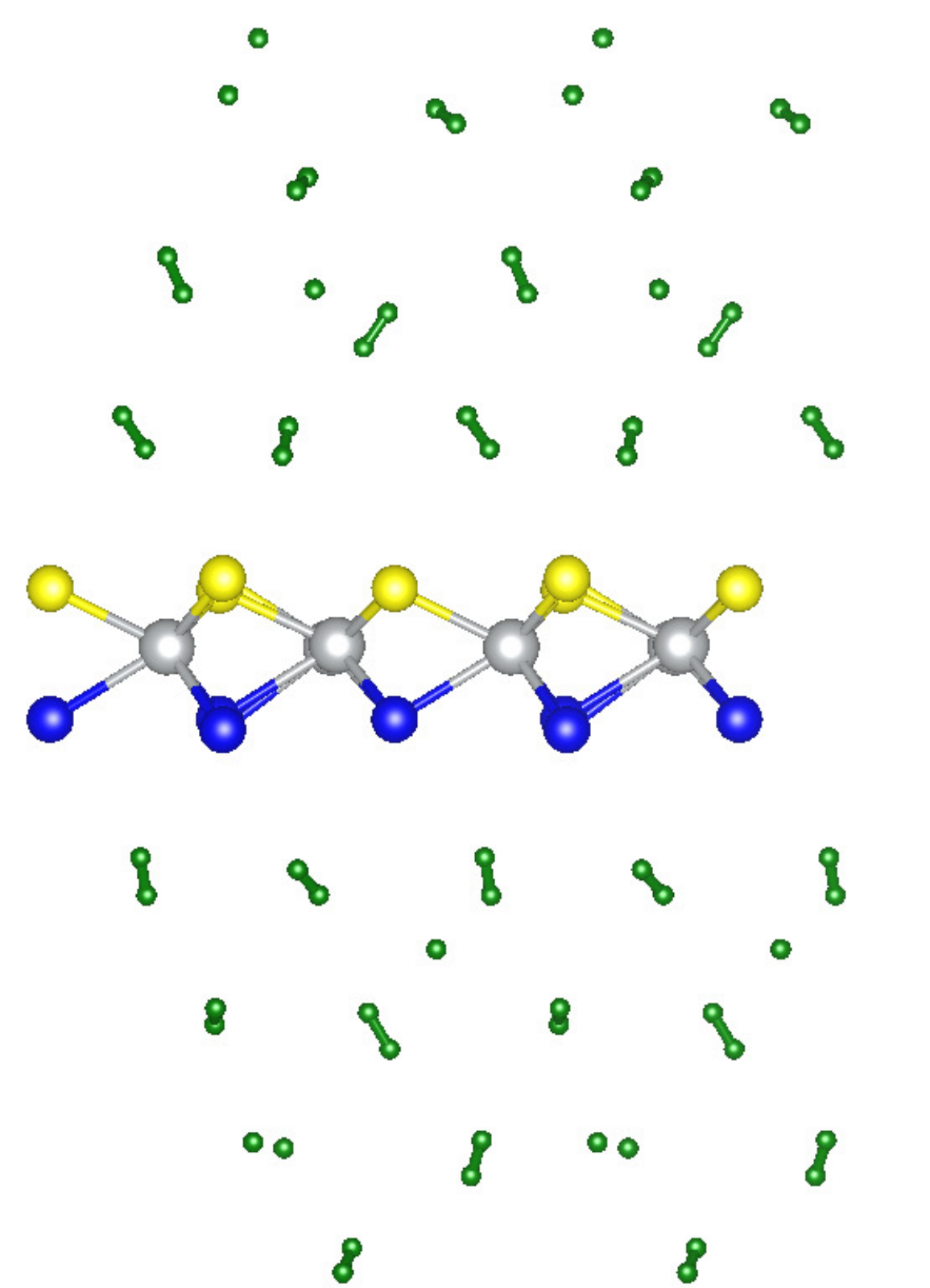}
\end{subfigure}
\caption{Ground state configuration for 32 H$_2$ molecules on a) 1T-NiSSe and b) 2H-NiSSe Janus structures.}
\label{fig:1T_2H_NiSSe_32H2}
\end{figure}

\begin{figure}[!ht]
    \centering
    \includegraphics[width=0.95\linewidth]{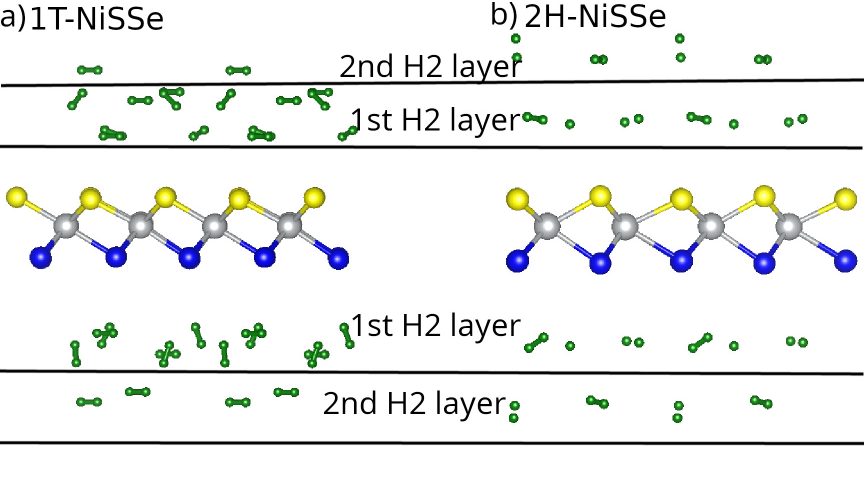}
    \caption{
    Side view of (a) 1T-NiSSe and (b) 2H-NiSSe monolayers decorated with H$_2$ molecules.
    The 1st H$_2$ layer and 2nd H$_2$ adsorption layers above and below the surface are indicated.  The maximum hydrogen uptake corresponds to 16 and 18 H$_2$ molecules for the 1T and 2H phases, respectively.}
    \label{fig:NiSSe_H2_layers}
\end{figure}

\begin{figure}[!ht]
		\centering
		\begin{subfigure}[]{0.45\columnwidth}
			\subcaption[]{}
			\includegraphics[width=\columnwidth,clip=true]{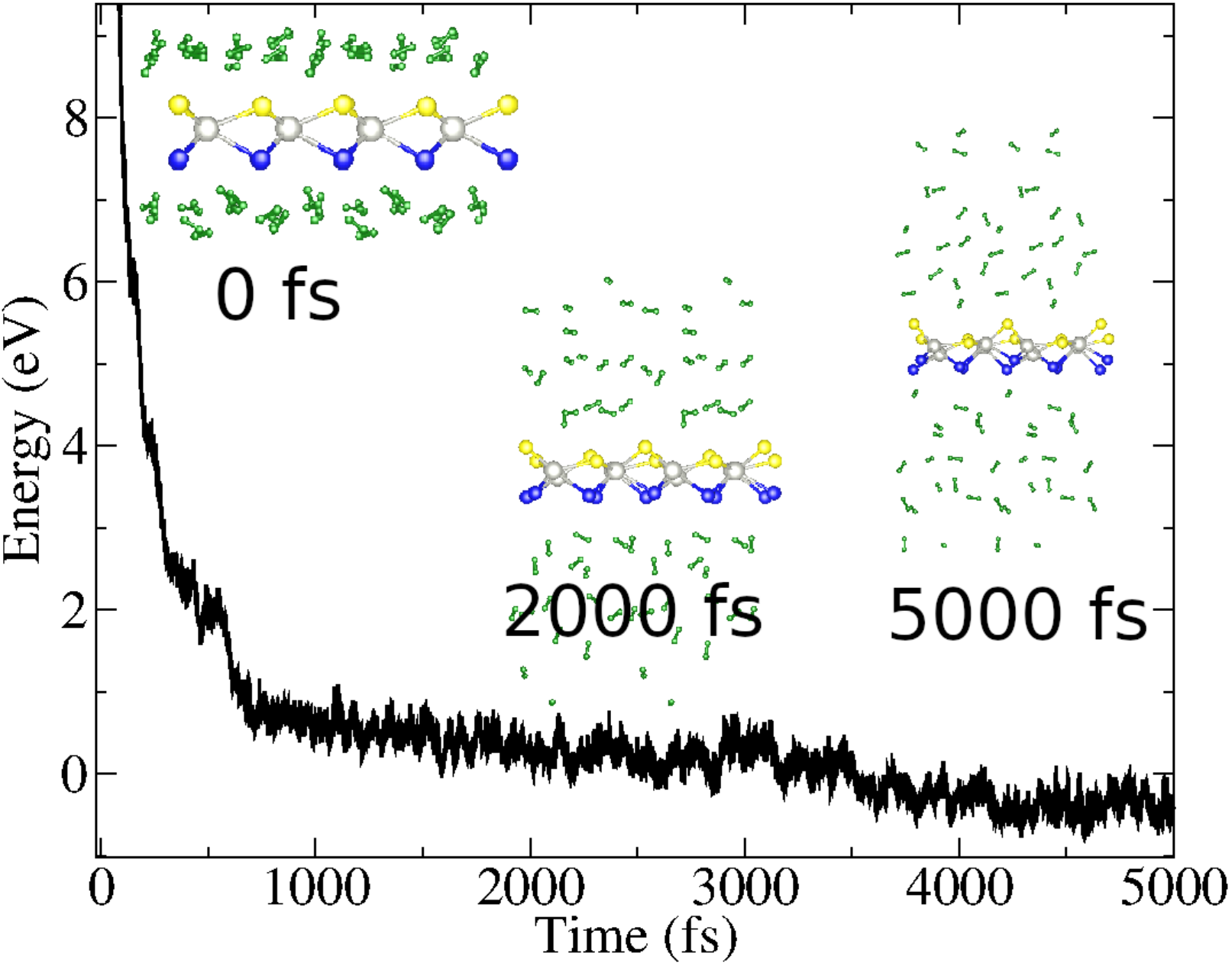}
		\end{subfigure}
		\begin{subfigure}[]{0.45\columnwidth}
			\subcaption[]{}
			\includegraphics[width=\columnwidth,clip=true]{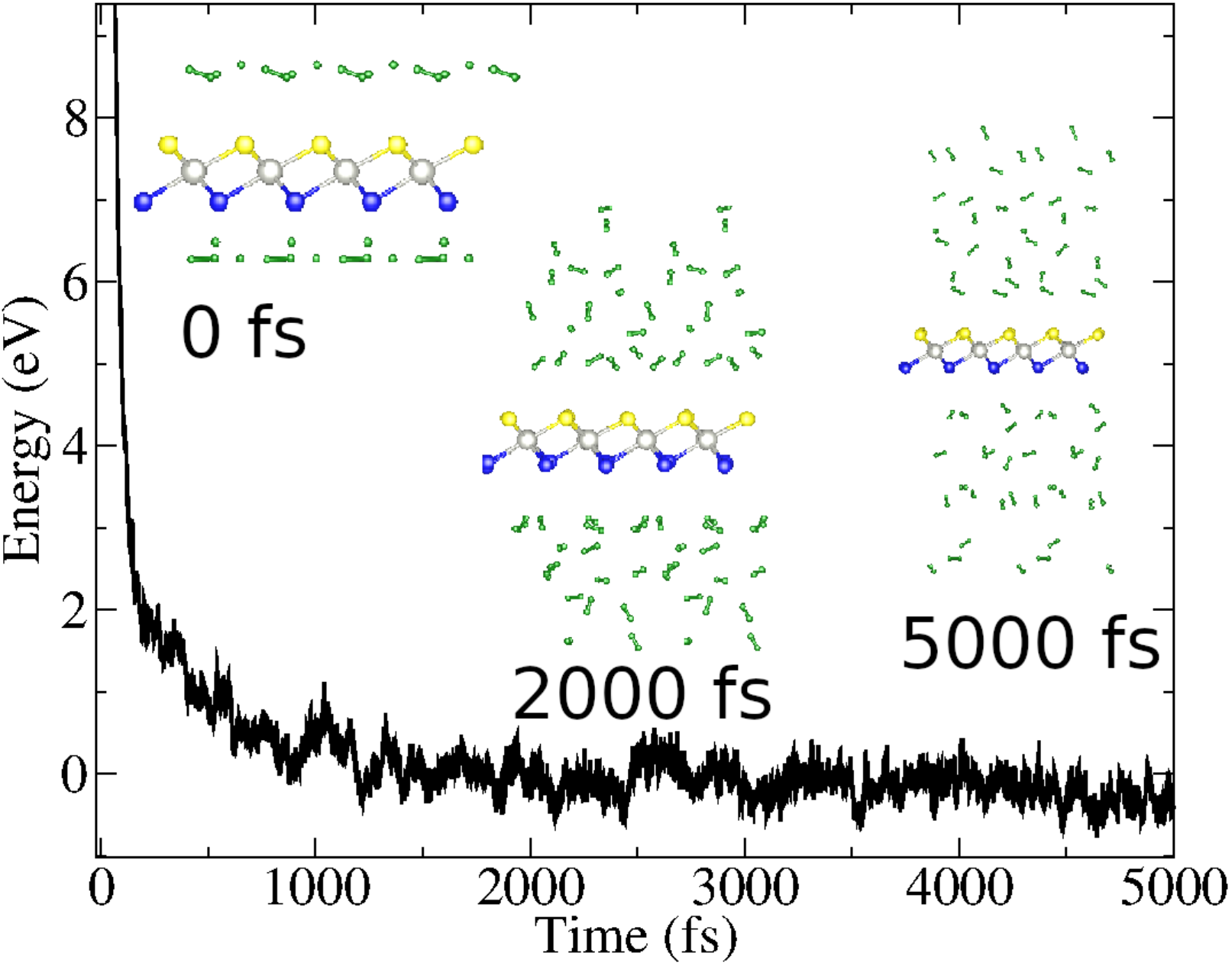}
		\end{subfigure}	\\
		\begin{subfigure}[]{0.45\columnwidth}
			\subcaption[]{}
			\includegraphics[width=\columnwidth,clip]{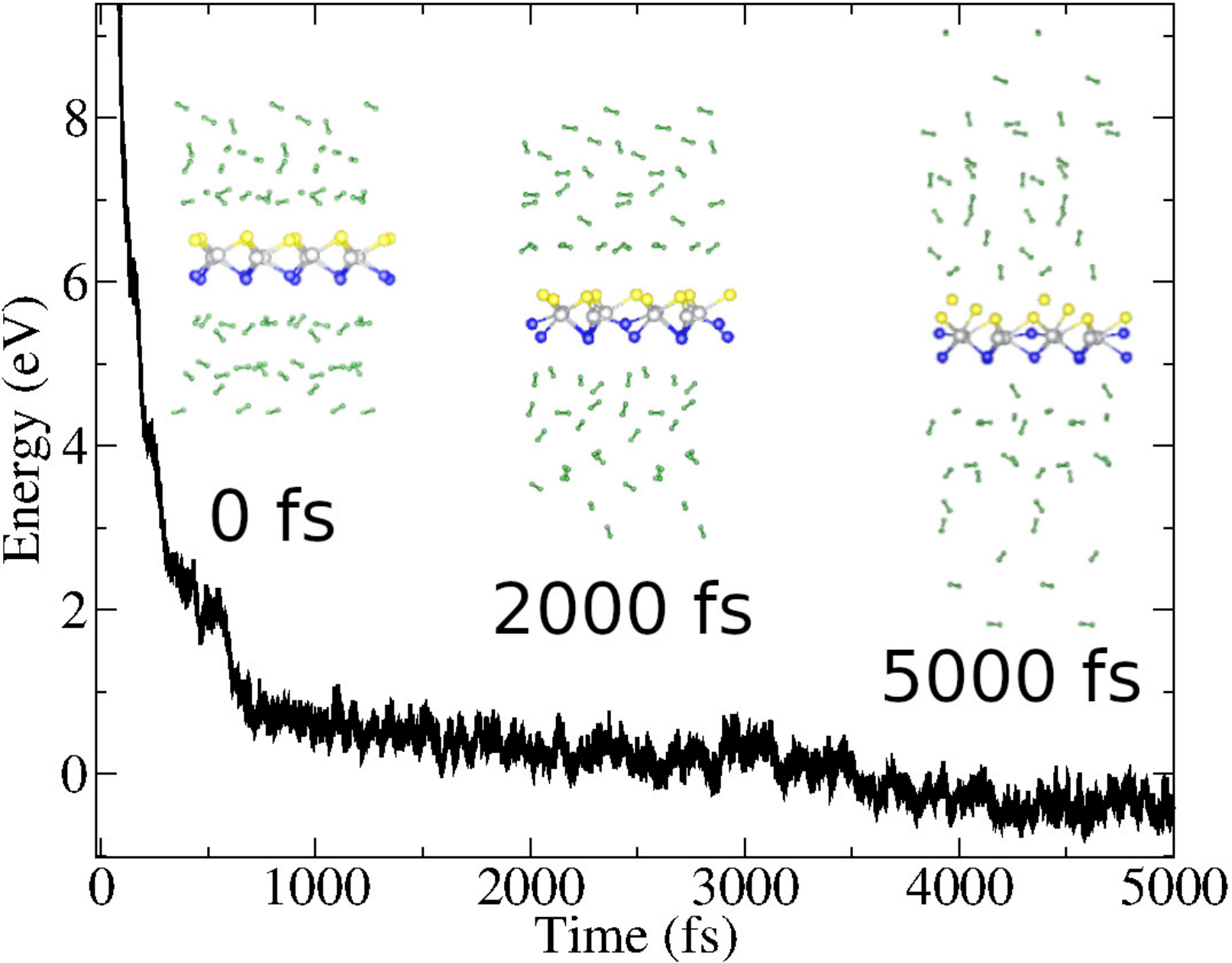}
		\end{subfigure}	
		\begin{subfigure}[]{0.45\columnwidth}
			\subcaption[]{}
			\includegraphics[width=\columnwidth,clip=true]{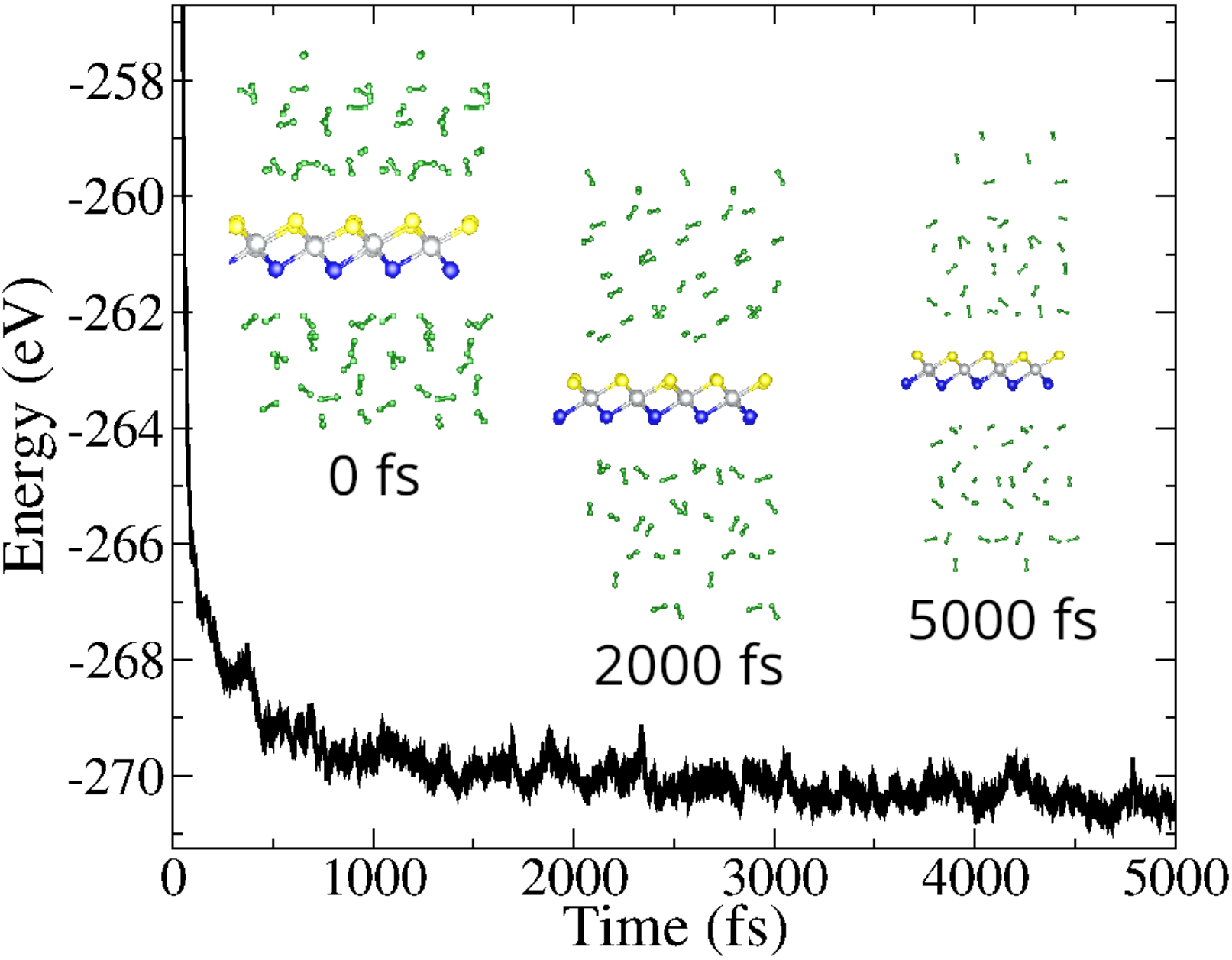}
		\end{subfigure}\\		
		\begin{subfigure}[]{0.45\columnwidth}
			\subcaption[]{}
			\includegraphics[width=\columnwidth,clip]{newmdenergyvstime1tpdsse.pdf}
		\end{subfigure}
			\begin{subfigure}[]{0.45\columnwidth}
			\subcaption[]{}
			\includegraphics[width=\columnwidth,clip]{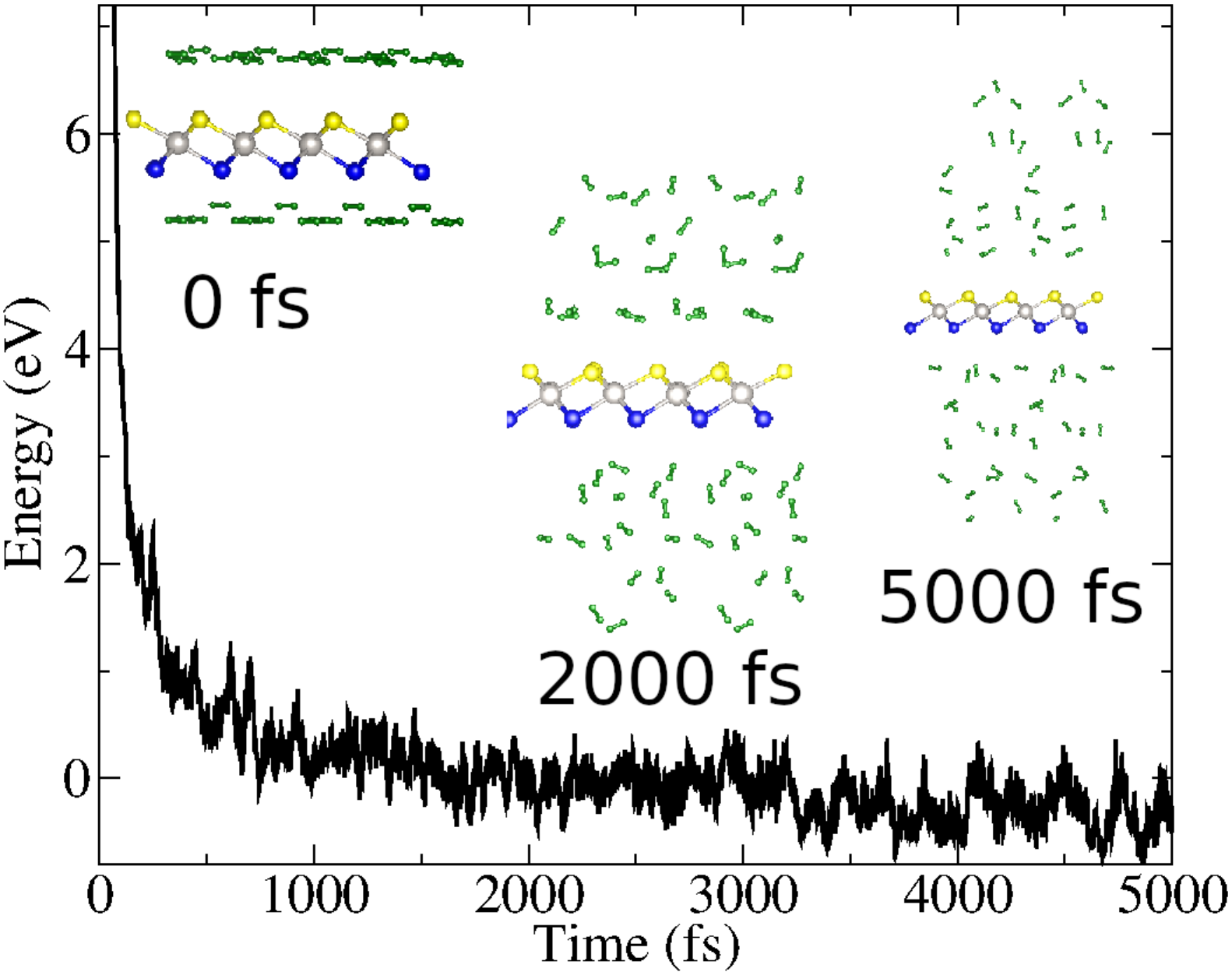}
		\end{subfigure}
		\caption{Total energy as a function of simulation time for hydrogen adsorption on a) 2H-NiSSe structure and b) 1-NiSSe structure, c) 2H-PdSSe, d) 1T-PdSSe, e) 2H-PtSSe and f) 1T-PtSSe for a concentration of 32 H$_2$ molecules.}
		\label{fig:mdenergyvstime32hnisse_Janus}
	\end{figure}

\begin{figure}[!ht]
\centering
\includegraphics[width=0.7\linewidth,clip]{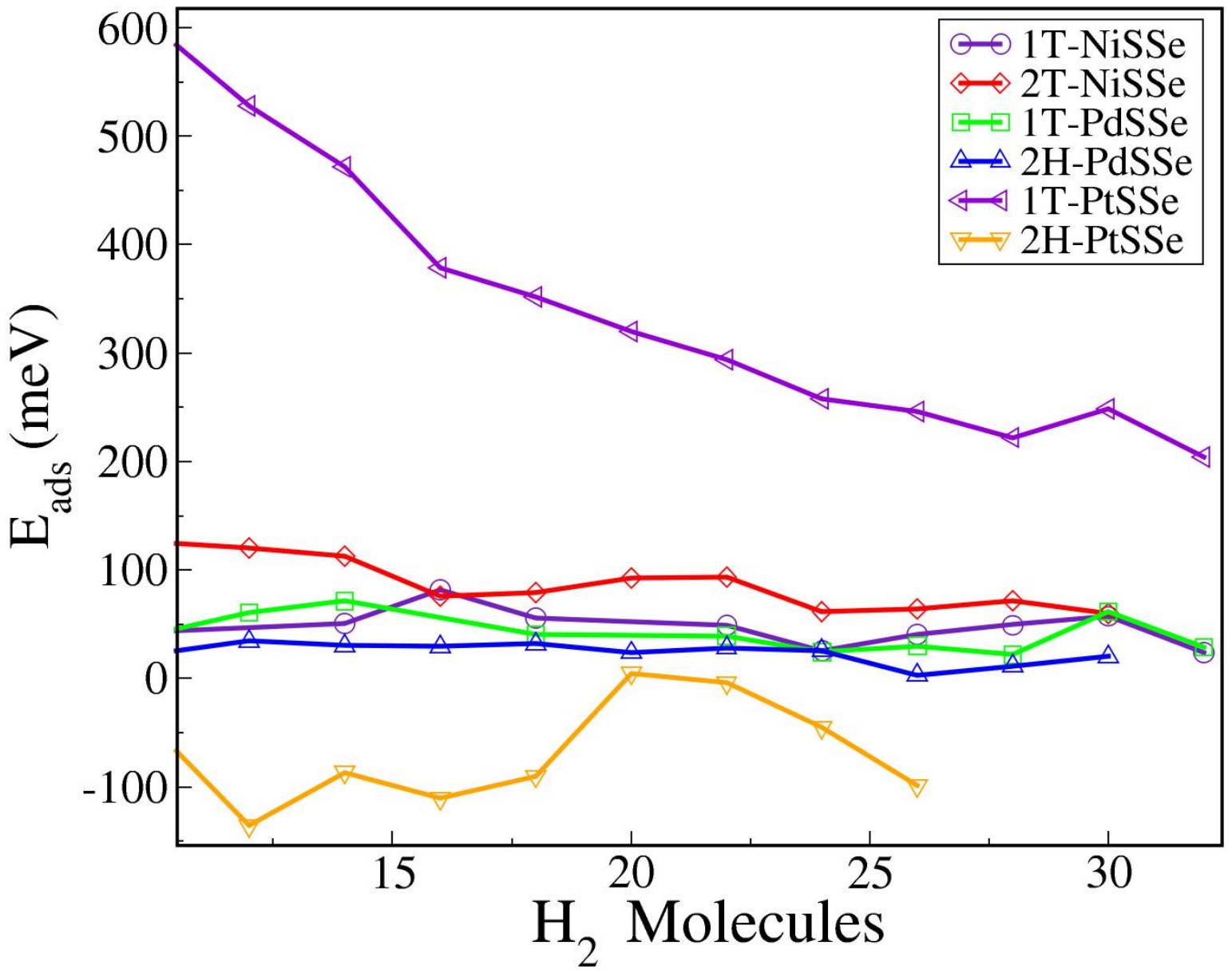}
\caption{Adsorption energy of hydrogen molecules on Janus structures as a function of the number of hydrogen molecules.}\label{fig:EadspermoleculesJANUS}
\end{figure}

To evaluate the practical hydrogen storage potential of Janus
dichalcogenides beyond the dilute regime, we examine adsorption at high
hydrogen loading, with up to 32 H$_2$ molecules per supercell. The
resulting relaxed configurations for representative Janus systems are
shown in Fig.~\ref{fig:1T_2H_NiSSe_32H2}, while the spatial distribution
and layering of hydrogen above the surface are illustrated in
Fig.~\ref{fig:NiSSe_H2_layers}. Quantitative metrics associated with
high-coverage adsorption and gravimetric performance are summarized in
Tables~\ref{tab:TMD_Janus_thickness} and \ref{tab:TMD_thickness}.

At high coverage, the adsorption behavior is governed less by the
binding strength of individual H$_2$ molecules and more by the ability
of the surface to accommodate multiple hydrogen layers without inducing
excessive intermolecular repulsion. In this regime, geometric factors
such as surface corrugation, adsorption volume, and layer spacing
become decisive. As a result, trends observed at low coverage are not
necessarily predictive of high-loading performance.

For NiSSe, Fig.~\ref{fig:1T_2H_NiSSe_32H2} reveals that both polymorphs
can accommodate multiple hydrogen layers while preserving molecular
integrity. However, clear differences emerge between the 1T and 2H
phases. In the 1T phase, hydrogen molecules are confined closer to the
surface, forming relatively compact layers. While this configuration
maximizes local binding strength, it also enhances H$_2$--H$_2$
repulsion, leading to a reduction in packing efficiency at very high
coverage. In contrast, the 2H phase provides a more open
trigonal-prismatic coordination environment, allowing hydrogen
molecules to distribute more uniformly and form well-separated layers,
as shown in Fig.~\ref{fig:NiSSe_H2_layers}.

This geometric advantage of the 2H phase is reflected in the
gravimetric capacities reported in Table~\ref{tab:TMD_Janus_thickness}. Despite
weaker single-molecule adsorption energies, 2H-NiSSe achieves a higher
effective hydrogen uptake than its 1T counterpart due to more efficient
multilayer formation. Similar behavior is observed for PdSSe, where the
2H polymorph balances moderate adsorption strength with sufficient
adsorption volume to support multilayer hydrogen storage.

The comparison across Janus systems further highlights the importance
of matching adsorption strength to geometric capacity. PtSSe, which
exhibits excessively strong adsorption at low coverage on the
S-terminated surface, performs poorly at high loading. The strong
confinement of hydrogen molecules near the surface leads to compressed
configurations with increased intermolecular repulsion, limiting the
maximum achievable hydrogen density. This behavior underscores that
overbinding is detrimental to high-capacity storage, even when initial
adsorption appears favorable.

The gravimetric performance summarized in Table~\ref{tab:TMD_Janus_thickness}
also illustrates the intrinsic trade-off between host mass and
adsorption energetics. While selenide-based Janus systems benefit from
enhanced polarizability and stronger binding, their higher atomic mass
reduces the overall hydrogen weight percentage. Conversely, lighter
sulfide-rich systems can achieve higher gravimetric capacities despite
weaker binding, provided that multilayer adsorption is geometrically
accessible.

Overall, the high-loading analysis demonstrates that optimal hydrogen
storage in Janus dichalcogenides arises from a balance between moderate
physisorption strength, sufficient adsorption volume, and low host
mass. Systems such as 2H-NiSSe and 2H-PdSSe best satisfy these competing
requirements, whereas excessively strong adsorption or excessive host
mass limits performance. These findings motivate an explicit assessment
of finite-temperature stability for high-coverage Janus systems, which
is addressed in the following section.

\begin{figure}[!ht]
\centering
\begin{subfigure}{0.3\columnwidth}
\subcaption[]{}
\includegraphics[width=\columnwidth,keepaspectratio,clip=true]{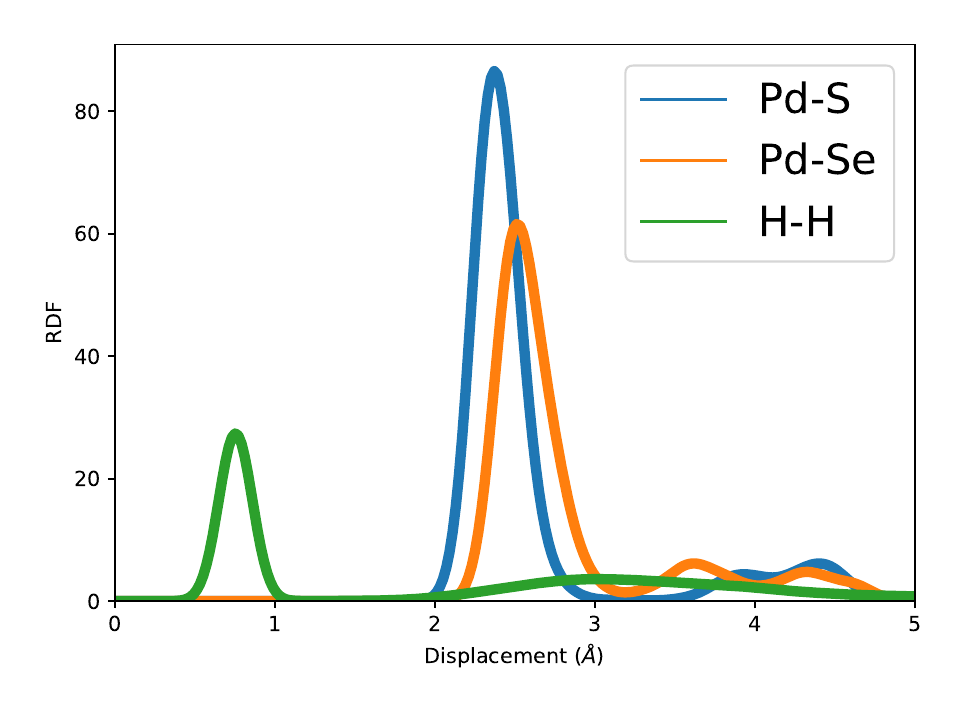}
\end{subfigure}%
\hfill
\begin{subfigure}{0.3\columnwidth}
\subcaption[]{}		\includegraphics[width=\columnwidth,keepaspectratio,clip=true]{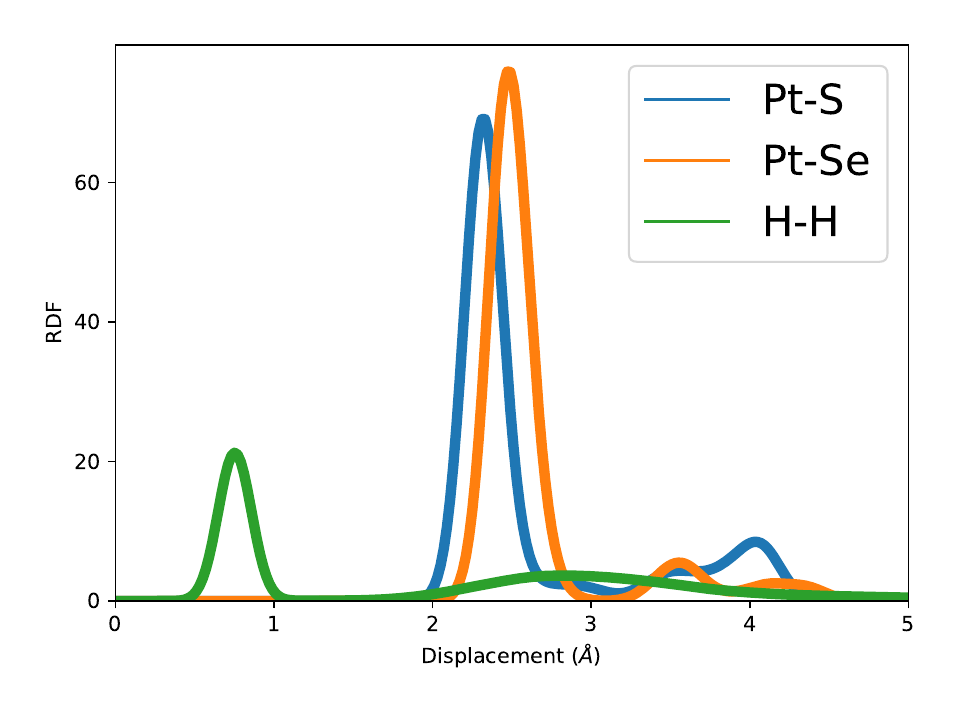}
\end{subfigure}%	
\hfill
\begin{subfigure}{0.3\columnwidth}
\subcaption[]{}	\includegraphics[width=\columnwidth,keepaspectratio,clip=true]{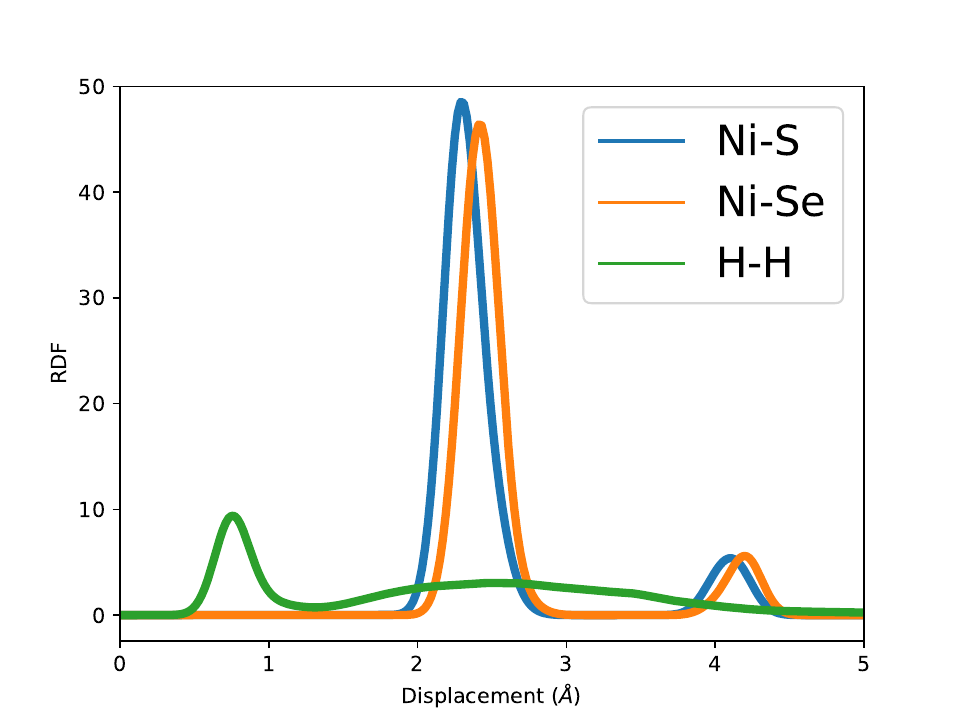}
\end{subfigure}
\hfill
\begin{subfigure}{0.3\columnwidth}
\subcaption[]{}			\includegraphics[width=\columnwidth,keepaspectratio,clip=true]{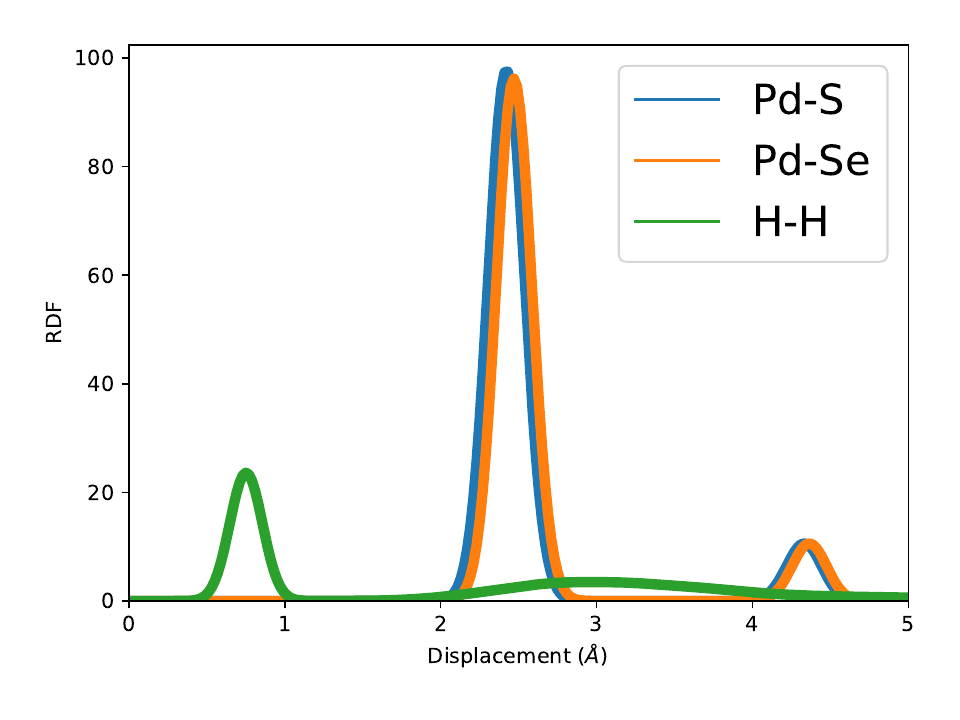}
\end{subfigure}%
\hfill
\begin{subfigure}{0.3\columnwidth}
\subcaption[]{}		\includegraphics[width=\columnwidth,keepaspectratio,clip=true]{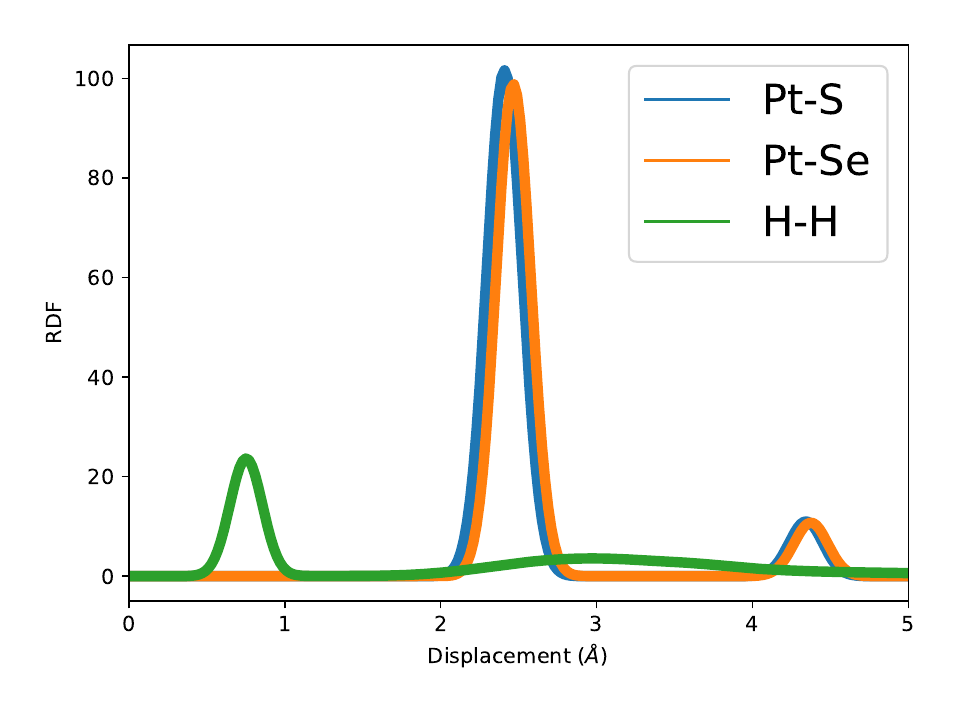}
\end{subfigure}%
\hfill	
\begin{subfigure}{0.3\columnwidth}
\subcaption[]{}			\includegraphics[width=\columnwidth,keepaspectratio,clip=true]{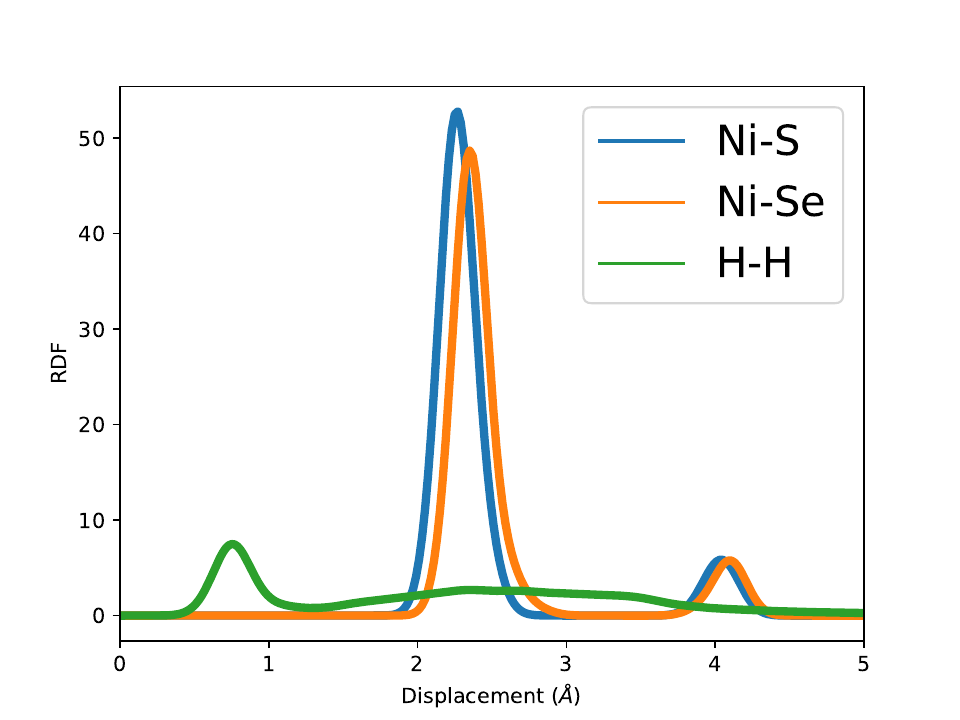}
\end{subfigure}%
\caption{RDF of MD simulations at \SI{300}{\kelvin} for 2H-phase of a) PdSSe, b) PtSSe and c) NiSSe and  1T-phase of d) PdSSe, e)  PtSSe and f) NiSSe.}
\label{fig:rdfMSSe}
\end{figure}

\begin{table}[h!]
\centering
\begin{tabular}{lcccc}
\hline
phase & surface H$_2$ & \multicolumn{2}{c}{$\Delta$ (\AA)} & gravimetric (\%)\\
\hline
& & DFT ($T=0$ K) & AIMD ($T=300$ K) & \\
\hline
1T-PdSSe       & Side A (B)     & $8.94 \, (8.84)$ & $14.81 \, (14.61)$ & 6.85\\
1T-PtSSe       & Side A (B)     & $8.15 \, (7.42)$ & $15.94 \, (14.19)$ & 4.97\\
1T-NiSSe       & Side A (B)     & $9.56 \, (9.83)$ & $11.07 \, (8.30)$ & 8.61 \\
\hline			
2H-PdSSe       & Side A (B)     & $6.55 \, (6.45)$ & $14.18 \, (13.18)$ & 6.85\\
2H-PtSSe       & Side A (B)     & $8.15 \, (7.42)$ & $16.71 \, (14.93)$ & 4.97\\
2H-NiSSe       & Side A (B)     & $7.26 \, (7.90)$ & $11.71 \, (8.29)$ & 8.61\\
\bottomrule
\end{tabular}
\caption{H$_2$ layer thickness $\Delta$ for DFT and AIMD calculations on side A (top layer) and B (bottom layer) and gravimetric capacity for Janus structures at concentration of 32 molecules.}
\label{tab:TMD_Janus_thickness}
\end{table}

\subsection{Finite-temperature behavior of Janus systems under high hydrogen coverage}

To assess the robustness of multilayer hydrogen adsorption in Janus
systems under realistic operating conditions, we performed
\textit{ab initio} molecular dynamics simulations at 300 K for Janus
monolayers loaded with up to 32 H$_2$ molecules per supercell. The time
evolution of the total energy for representative systems is shown in
Fig.~\ref{fig:mdenergyvstime32hnisse_Janus}, while additional snapshots
and trajectory analyses are presented in Figs.~16 and 17.

For all investigated Janus systems, the AIMD trajectories exhibit a
rapid equilibration phase within the first few hundred femtoseconds,
followed by stable energy fluctuations around a constant mean value.
The absence of systematic energy drift confirms that the multilayer
hydrogen configurations identified at zero temperature correspond to
thermally stable states rather than metastable artifacts of static
relaxation.

Throughout the simulations, hydrogen remains strictly molecular. The
H--H bond length fluctuates around its equilibrium value, and no
dissociation or surface hydrogenation events are observed. This
behavior persists even at the highest hydrogen loading, demonstrating
that Janus-induced polarity and multilayer confinement do not promote
chemical activation of H$_2$. Instead, adsorption remains governed by
physisorption and is therefore intrinsically reversible at room
temperature.

The dynamical behavior of hydrogen molecules reveals clear differences
between Janus systems and structural polymorphs. In 2H-NiSSe and
2H-PdSSe, hydrogen molecules retain a layered distribution above the
surface while exhibiting moderate lateral mobility within each layer.
This combination of confinement and mobility is indicative of an
optimal physisorption regime, where hydrogen is sufficiently stabilized
to prevent rapid desorption but not so strongly bound as to suppress
diffusion or reversibility.

In contrast, PtSSe exhibits markedly reduced hydrogen mobility during
the AIMD simulations. Hydrogen molecules remain tightly confined near
the surface, forming compact and partially compressed layers. While
this strong confinement enhances instantaneous adsorption strength, it
also amplifies intermolecular repulsion and suppresses molecular
rearrangement. As a result, PtSSe displays reduced dynamical
reversibility, consistent with its inferior high-coverage performance
identified in Sec.~3.7.

The comparison between 1T and 2H polymorphs further highlights the
importance of surface geometry at finite temperature. In the 1T phase,
hydrogen molecules are confined closer to the surface, leading to
stronger instantaneous interactions but reduced configurational
entropy. In contrast, the more open trigonal-prismatic coordination of
the 2H phase allows hydrogen molecules to sample a broader region of
configuration space while remaining bound. This enhanced configurational
flexibility contributes to the superior thermal stability and storage
performance of 2H-based Janus systems at high coverage.

Taken together, the AIMD results confirm that the most promising Janus
candidates identified from static calculations—namely 2H-NiSSe and
2H-PdSSe—retain their favorable adsorption characteristics under
ambient thermal conditions. Multilayer hydrogen adsorption in these
systems is dynamically stable, fully reversible, and robust against
thermal fluctuations. These findings demonstrate that the optimal
balance between adsorption strength, geometric openness, and molecular
mobility identified in earlier sections persists at finite temperature,
providing a consistent and physically grounded picture of hydrogen
storage in Janus dichalcogenides.

\section{Conclusions}

In this work, we performed a comprehensive first-principles investigation
of hydrogen adsorption and storage in pristine and Janus transition-metal
dichalcogenide monolayers MX$_2$ and MSSe (M = Ni, Pd, Pt; X = S, Se),
combining static density-functional theory calculations with finite-temperature
\textit{ab initio} molecular dynamics simulations. By systematically comparing
structural polymorphs, chemical composition, and surface asymmetry, we
established a clear microscopic understanding of how electronic structure
and lattice geometry govern hydrogen adsorption in two-dimensional
dichalcogenides.

Electronic-structure analysis reveals that the adsorption behavior is
controlled primarily by the participation of metal $d$ states near the
Fermi level. Moving from Pd and Pt to Ni increases $d$-orbital localization
and enhances metal–chalcogen hybridization, which systematically strengthens
hydrogen–surface interactions. Chalcogen substitution from S to Se raises
the valence-band manifold and increases polarizability, further modulating
adsorption strength. These trends explain the monotonic progression from
weakly interacting Pd- and Pt-based systems to strongly binding Ni-based
dichalcogenides.

Structural phase plays an equally decisive role. The metallic 1T polymorph
consistently promotes stronger hydrogen adsorption than the semiconducting
2H phase due to enhanced electronic screening and a higher density of states
at the Fermi level. However, at high hydrogen coverage, the more open
trigonal-prismatic coordination of the 2H phase provides a larger accessible
adsorption volume and greater configurational freedom for hydrogen molecules.
As a result, 2H systems exhibit superior multilayer stability and improved
reversibility under realistic loading conditions, despite their weaker
single-molecule adsorption energies.

Janus functionalization introduces an intrinsic out-of-plane dipole that
breaks mirror symmetry and induces chemically inequivalent S- and Se-terminated
surfaces. This asymmetry produces measurable differences in adsorption energy,
layer thickness, and hydrogen distribution between the two sides of the
monolayer. While Janus engineering enhances adsorption selectivity and enables
fine tuning of binding strength, it does not fundamentally alter the
physisorption-dominated nature of hydrogen uptake. Instead, its primary role
is to optimize the balance between confinement and reversibility, particularly
in Ni- and Pd-based systems.

Finite-temperature \textit{ab initio} molecular dynamics simulations at 300 K
confirm that hydrogen remains strictly molecular across all investigated
materials, even at the highest loadings considered. No dissociation or
irreversible chemisorption is observed. Among all systems, 2H-NiSSe and
2H-PdSSe emerge as the most robust candidates, combining moderate adsorption
energies, stable multilayer hydrogen configurations, and excellent thermal
stability. In contrast, Pt-based Janus systems exhibit overly strong confinement
and reduced reversibility, limiting their suitability for practical hydrogen
storage.

Overall, our results demonstrate that efficient hydrogen storage in
two-dimensional dichalcogenides is governed by a delicate interplay between
electronic structure, surface geometry, and adsorption-induced entropy.
Optimal performance is achieved not by maximizing adsorption strength, but by
maintaining hydrogen binding within the intermediate physisorption regime
that enables reversible, thermally stable uptake. These insights establish
clear design principles for engineering low-dimensional materials for hydrogen
storage and highlight Janus Ni- and Pd-based dichalcogenides as promising
platforms for reversible molecular hydrogen adsorption.

	\section{Acknowledgments}
	All authors acknowledge the financial support from the Brazilian
funding agency CNPq under grant numbers 305174/2023-1, 408144/2022-0
and 305952/2023-4. G.E.G.A. thanks CNPq for a undergraduate
scholarship. We thank computational resources from Supercomputers LaMCAD at UFG, Santos
Dumont at LNCC and CENAPAD-SP at Unicamp.

	\section*{Authors contributions}
	Flavio Bento de Oliveira, Gabriel Elyas Araujo and Andreia Luisa da Rosa and equally contributed to the calculations, validation of the results, and writing and revising the manuscript.
	\section*{Declaration of competing interest}
	The authors declare that they have no known competing financial interests or personal relationships that could have appeared to
	influences the work reported in this study.
	\section*{Data Availability}
	Data supporting this study are available upon request.
	
%\bibliographystyle{iopart-num}
%\bibliography{ref}

\clearpage
\newpage

\centering{Supporting material\\Hydrogen storage in pristine and Janus transition-metal dichalcogenide monolayers: electronic origins, coverage effects, and finite-temperature stability}

\setcounter{section}{0}

\section{Lattice parameters}

\begin{figure}[!ht]
\centering
		\begin{subfigure}[b]{0.32\columnwidth}
			\subcaption[]{}
\includegraphics[width=\columnwidth,clip=true]{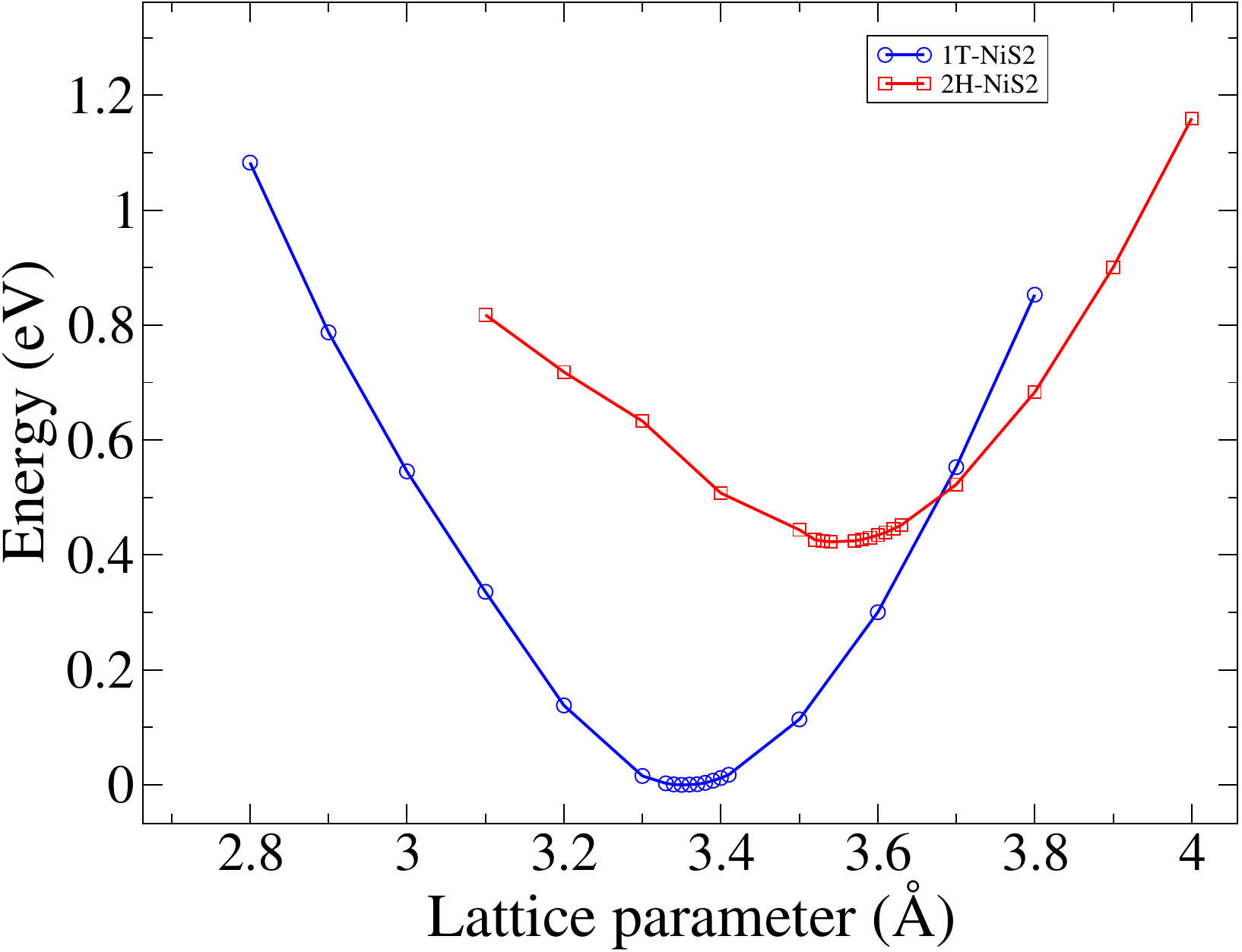}
		\end{subfigure}\hfill
		\begin{subfigure}[b]{0.32\columnwidth}
			\subcaption[]{}
			\includegraphics[width=\columnwidth,clip=true]{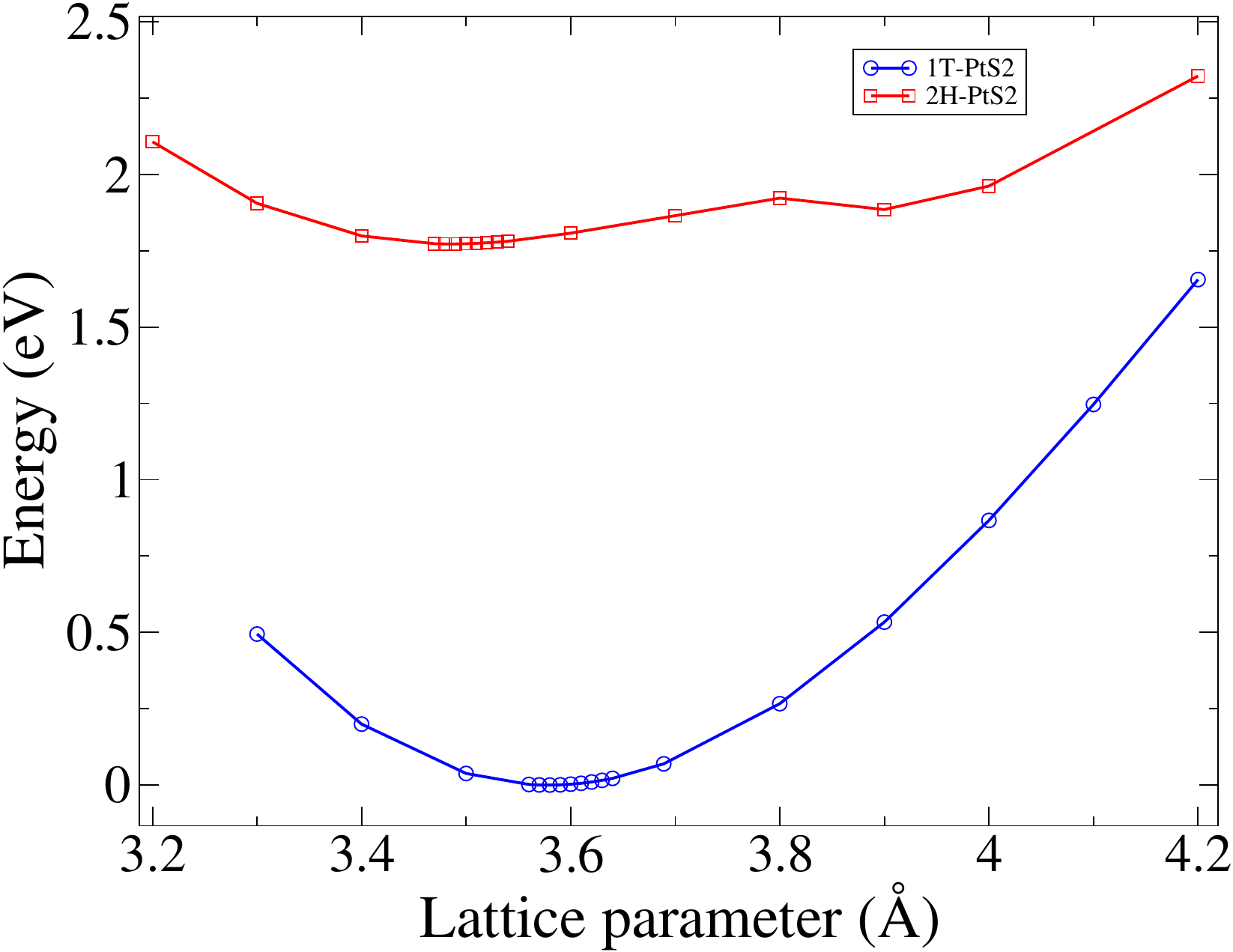}
		\end{subfigure}\hfill
		\begin{subfigure}[b]{0.32\columnwidth}
			\subcaption[]{}
			\includegraphics[width=\columnwidth,clip=true]{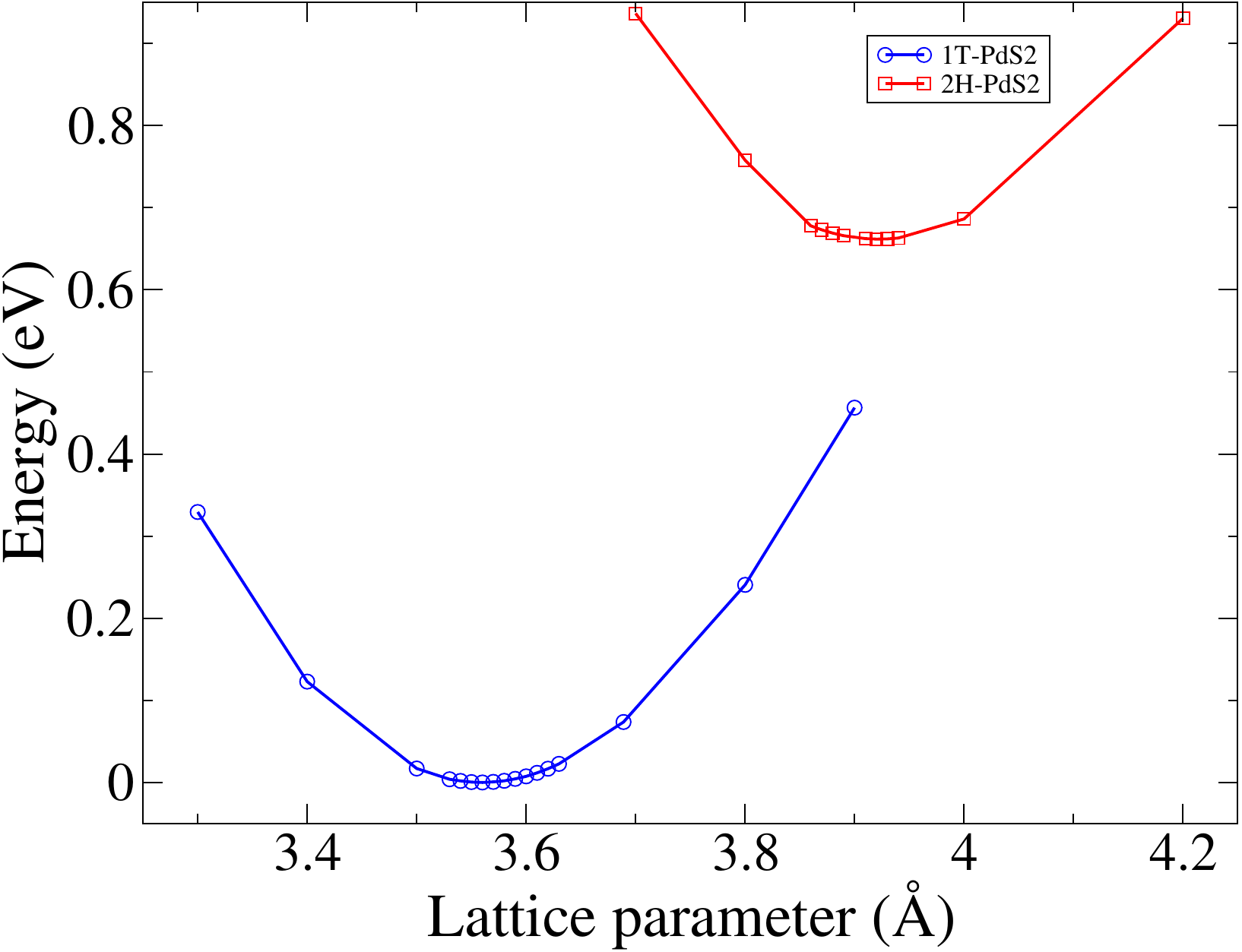}
		\end{subfigure}
		\\
		\begin{subfigure}[b]{0.32\columnwidth}
			\subcaption[]{}
			\includegraphics[width=\textwidth,clip=true]{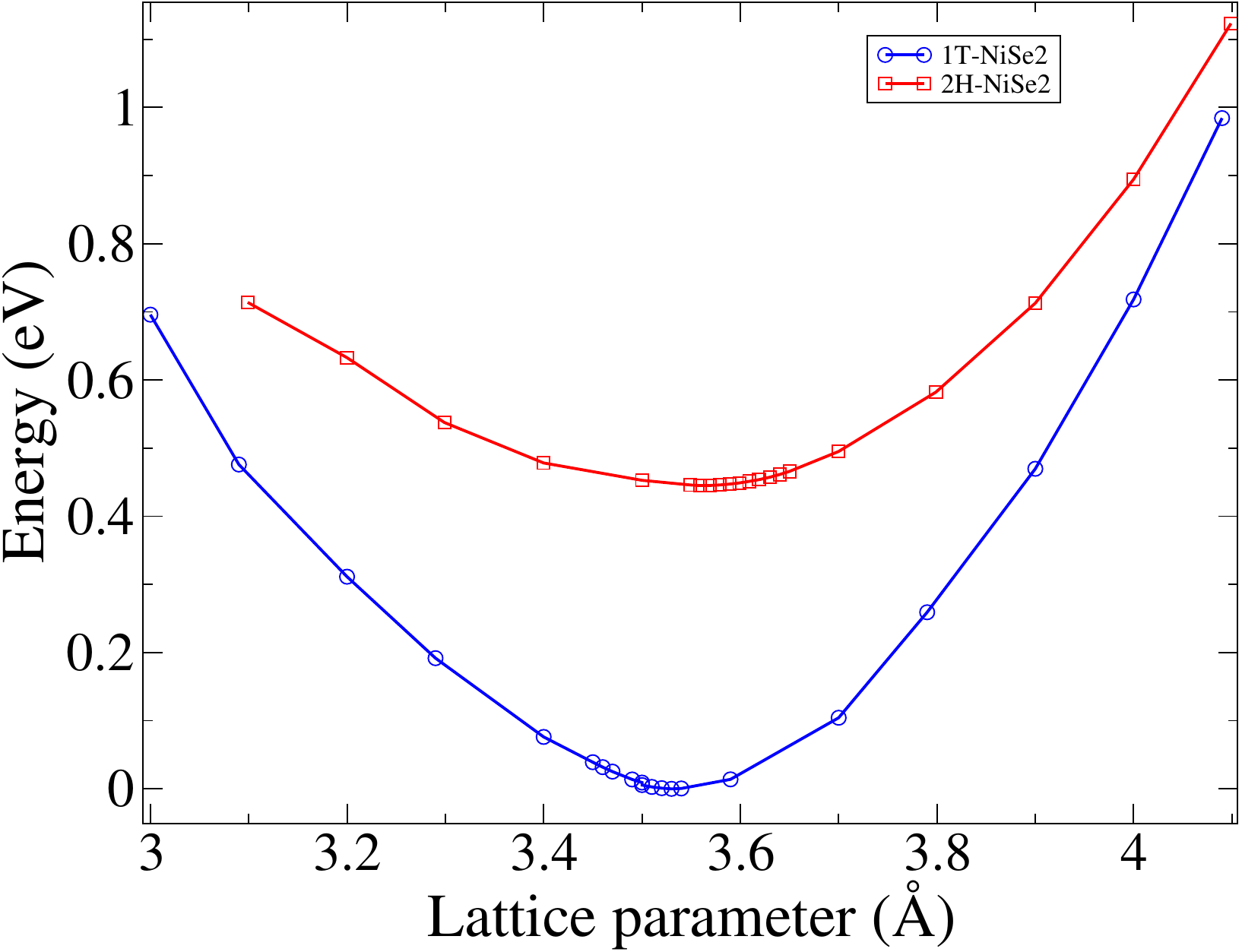}
		\end{subfigure}\hfill
		\begin{subfigure}[b]{0.32\columnwidth}
			\subcaption[]{}
			\includegraphics[width=\textwidth,clip=true]{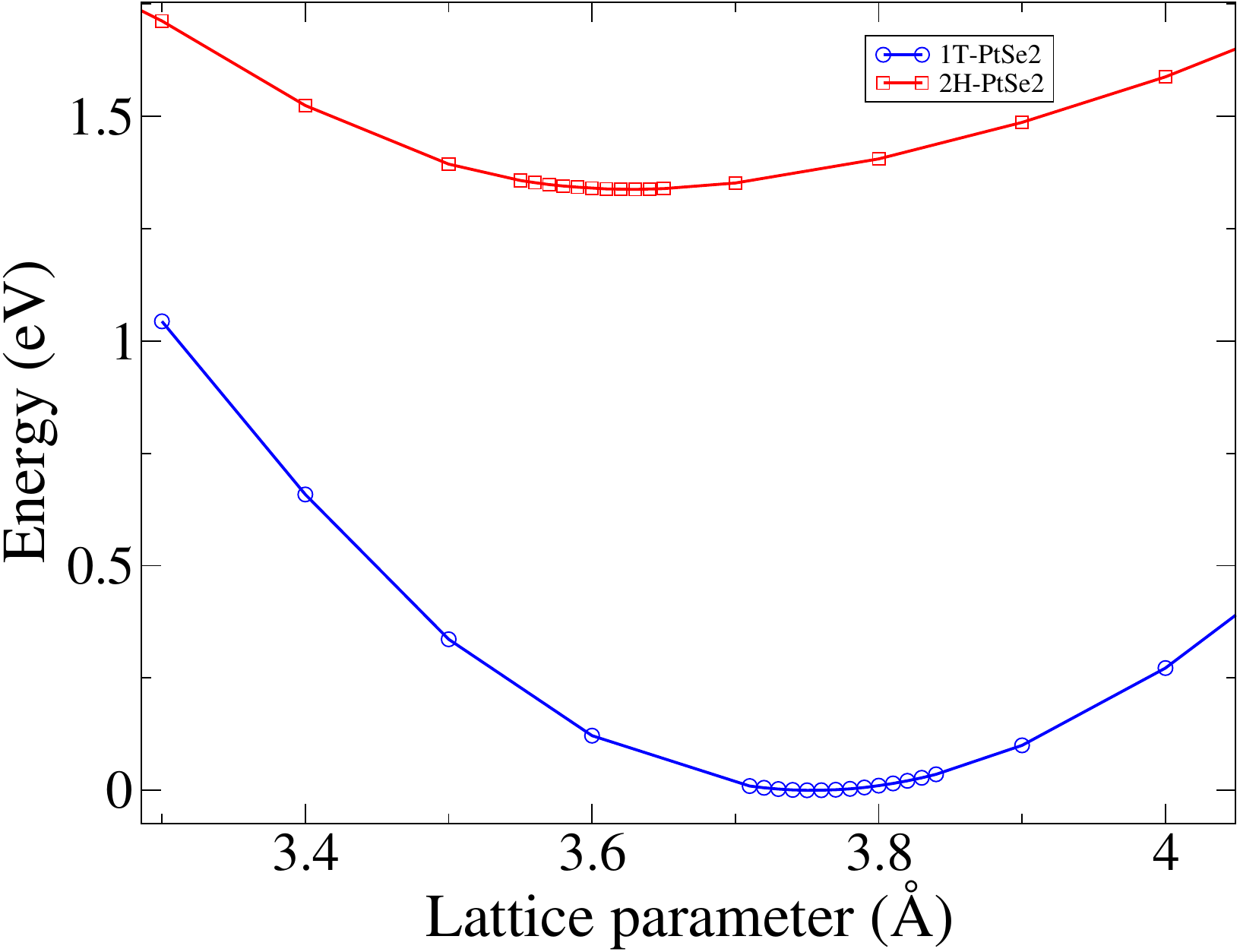}
		\end{subfigure}\hfill
		\begin{subfigure}[b]{0.32\columnwidth}
			\subcaption[]{}
			\includegraphics[width=\columnwidth,clip=true]{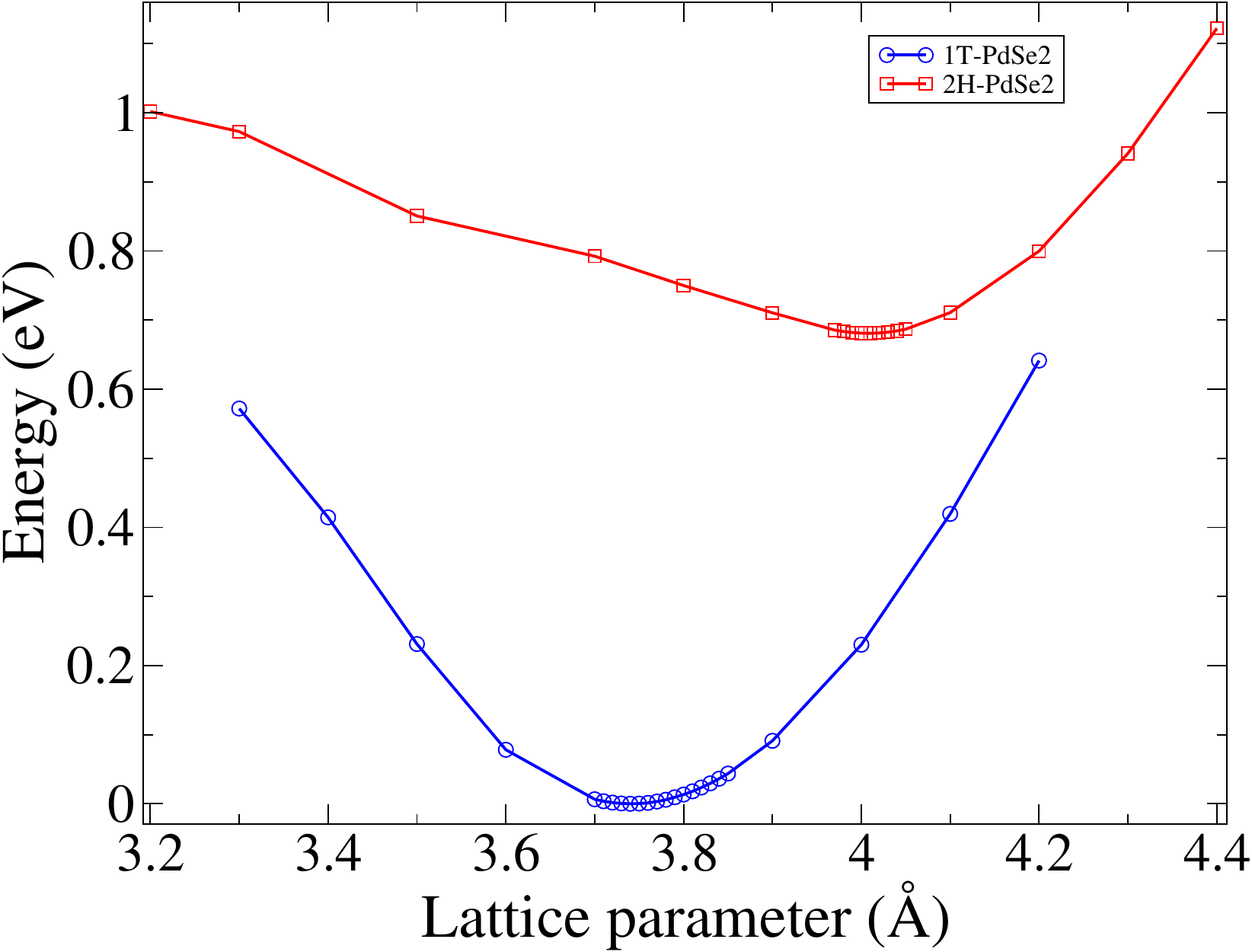}
		\end{subfigure}
		\caption{Total energy as a function of lattice parameter for a) NiS$_2$. b) PtS$_2$, c) PdS$_2$, d) NiSe$_2$, e) PtSe$_2$, f) PdSe$_2$.}	
	\end{figure}

	\begin{figure}
		\centering
		\begin{subfigure}[b]{0.32\columnwidth}
			\includegraphics[width=\columnwidth]{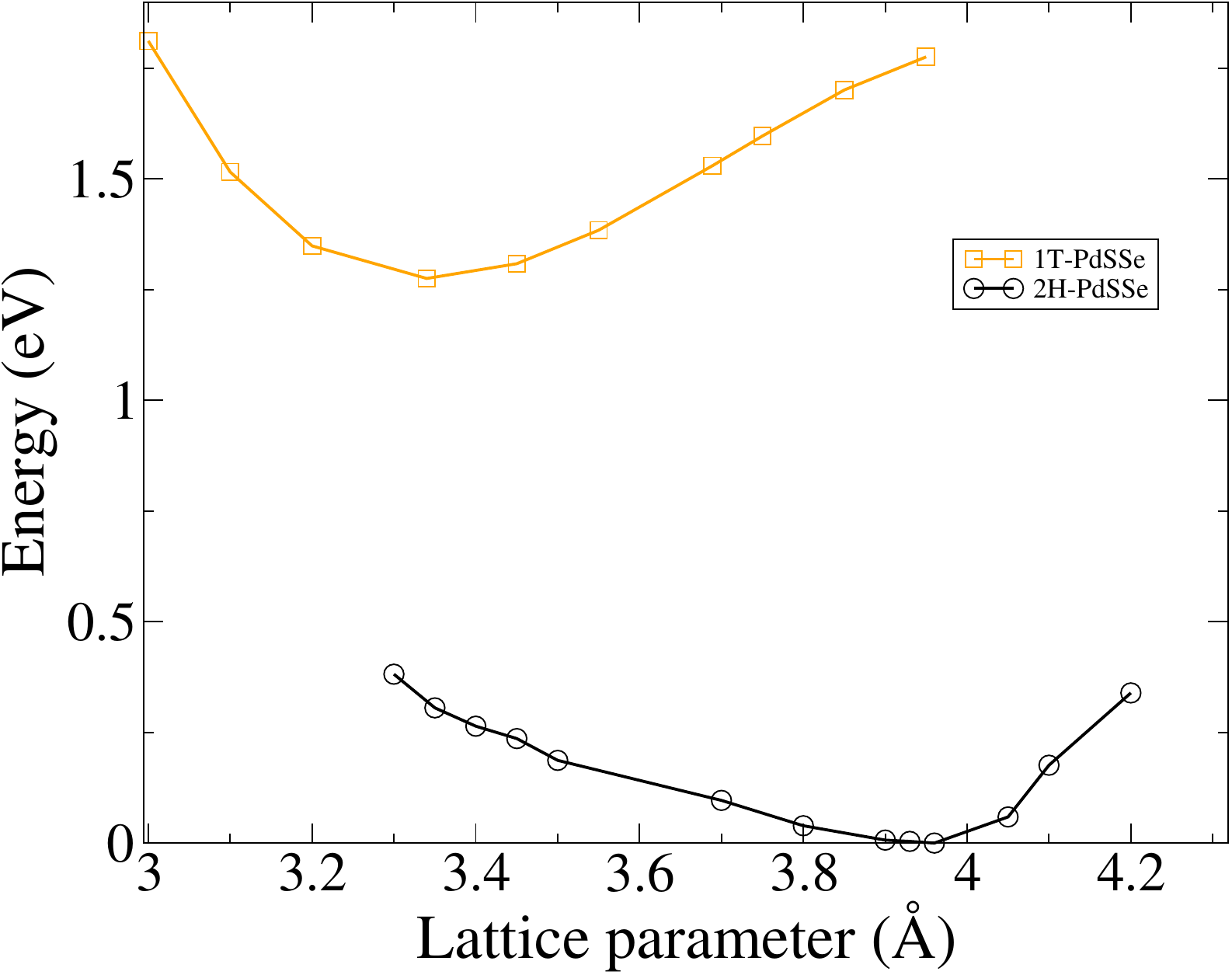}
			\subcaption[]{}
		\end{subfigure}\hfill
		\begin{subfigure}[b]{0.32\columnwidth}
			\includegraphics[width=\columnwidth]{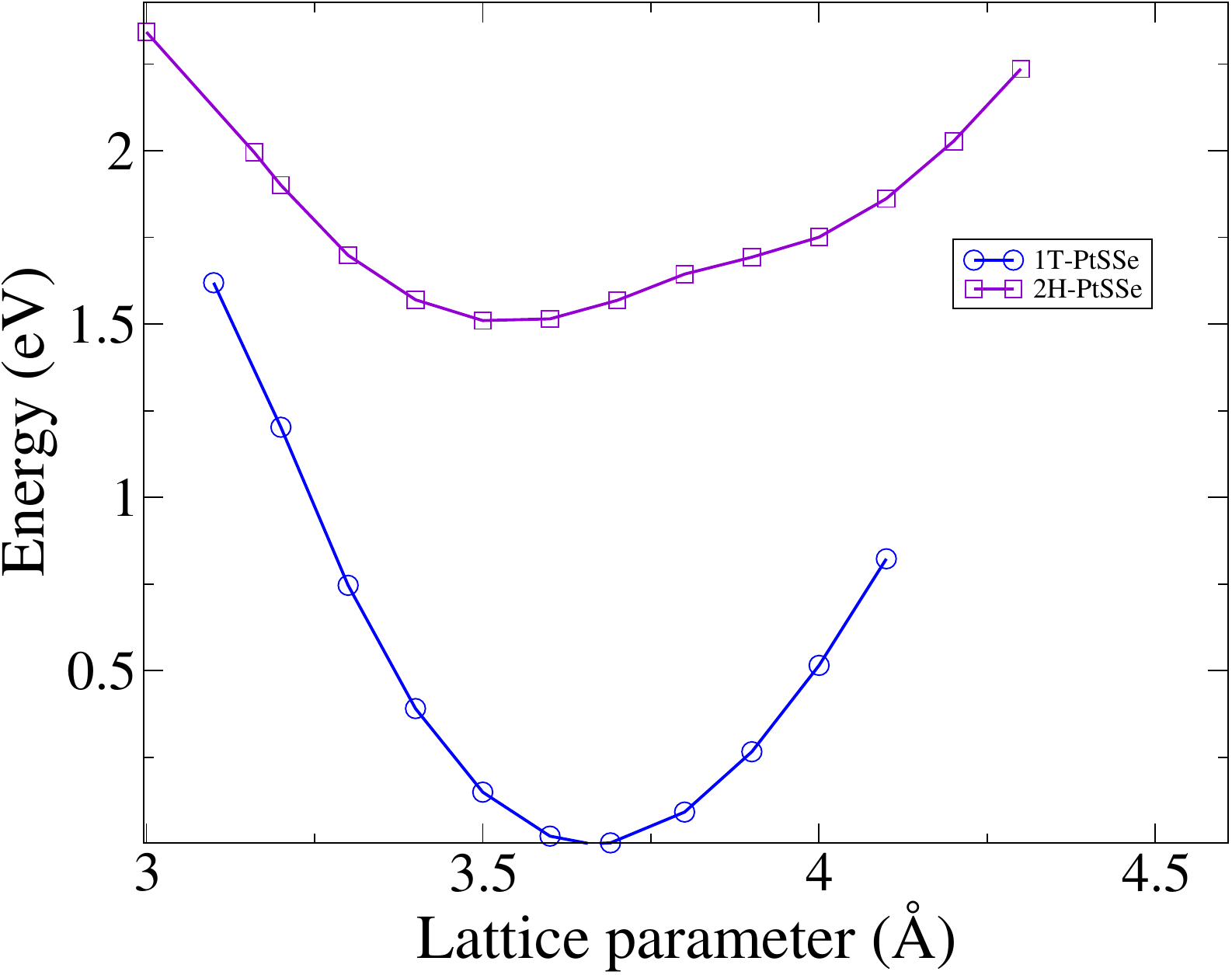}
			\subcaption[]{}
		\end{subfigure}\hfill
		\begin{subfigure}[b]{0.32\columnwidth}
			\includegraphics[width=\columnwidth]{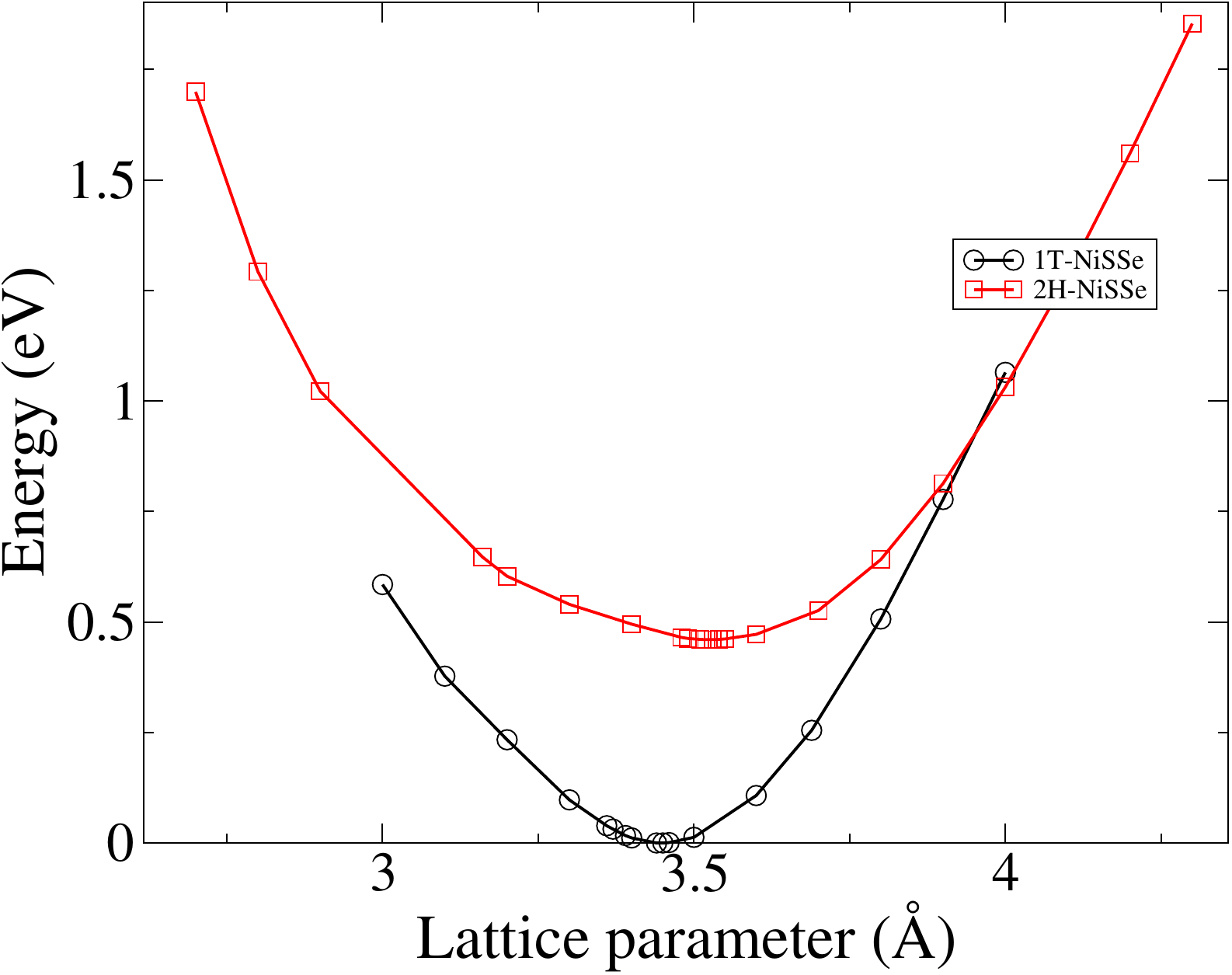}
			\subcaption[]{}
		\end{subfigure}
		\caption{Total energy as a function of lattice parameter $a$ for 2H and 1T phases of a) PdSSe, b) PtSSe and c) NiSSe.}\label{fig:relaxationxsse}
	\end{figure}

\section{AIMD simulations}

\begin{figure}
\centering
\includegraphics[width=0.5\columnwidth,clip=true]{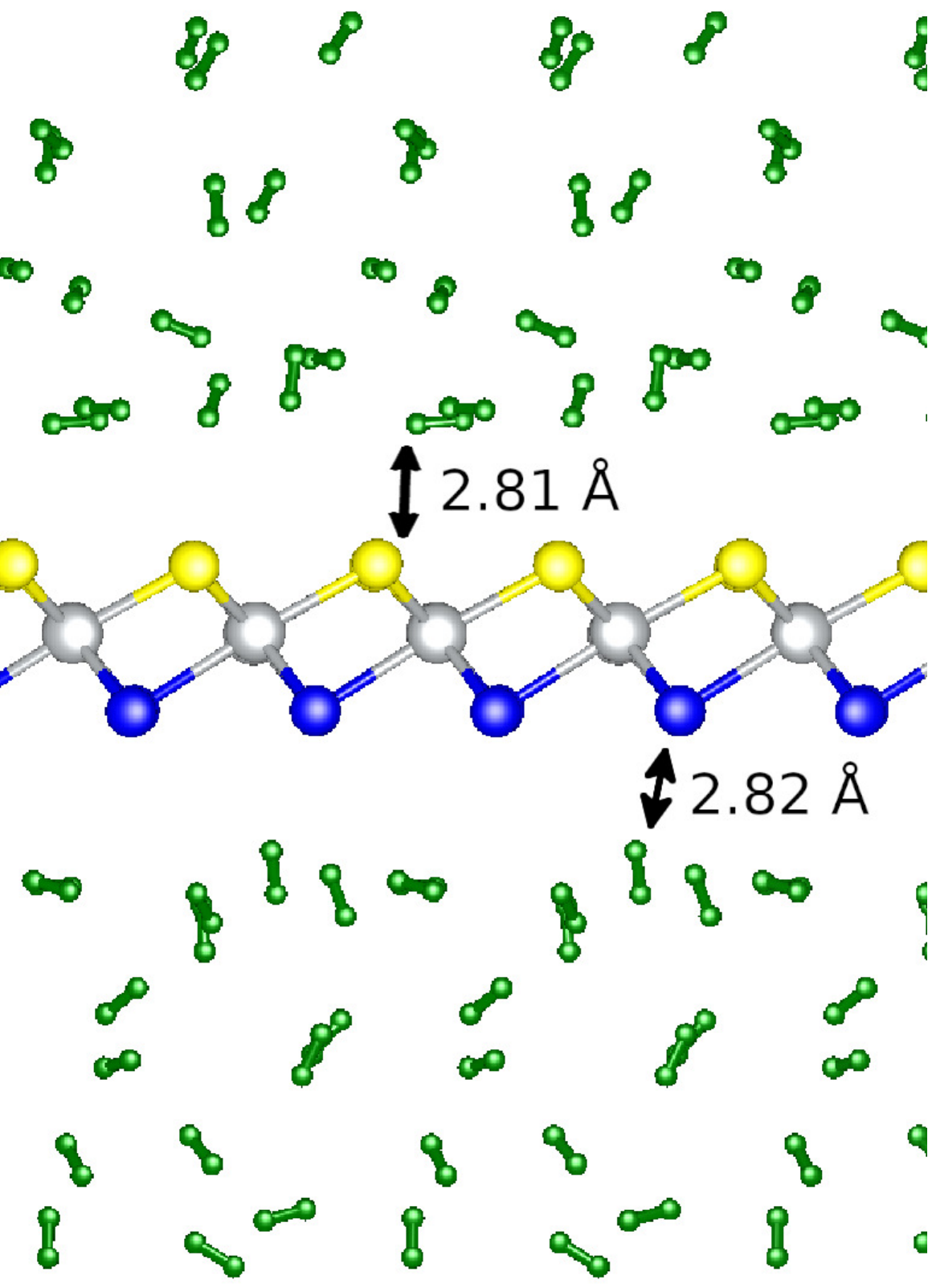}      \caption{Formation of multilayer hydrogen on Janus NiSSe monolayers. The first hydrogen layer distance to the surface is shown.}          
\end{figure}

\begin{figure}[!ht]
\centering
\begin{subfigure}[b]{0.32\columnwidth}
                        \subcaption[]{}
\includegraphics[width=\columnwidth,clip=true]{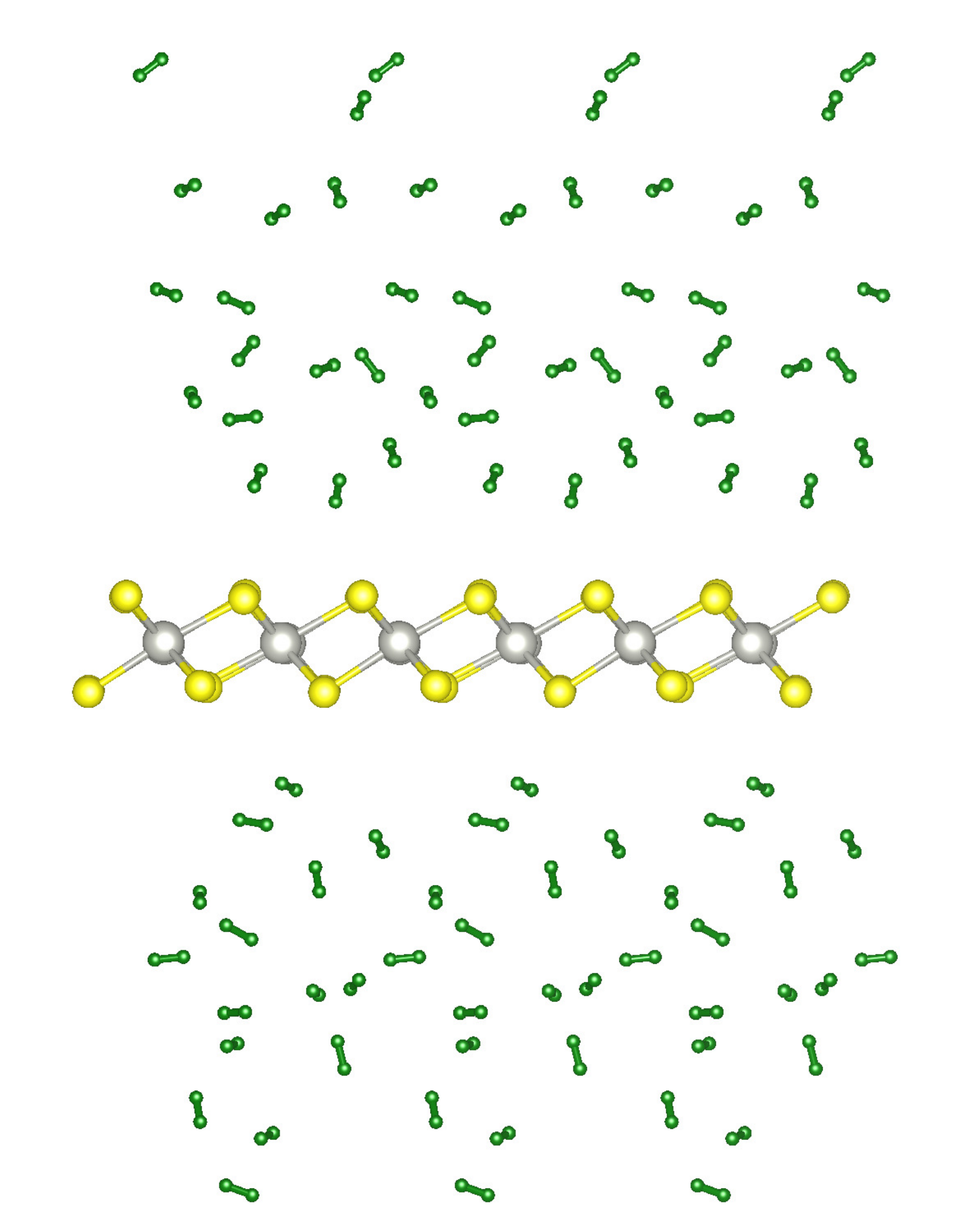}
 \end{subfigure}\hfill
\begin{subfigure}[b]{0.32\columnwidth}
                        \subcaption[]{}
\includegraphics[width=\columnwidth,clip=true]{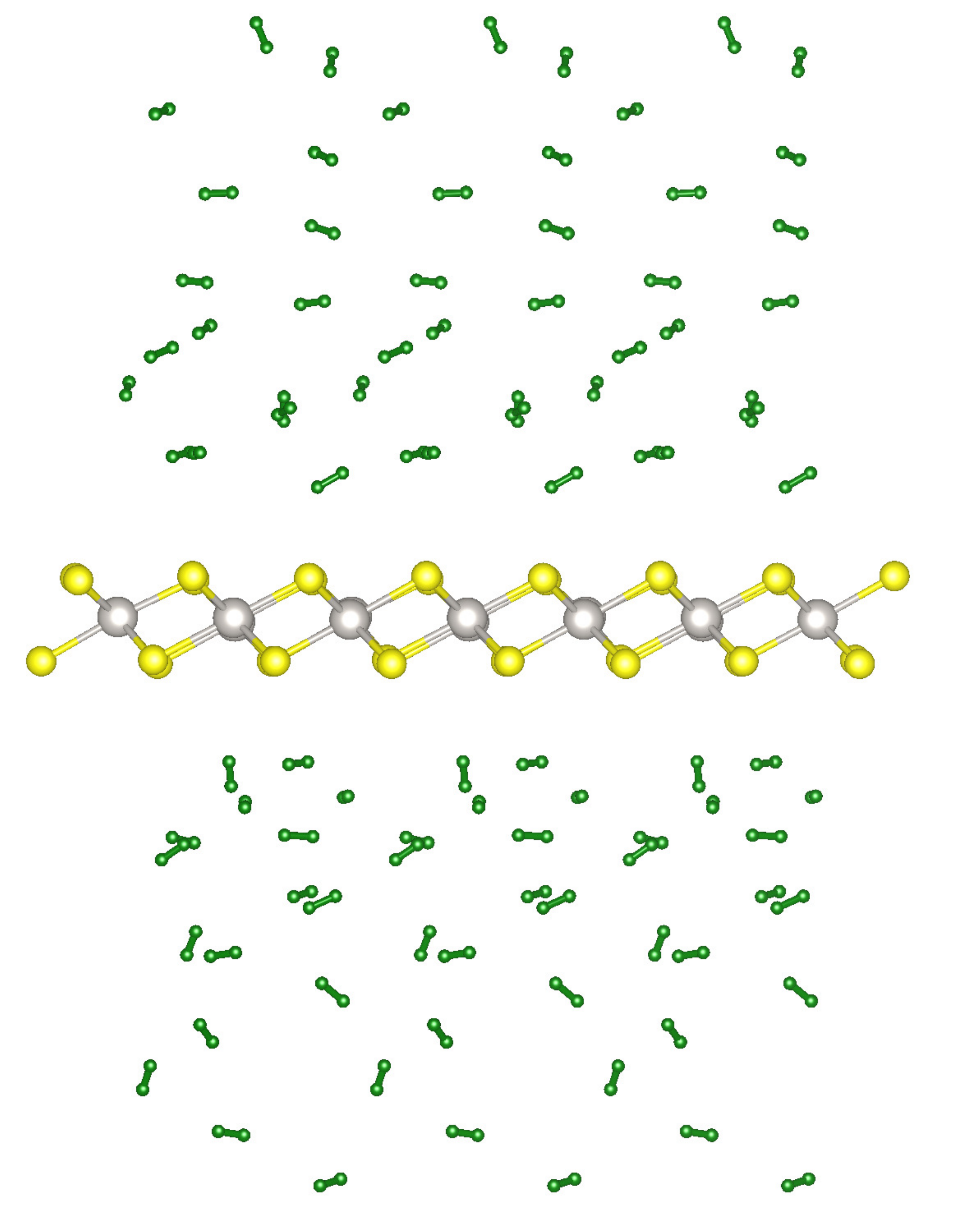}
 \end{subfigure}\hfill
 \begin{subfigure}[b]{0.32\columnwidth}
                        \subcaption[]{}
\includegraphics[width=\columnwidth,clip=true]{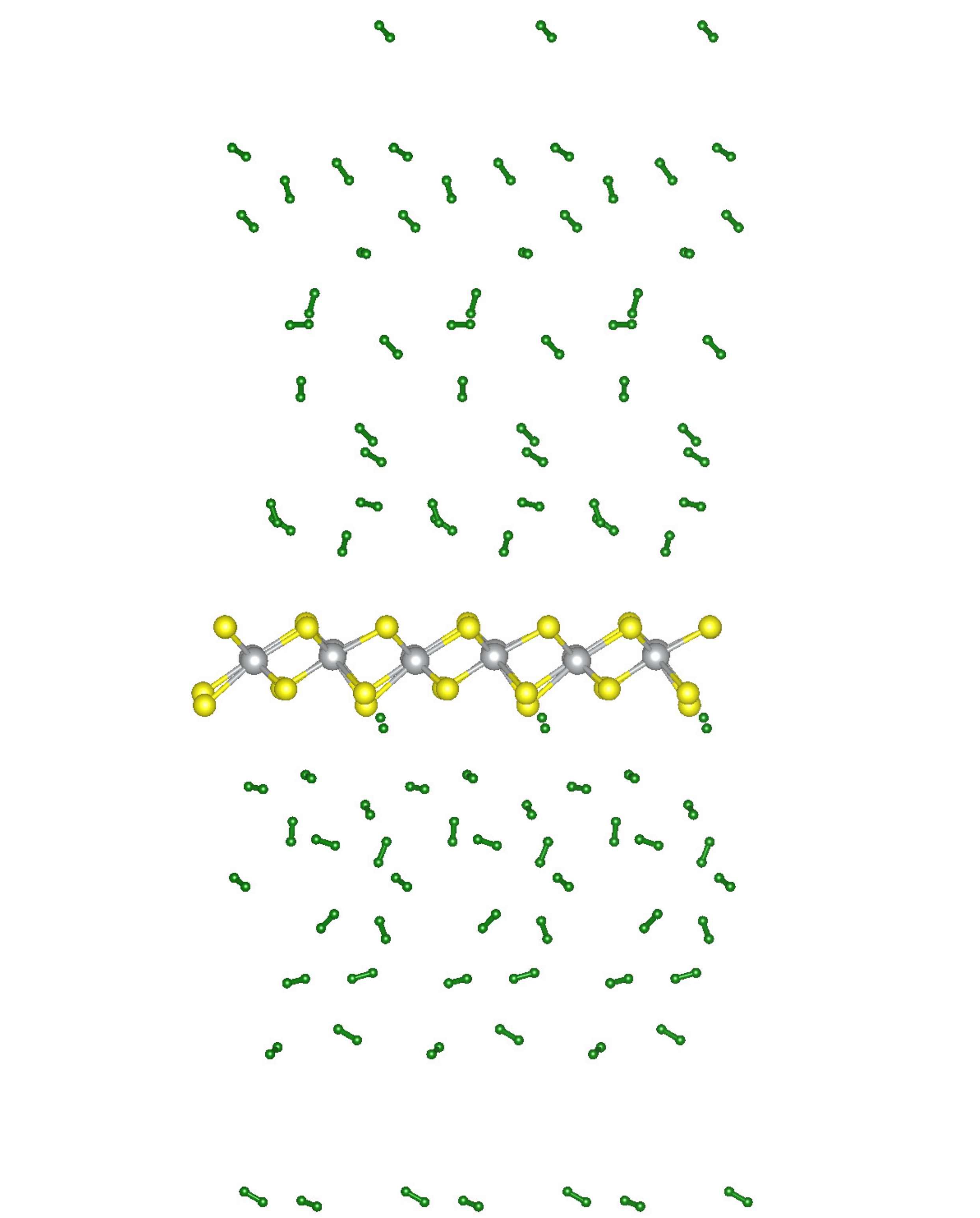}
\end{subfigure}
 \begin{subfigure}[b]{0.32\columnwidth}
                        \subcaption[]{}
\includegraphics[width=\columnwidth,clip=true]{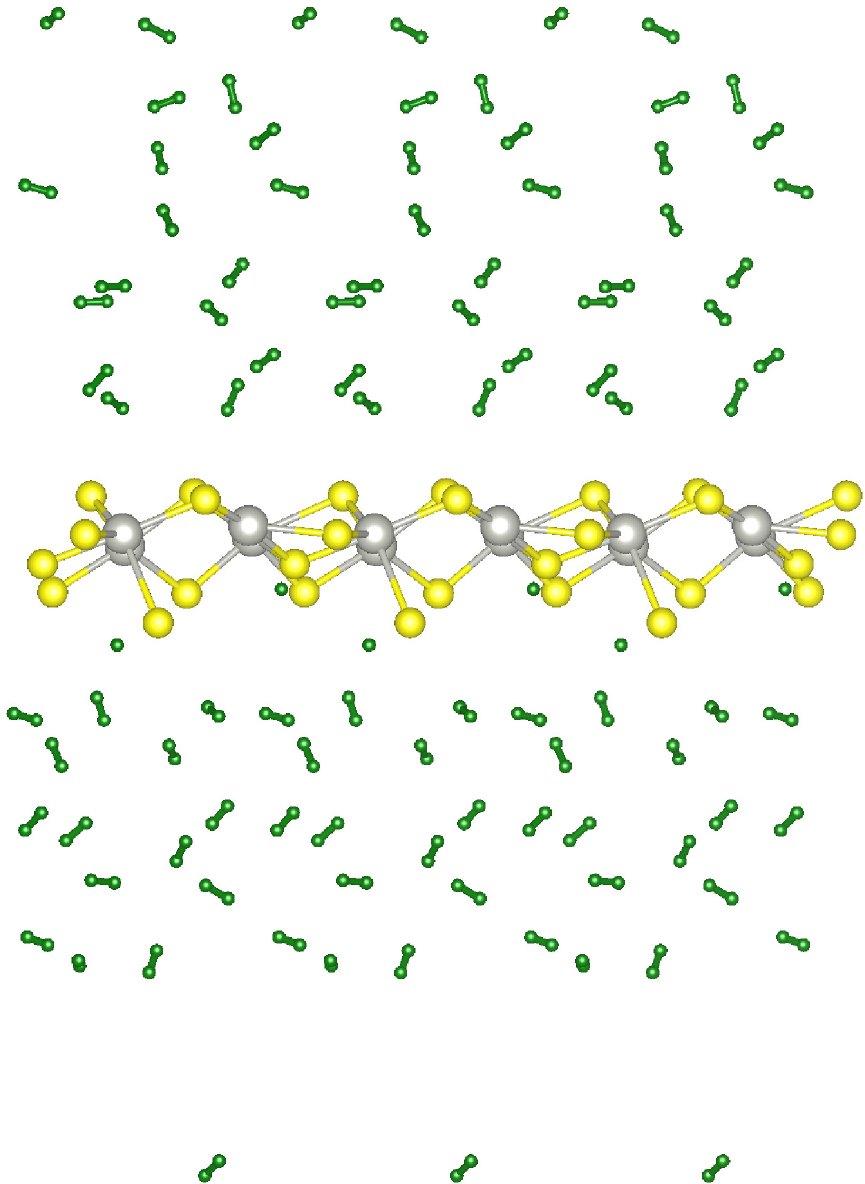}
\end{subfigure}\hfill
 \begin{subfigure}[b]{0.32\columnwidth}
                        \subcaption[]{}
\includegraphics[width=\columnwidth,clip=true]{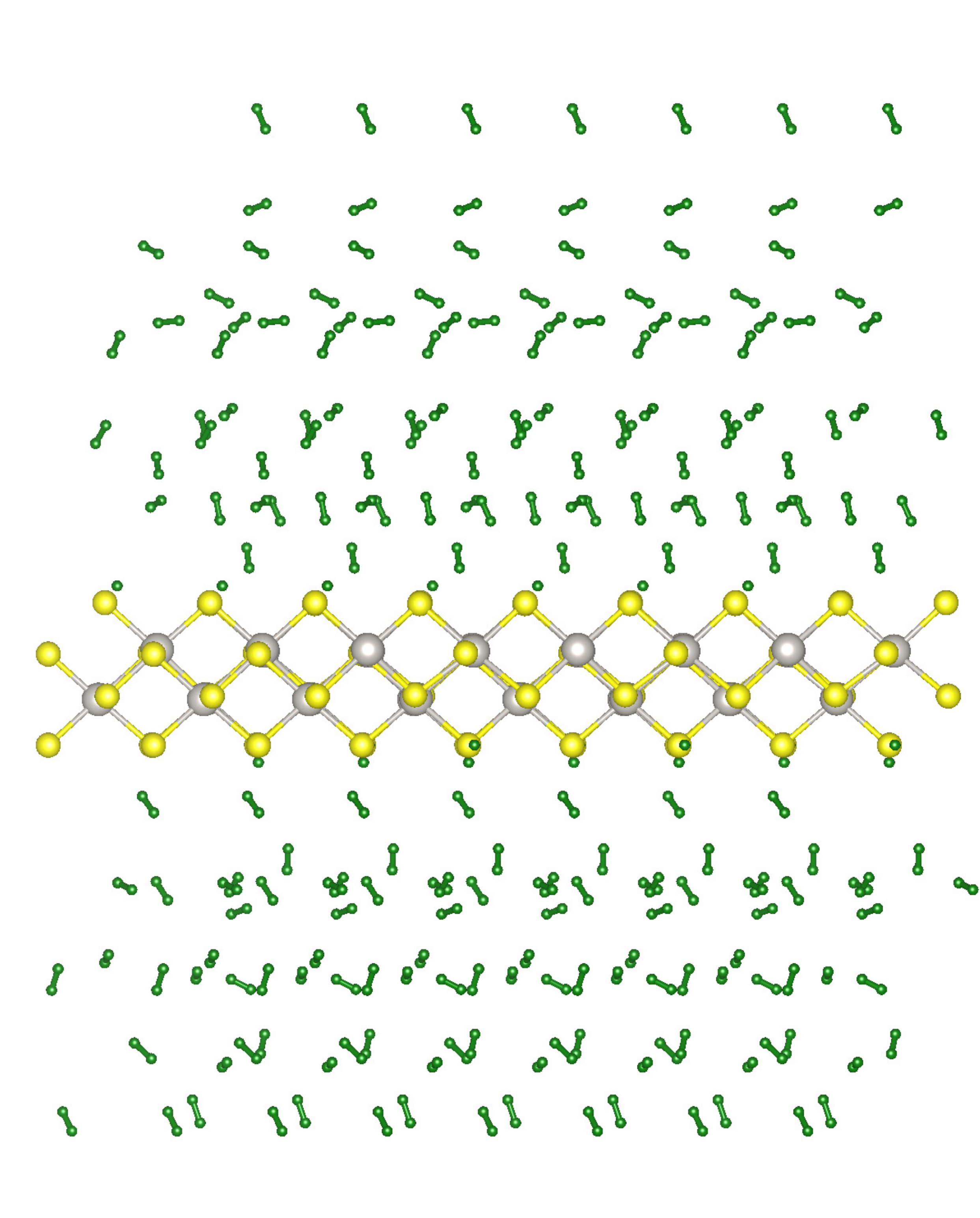}
\end{subfigure}\hfill
 \begin{subfigure}[b]{0.32\columnwidth}
                        \subcaption[]{}
\includegraphics[width=\columnwidth,clip=true]{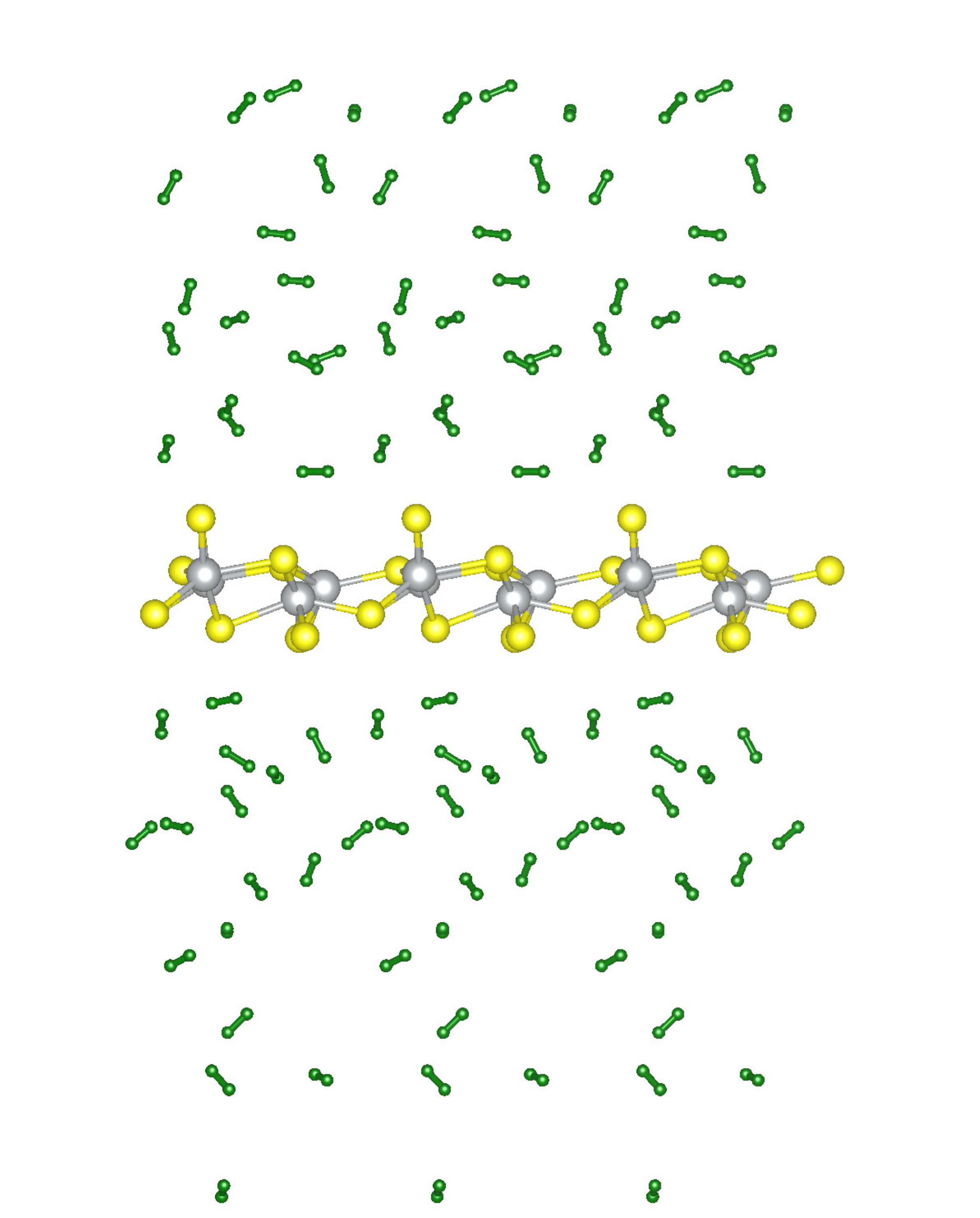}
\end{subfigure}
\caption{AIMD simulations at 300\,K of 32 H$_2$ molecules on a) 1T-$PdS_2$, b) 1T-$PtS_2$, c) 1T-$NiS_2$, d) 2H-$PdS_2$, e) 2H-$PtS_2$ and f) 2H-$PdS_2$.}          
\end{figure}

\begin{figure}
\centering
\begin{subfigure}[b]{0.32\columnwidth}
\subcaption[]{}
\includegraphics[width=\columnwidth,clip=true]{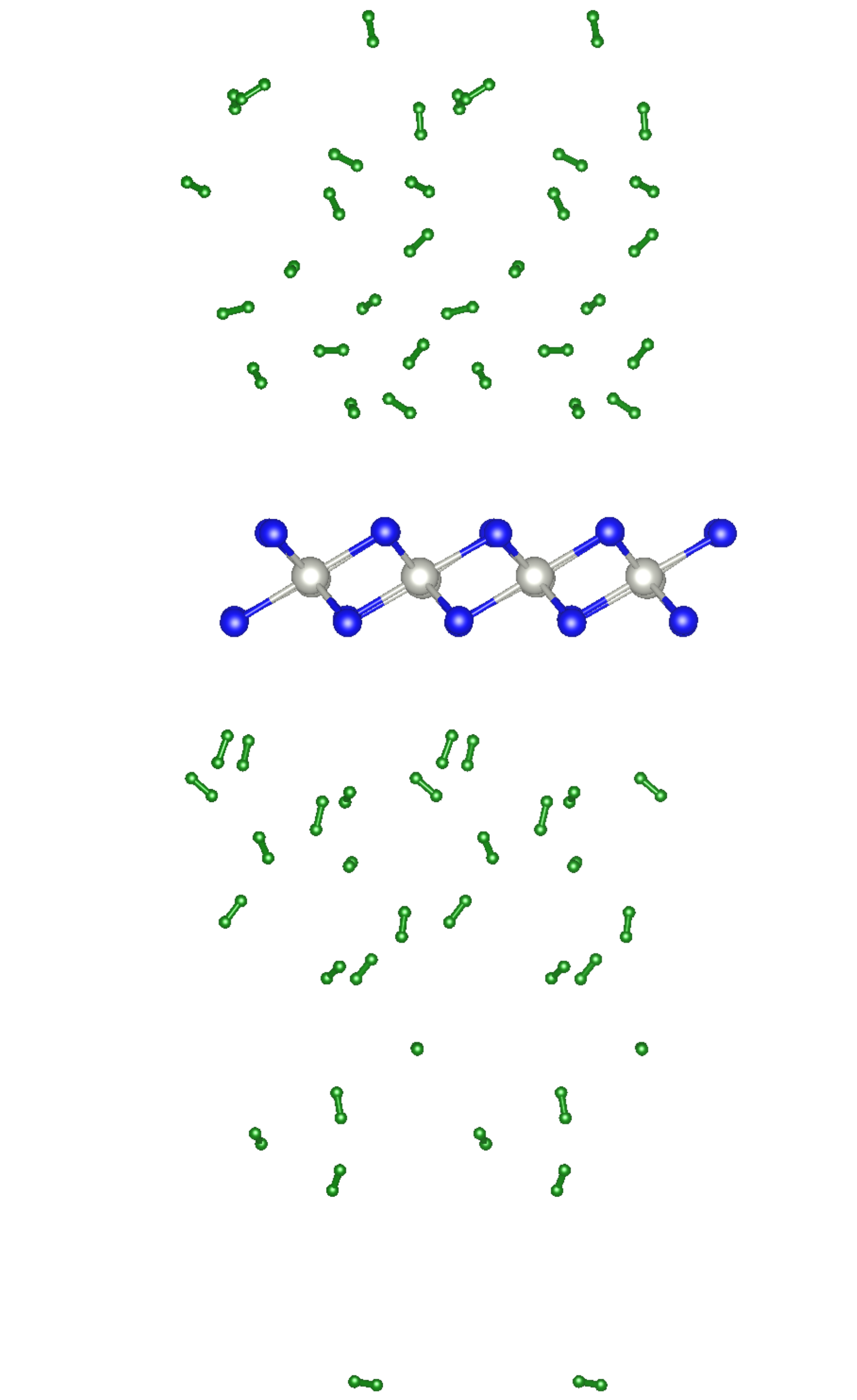}
 \end{subfigure}
 \begin{subfigure}[b]{0.32\columnwidth}
                        \subcaption[]{}
\includegraphics[width=\columnwidth,clip=true]{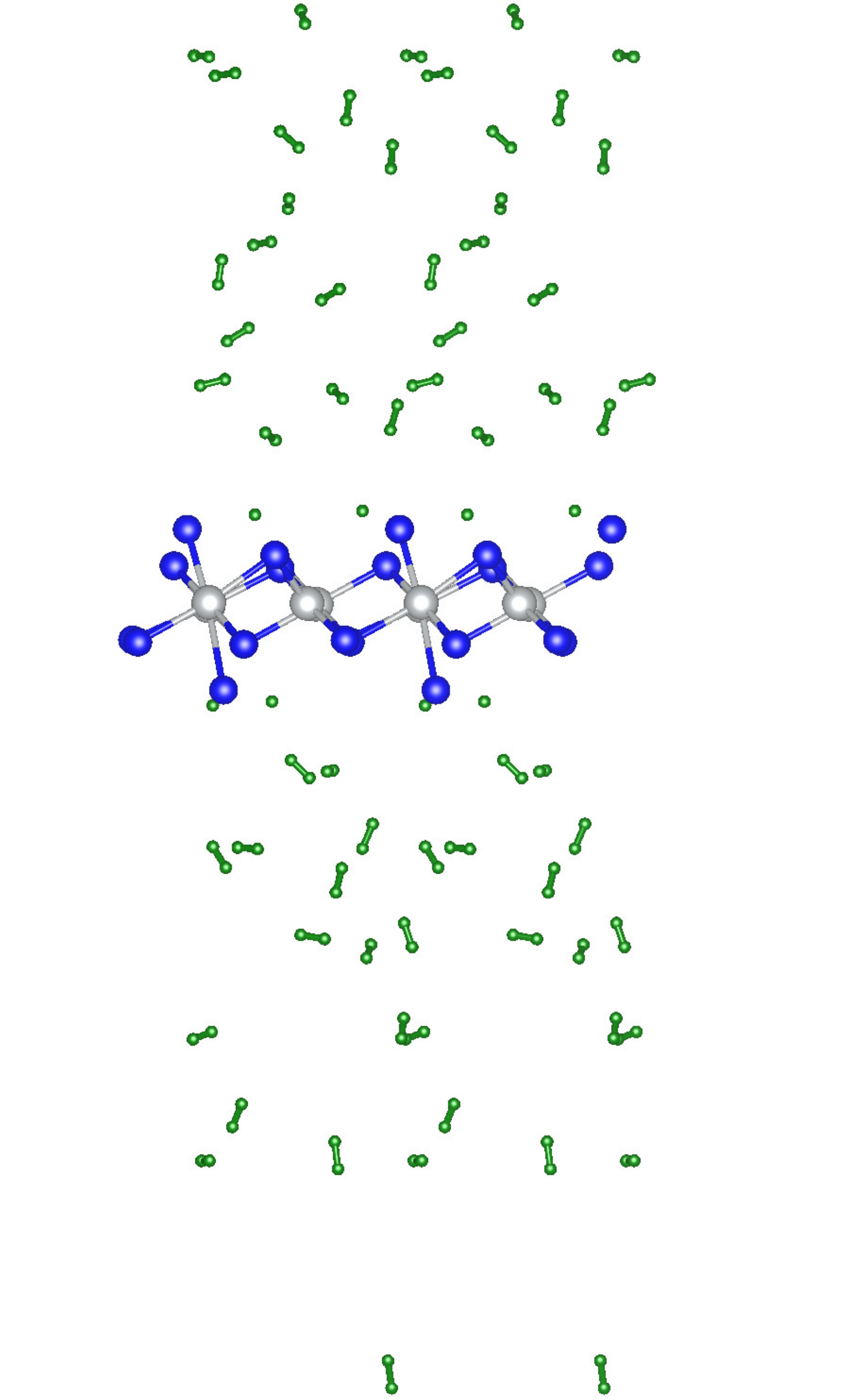}
\end{subfigure}
 \begin{subfigure}[b]{0.32\columnwidth}
                        \subcaption[]{}
\includegraphics[width=\columnwidth,clip=true]{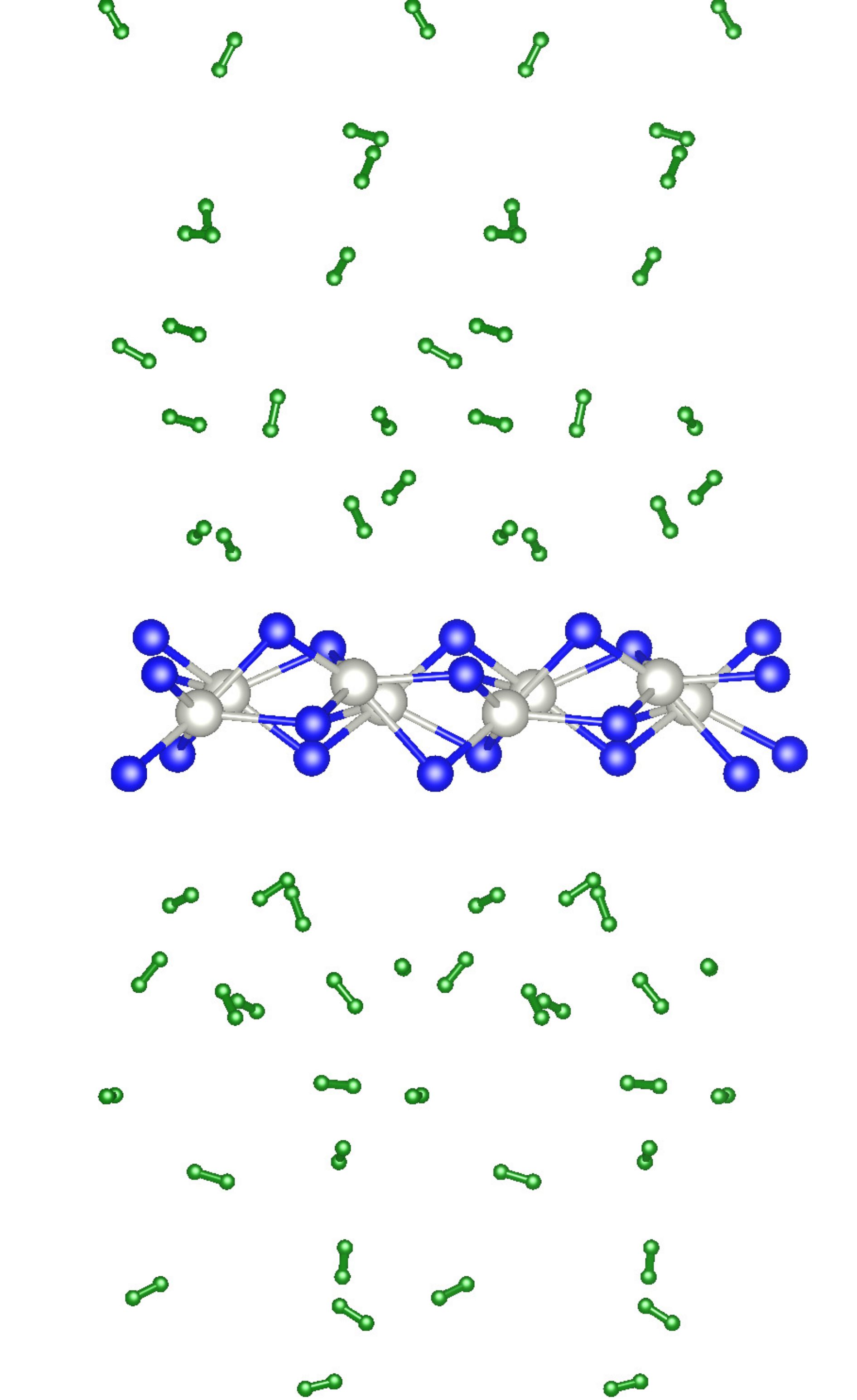}
\end{subfigure}\\
 \begin{subfigure}[b]{0.32\columnwidth}
                        \subcaption[]{}
\includegraphics[width=\columnwidth,clip=true]{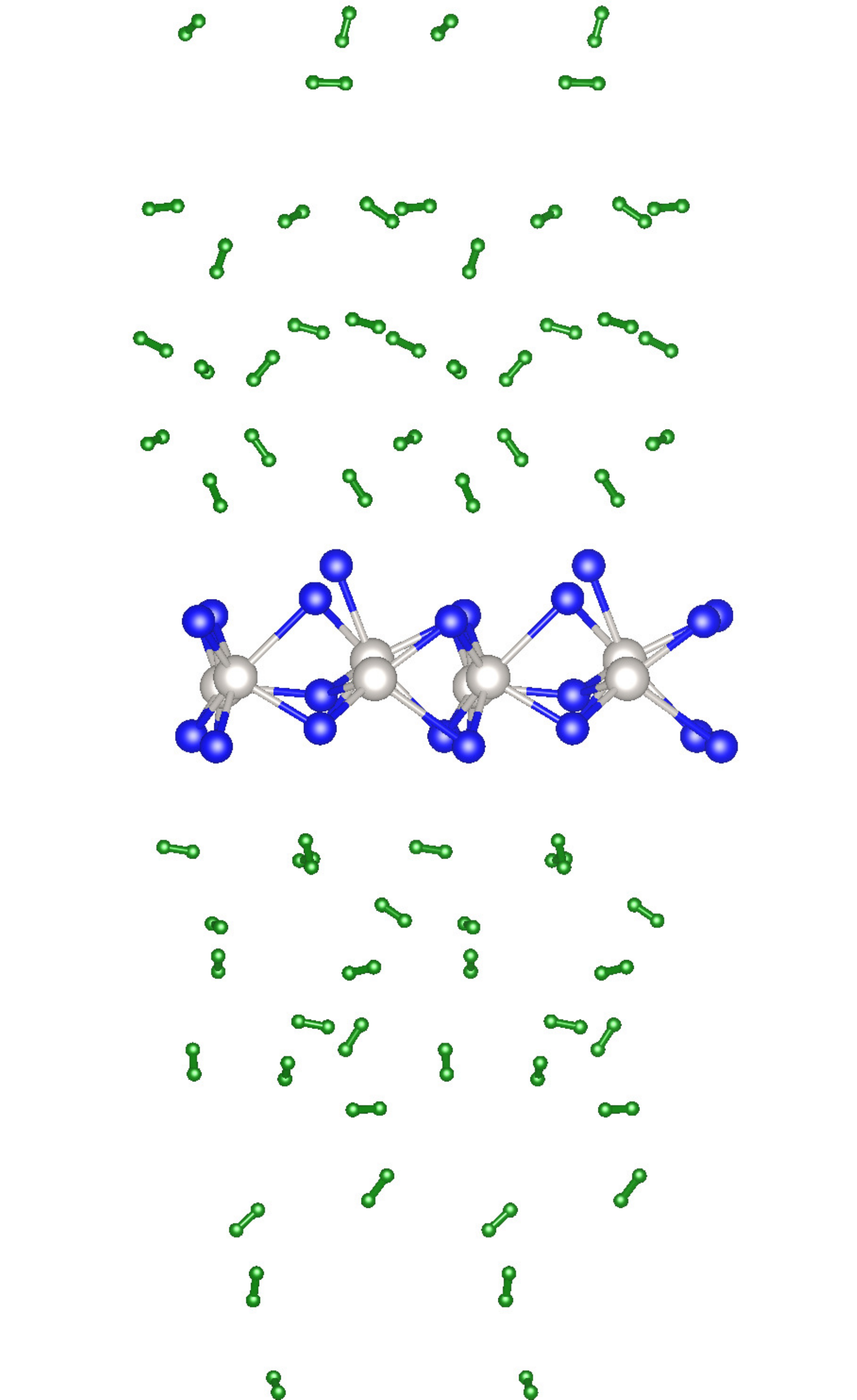}
\end{subfigure}
 \begin{subfigure}[b]{0.32\columnwidth}
                        \subcaption[]{}
\includegraphics[width=\columnwidth,clip=true]{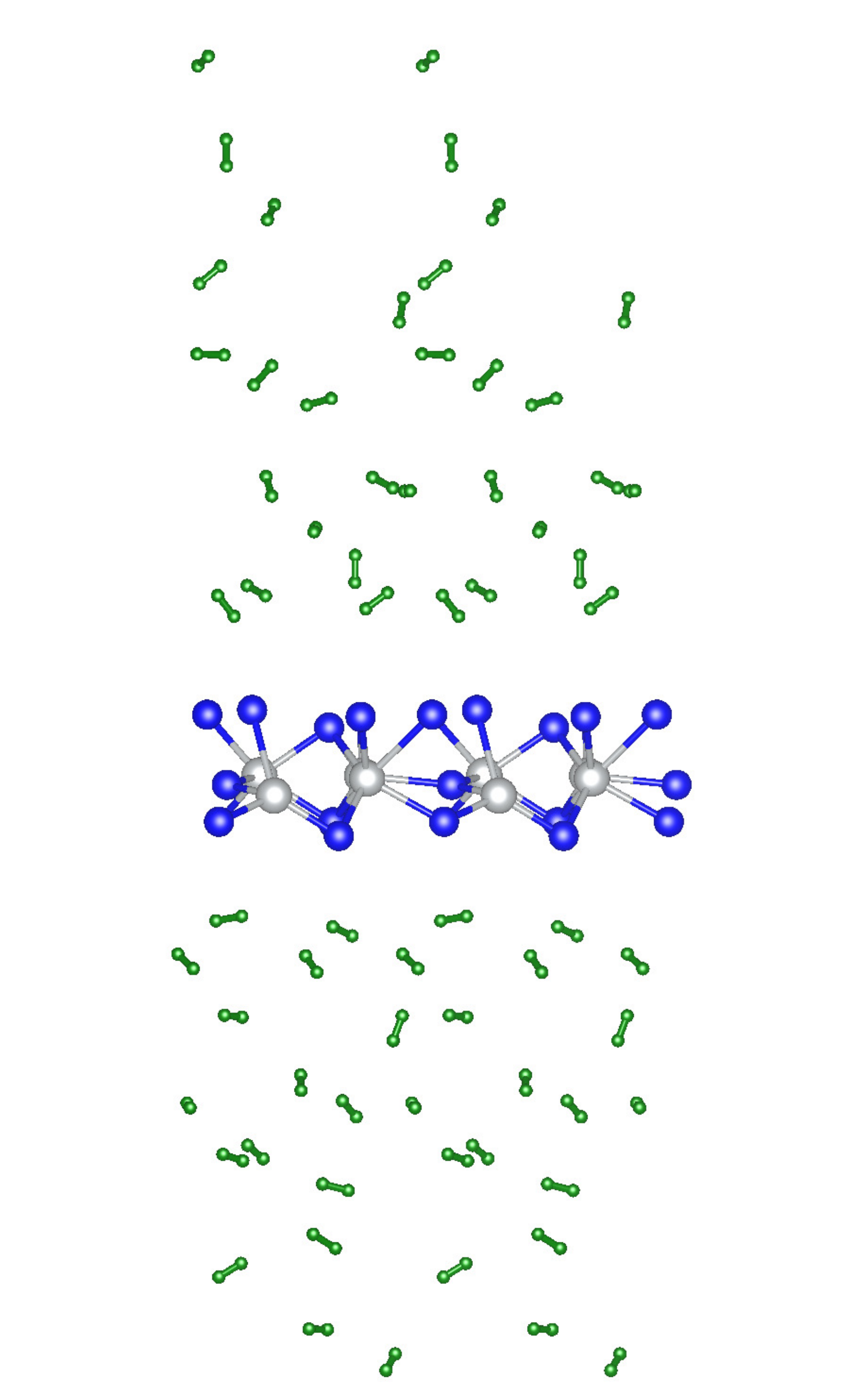}
\end{subfigure}
\caption{AIMD simulations at 300\,K of 32 H$_2$ molecules on a) 1T-$PdSe_2$,  b) 1T-$NiSe_2$, c) 2H-$PdSe_2$, d) 2H-$PtSe_2$ and e) 2H-$PdSe_2$.}   \end{figure}

\begin{figure}
\centering
\begin{subfigure}[b]{0.45\columnwidth}
\subcaption[]{}
\includegraphics[width=\columnwidth,clip=true]{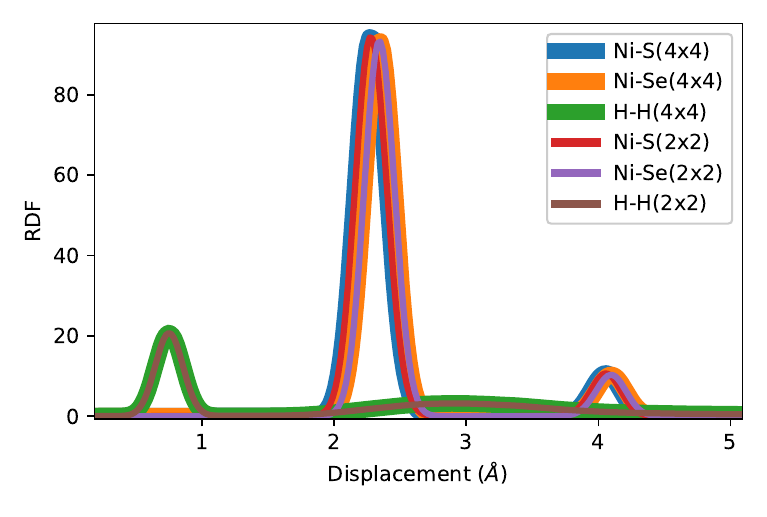}
\end{subfigure}
\begin{subfigure}[b]{0.45\columnwidth}
\subcaption[]{}
\includegraphics[width=\columnwidth,clip=true]{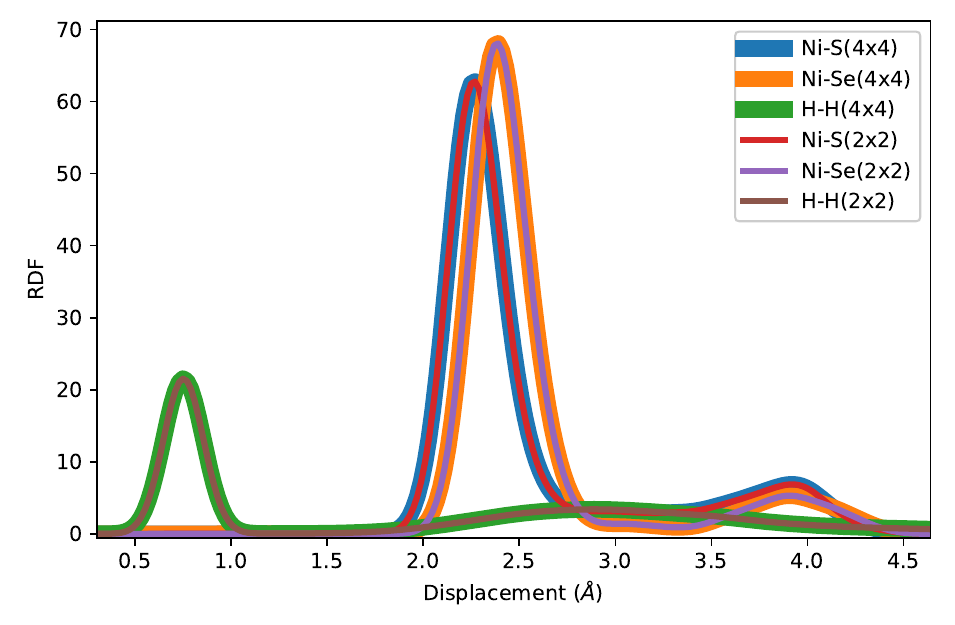}
\end{subfigure}
\caption{RDF of AIMD simulations at 300\,K of 32 H2$_2$ molecules on a) 1T-NiSSe and b) 2H-NiSSe for (4x4) and (2x2) supercells.}
\end{figure}

\clearpage
\newpage

\providecommand{\newblock}{}

\end{document}